\documentclass[aps,prd,twocolumn,
superscriptaddress,floatfix,showpacs,showkeys,preprintnumbers]{revtex4}

\usepackage[utf8]{inputenc}
\usepackage[T1]{fontenc}
\usepackage{slashed}
\usepackage{times}

\usepackage{amssymb,amsfonts,amsmath,amsthm}
\usepackage{dsfont,bbm}
\usepackage{dcolumn}

\usepackage{epsf}
\usepackage{graphicx}
\usepackage[caption=false]{subfig}
\usepackage{dsfont}
\usepackage{slashed}
\usepackage[active]{srcltx}
\usepackage[usenames]{color}

\newcommand{\be}{\begin{equation}}
\newcommand{\ee}{\end{equation}}

\DeclareMathOperator{\Li}{Li}

\newcommand \vev [1] {\langle{#1}\rangle}

\def\II{\hbox{{1}\kern-.25em\hbox{l}}}
\begin{document}


\newcommand{\gpdH}{H}
\newcommand{\gpdE}{E}
\newcommand{\gpdG}{G}
\newcommand{\gpdF}{F}
\newcommand{\gpdtH}{\widetilde{H}}
\newcommand{\gpdtE}{\widetilde{E}}
\newcommand{\gpdtG}{\widetilde{G}}

\newcommand{\cffH}{{\mathcal H}}
\newcommand{\cffE}{{\mathcal E}}
\newcommand{\cffG}{{\mathcal G}}
\newcommand{\cffF}{{\mathcal F}}
\newcommand{\cfftH}{\widetilde{\mathcal H}}
\newcommand{\cfftE}{\widetilde{\mathcal E}}
\newcommand{\cfftG}{\widetilde{\mathcal G}}
\newcommand{\cffbE}{\overline{\mathcal E}}

\newcommand{\cffHbmp}{{\mathfrak H}}
\newcommand{\cffEbmp}{{\mathfrak E}}
\newcommand{\cffGbmp}{{\mathfrak G}}
\newcommand{\cffFbmp}{{\mathfrak F}}
\newcommand{\cfftHbmp}{\widetilde{\mathfrak H}}
\newcommand{\cfftEbmp}{\widetilde{\mathfrak E}}
\newcommand{\cfftGbmp}{\widetilde{\mathfrak G}}
\newcommand{\cffbEbmp}{\overline{\mathfrak E}}

\newcommand{\Csigma}{T^\sigma}
\newcommand{\Cpl}{T^+}
\newcommand{\Cmi}{T^-}
\newcommand{\C}{T}
\newcommand{\bu}{\bar{u}}

\newcommand{\xB}{x_{\rm B}}
\newcommand{\Q}{Q}
\newcommand{\M}{m}
\newcommand{\tK}{\widetilde{K}}
\newcommand{\GeV}{{\rm GeV}}

\newcommand{\GK}{{\it GK12 }}

\newcommand{\alphaRad}{z}
\newcommand{\betaRad}{y}


\title{Deeply Virtual Compton Scattering to the twist-four accuracy:\\
Impact of finite-$t$ and target mass corrections}


\date{\today}

\author{V.M.~Braun}
\affiliation{Institut f\"ur Theoretische Physik, Universit\"at
   Regensburg,D-93040 Regensburg, Germany}
\author{A.N.~Manashov}
\affiliation{Institut f\"ur Theoretische Physik, Universit\"at
   Regensburg,D-93040 Regensburg, Germany}
\affiliation{Department of Theoretical Physics,  St.-Petersburg 
University, 199034, St.-Petersburg, Russia}
\author{D.~M{\"u}ller}
\affiliation{Institut f\"ur Theoretische Physik II, Ruhr-Universit\"at
   Bochum,D-44780 Bochum, Germany}
\author{B.M.~Pirnay}
\affiliation{Institut f\"ur Theoretische Physik, Universit\"at
   Regensburg,D-93040 Regensburg, Germany}

\date{\today}

\begin{abstract}
  \vspace*{0.3cm}
\noindent We carry out the first complete calculation of kinematic power corrections $\sim
t/Q^2$ and $\sim m^2/Q^2$ to several key observables in Deeply Virtual Compton Scattering.
The issue of convention dependence of the leading twist approximation is discussed in detail.
In addition we work out representations for the higher twist corrections in terms of double distributions,
Mellin-Barnes integrals and also within a dissipative framework.
This study removes an important source of uncertainties in the QCD predictions for
intermediate photon virtualities $Q^2\sim 1$-$5\,\GeV^2$ that are accessible in the existing
and planned experiments. In particular the finite-$t$ corrections are significant and
must be taken into account in the data analysis.
 \end{abstract}



\maketitle


\section{Introduction}

Deeply Virtual Compton Scattering (DVCS) is the cleanest process that gives access to
generalized parton distributions (GPDs) \cite{Mueller:1998fv,Radyushkin:1996nd,Ji:1996nm} and is receiving a lot of
attention, see, e.g., the reviews~\cite{Diehl:2003ny,Belitsky:2005qn}. In this process the photon virtuality $Q$  is taken to be the largest scale
which is at least of the order of $1$-$2\,\GeV$.
The existing experimental results come from HERA (H1 \cite{Adloff:2001cn,Aktas:2005ty,Aaron:2007ab,Aaron:2009ac},
ZEUS \cite{Chekanov:2003ya, Chekanov:2008vy}, HERMES \cite{Airapetian:2001yk,Airapetian:2006zr,Airapetian:2008aa,Airapetian:2009aa,Airapetian:2010ab,Airapetian:2011uq,Airapetian:2012mq,Airapetian:2012pg}) at DESY and Jefferson Lab
(CLAS \cite{Stepanyan:2001sm,Chen:2006na,Girod:2007aa,Gavalian:2008aa} and Hall A \cite{Munoz_Camacho:2006hx,Mazouz:2007aa})   and many more measurements are planned after the Jefferson Lab $12$~GeV upgrade and at COMPASS-II at CERN. DVCS plays also a virtual role in the physics case of
proposed collider experiments, the Electron-Ion-Collider at RHIC or JLAB \cite{Deshpande:2012bu} and the Large-Hadron-Electron-Collider at CERN \cite{AbelleiraFernandez:2012cc}.

The standard theoretical framework is based on collinear factorization which is proven in QCD to the
leading power accuracy in the photon virtuality $Q$ \cite{Collins:1998be}. In this approach the DVCS amplitudes are written
as convolutions of perturbatively calculable coefficient functions
and nonperturbative GPDs that represent the nontrivial nucleon structure.
The DVCS coefficient functions have been calculated including the next-to-leading-order (NLO) $\mathcal{O}(\alpha_s)$ corrections
\cite{Belitsky:1997rh,Ji:1997nk,Ji:1998xh,Mankiewicz:1997bk,Pire:2011st},
and the scale-dependence of GPDs is known to the two-loop accuracy \cite{Belitsky:1998gc,Belitsky:1999hf} so that the complete NLO renormalization-group
improved calculation of the amplitudes is possible \cite{Belitsky:1999sg,Freund:2001rk,Kumericki:2007sa}.
Experimental observables --- cross sections and asymmetries --- are obtained from
the amplitudes (either leading order (LO) or NLO) taking into account the interference with purely electromagnetic
Bethe-Heitler (BH) bremsstrahlung process and including the relevant kinematic factors that are usually
taken at face value (not expanded in powers of $1/Q$). This approach, commonly referred to as the
leading twist approximation, appears to be sufficient to describe
unpolarized proton DVCS data \cite{Kumericki:2009uq,Kumericki:2010fr,Kumericki:2011zc}, raising the
hope that a fully quantitative description is within reach \cite{Kumericki:2013br}. The future data will have much higher statistics and
allow one to extract at least some GPDs with controllable precision.

The leading-twist approximation is, however incomplete and in fact convention-dependent.
It is well known that the leading twist DVCS amplitudes do not satisfy electromagnetic Ward identities.
The Lorentz (translation) invariance is violated as well: The results depend on the frame of reference chosen
to define the skewedness parameter and the helicity amplitudes.
In all cases, the required symmetries are restored by contributions that are formally suppressed by  powers
of $1/Q$, dubbed higher-twist corrections.

Such power corrections can be called kinematic as they are expressed in terms of the same GPDs that enter the
leading-twist amplitudes, i.e.~do not involve new nonperturbative input. Their role, from the theory point of view,
is to restore exact symmetries of the theory that are broken in the leading twist approximation and make the
calculation unambiguous. By this reason one can expect that the subset of kinematical power corrections is factorizable
for arbitrary twist.

The relevant twist-three contributions $1/Q$ have been studied in some detail \cite{Anikin:2000em,Penttinen:2000dg,Belitsky:2000vx,Kivel:2000cn,Radyushkin:2000ap}
and it has been shown that kinematic twist-three corrections also restore the invariance under
Lorentz rotations to the $1/Q$ accuracy~\cite{Radyushkin:2001fc}. Such corrections have been evaluated partially also at the NLO~\cite{Kivel:2003jt}.
Phenomenological studies of the size of twist-three effects were attempted by various authors with the generic conclusion
that these corrections are not negligible in the experimental accessible phase space.

Kinematic twist-four effects $1/Q^2$ appear to be more complicated and their structure has been understood
only recently. These contributions correspond to corrections to the DVCS amplitudes of the type
$m^2/Q^2,\,\, t/Q^2$, where $m$  is the target (nucleon) mass and $t=(p'-p)^2$ is
the momentum transfer to the target.
Since the bulk of the existing and expected data is for photon virtualities $Q^2<5$~GeV$^2$, such corrections
may have significant impact on the data analysis and should be taken into account.
The finite-$t$ corrections are of special importance if one wants to access the
three-dimensional picture of the proton in longitudinal and transverse
planes~\cite{Burkardt:2002hr} in which case the $t$--dependence has to be measured in a
sufficiently broad range.

The necessity of taking into account $1/Q^2$ kinematic power corrections to DVCS is widely acknowledged%
~\cite{Belitsky:2005qn,Anikin:2000em,Blumlein:2000cx,Kivel:2000rb,Radyushkin:2000ap,
Belitsky:2000vx,Belitsky:2001hz,Geyer:2004bx,Belitsky:2010jw}.
This task proves to be nontrivial because in addition to Nachtmann-type contributions related to
subtraction of traces in the leading-twist operators $\mathcal{O}_{\mu_1\ldots\mu_n}$ one must take into account their
higher-twist descendants obtained by adding total derivatives: $ \mathcal{O}_1 \sim
\partial^2 \mathcal{O}_{\mu_1\ldots\mu_n}$, and $\mathcal{O}_2 \sim \partial^{\mu_1}\mathcal{O}_{\mu_1\ldots\mu_n}$.
The problem arises because matrix elements of the
operator $\mathcal{O}_2$ on free quarks vanish~\cite{Ferrara:1972xq}. Thus in order to
find its LO coefficient function in the operator product expansion
of two electromagnetic currents one is forced to consider either more complicated
(quark-antiquark-gluon) matrix elements, or stay with the quark-antiquark operators but go over
to the next-to-leading order in $\alpha_s$.
Either way the main challenge is the separation of the contribution of interest from the `genuine'
quark-gluon twist-four operators.

The guiding principle suggested in Ref.~\cite{Braun:2011zr} is that a self-consistent separation
can only be achieved if `genuine', or `dynamical' contributions do not get mixed with the
descendants of the leading-twist operators by the QCD evolution.
Explicit diagonalization of the twist-four mixing matrix (which is a formidable task) can be avoided~\cite{Braun:2011zr,Braun:2011dg}
using conformal symmetry  which implies that LO coefficient functions of
kinematic and genuine twist-four operators are mutually
orthogonal with a proper weight function~\cite{Braun:2009vc}.
Using this approach Braun, Manashov and Pirnay (BMP) calculated
the finite-$t$ and target-mass corrections to DVCS for a
scalar target~\cite{Braun:2012bg} and for a spin-1/2 (nucleon) target~\cite{Braun:2012hq}.
In both cases the restoration of gauge- and translation-invariance to the $\mathcal{O}(1/Q^2)$ accuracy
has been verified and also found that the structure of kinematic corrections proves to be consistent
with collinear factorization.

In a parallel development, following or extending the work in Refs.~\cite{Belitsky:2001ns,Belitsky:2008bz,Belitsky:2010jw}, Belitsky, M{\"u}ller and Ji (BMJ)~\cite{Belitsky:2012ch} suggested
a new decomposition of the Compton hadronic tensor in terms of photon helicity-dependent
Compton Form Factors (CFFs) that are free from kinematical singularities at the edges
of the available phase space. Although the main motivation for this study has been different, namely
to establish the connection of large-$Q^2$ description in terms of GPDs and small-$Q^2$ description in
terms of generalized polarizabilities, the BMJ basis seems to be well suited for the study
of higher twist effects.

In this paper we present the results of the first study of the numerical impact of kinematic twist-three
and twist-four corrections on several key experimental observables in DVCS for the kinematics of the existing (and planned)
measurements. Our calculation incorporates the BMP helicity amplitudes~\cite{Braun:2012hq} and uses
the BMJ CFF decomposition. Convention-dependence of the standard leading twist approximation is emphasized and
illustrated on a few examples.

The presentation is organized as follows. In Sec.~\ref{Sec:electroproduction} we express the electroproduction cross section
in terms of an exact BMJ parametrization of the DVCS amplitude and provide the formulae for some key observables.  Sec.~\ref{Sec:Compton}
contains an analysis of the generic structure of kinematical twist-three and twist-four corrections and the expected size of various contributions.
We also explain and discuss the convention dependence of the leading-twist results.
In Sec.~\ref{Sec:corrections} we present an analysis of kinematic higher twist corrections for a selected set of measured observables,
making use of a popular GPD model \cite{Radyushkin:1998es,Musatov:1999xp}, refined by Goloskokov and Kroll \cite{Goloskokov:2007nt,Goloskokov:2009ia}.
The final Sec.~\ref{Sec:summary} is reserved for a summary and conclusions.

One appendix contains the original result of Ref.~\cite{Braun:2012hq} and explains how to translate it in the conventions of Ref.~\cite{Belitsky:2012ch}.
In the three further appendices we give analytic expressions for the
higher twist contributions  in the double distribution and Mellin-Barnes
integral representations, and also within a dissipative framework.

\section{Electroproduction of photons}
\label{Sec:electroproduction}

The electroproduction of a photon, e.g.,
 off a nucleon target,
\begin{align}
e^\pm(k_1,\lambda_1)  N(p_1,s_1) \to e^\pm(k_2,\lambda_2)  N(p_2,s_2) \gamma(q_2,h_2)\,,
\end{align}
receives contributions of the Bethe-Heitler (BH) brems\-strah\-lung process,
whose amplitude ${\cal T}^{\rm BH}$ is parameterized in terms of {\em two} electromagnetic nucleon form factors,
and the DVCS process
\begin{align}
\gamma^*(q_1,h_1)+ N(p_1,s_1) \rightarrow \gamma(q_2,h_2)+ N(p_2,s_2)\,,
\label{DVCS-process}
\end{align}
described by {\em twelve} complex valued helicity amplitudes ${\cal T}^{\rm DVCS}$, specified below.
The photons have momenta $q_i$ and helicities $h_i$ and the nucleon states the momenta $p_i$ and  polarization vectors $s_i$,
where $i=1(2)$ refers to the initial (final) state.
The full electroproduction amplitude is given by the sum
\begin{align}
{\cal T}  =  {\cal T}^{\rm BH} + {\cal T}^{\rm DVCS}\,.
\end{align}

The five-fold differential cross section in the laboratory frame, where
the incoming electron momentum has a positive $x$-component and the virtual photon
travels along the negative $z$-direction~\cite{Belitsky:2001ns,Belitsky:2008bz,Belitsky:2010jw,Belitsky:2012ch}, can be written as
\begin{align}
d \sigma
=
\frac{\alpha_{\rm em}^3  \xB y^2 } {16 \, \pi^2 \,  \Q^4 \sqrt{1 + \gamma^2}}
\left| \frac{\cal T}{e^3} \right|^2
d \xB d \Q^2 d |t| d \phi d \varphi
\,.
\label{dsigma}
\end{align}
Here $\alpha_{\rm em}=e^2/4\pi$ is the electromagnetic fine structure constant, $\Q^2=-q_1^2$ is the (initial) photon virtuality,
$\xB=\Q^2/2(p_1\cdot q_1)$ the Bjorken scaling variable and $t=(p_2-p_1)^2$ the momentum transfer.
The angle $\phi$ is defined as the azimuthal angle between the leptonic and reaction planes and, in
the case of a transversely polarized nucleon, $\varphi$ is the azimuthal angle of the polarization vector.
Hereafter we use the notation
\begin{align}
       \gamma=2 m\xB/Q\,,
\end{align}
where $m$ is the nucleon mass.
The usual electron energy loss variable $y=p_1\cdot q_1/p_1\cdot k_1$ is related to
the other kinematical variables as $\Q^2= y \xB (s-m^2)$ where $s$ is the center-of-mass energy.
We add that nowadays often another laboratory frame is used, so-called Trento convention, where the azimuthal
angle $\phi^{\rm Trento}$ is related to the adopted here by
\begin{align}
\phi^{\rm Trento}  = \pi - \phi\,.
\label{Trento}
\end{align}

The BH amplitude ${\cal T}^{\rm BH}$ is electron charge even and real valued to the leading order in QED.
The electroproduction amplitude squared appearing in Eq.~(\ref{dsigma}) can therefore be decomposed as
\begin{align}
|{\cal T}|^2= |{\cal T}^{\rm BH}|^2 +  2{\cal T}^{\rm BH} \,{\Re\text{e}}\left[ {\cal T}^{\rm DVCS}\right]  +  |{\cal T}^{\rm DVCS}|^2.
\end{align}
The $|{\cal T}^{\rm BH}|^2$ term is written in terms of the nucleon form factors. The corresponding expression can be found, e.g., in Ref.~\cite{Belitsky:2001ns}.
Most interesting for phenomenology is the interference term that is linear in DVCS amplitudes:
\begin{align}
  {\cal I} =  2{\cal T}^{\rm BH} \,{\Re\text{e}} \left[{\cal T}^{\rm DVCS}\right]\,.
\label{InterferenceTerm2}
\end{align}
${\cal T}^{\rm DVCS}$ is electric charge odd, i.e.~this  contribution has different sign for electron vs.~positron scattering.
The interference term  has a rich angular structure and can be decomposed in unpolarized, longitudinal, and two transversely polarized parts as
\begin{align}
\label{InterferenceTerm}
{\cal I} &=  {\cal I}_{\rm unp}(\phi) + {\cal I}_{\rm LP}(\phi) \cos\theta \notag
\\
&\phantom{={}}+\bigl[{\cal I}_{{\rm TP}+}(\phi) \cos\varphi + {\cal I}_{{\rm TP}-}(\phi) \sin\varphi\bigr] \sin\theta\,,
\end{align}
where $\theta$ is the polar angle of the nucleon polarization vector.
The separate terms ${\cal I}_{\rm S}(\phi)$ for the four polarization options
${\rm S}\in \{{\rm unp},{\rm LP}, {\rm TP}_+,  {\rm TP}_-\}$
are usually written as the harmonic expansion w.r.t.~azimuthal angle $\phi$
of the form
\begin{align}
\label{InterferenceTerm1}
{\cal I}_{\rm S}(\phi)
=
\frac{\pm e^6}{\xB y^3 t {\cal P}_1 (\phi) {\cal P}_2 (\phi)}
\biggl\{&
\sum_{n = 0}^3
c_{n,{\rm S}}^{\cal I}\, \cos(n \phi)
\nonumber\\
+
&\sum_{n = 1}^3  s_{n,{\rm S}}^{\cal I}\, \sin(n \phi)
\biggr\},
\end{align}
where the  $\phi$-dependence of the electron propagators in the BH amplitude is contained in the prefactor $1/{\cal P}_1(\phi) {\cal P}_2(\phi)$ (see e.g.~\cite{Belitsky:2001ns})
and the sign $+(-)$ refers to an electron (positron) beam.
It is usually assumed  that
the lowest $n\in \{0,1\}$ harmonics come from photon helicity conserved processes related to the twist-two CFFs,
the $n=2$ harmonics from longitudinal-to-transverse spin flip contributions that give access to twist-three CFFs,
and the $n=3$ ones from transverse photon helicity flip contributions~\cite{Diehl:1997bu,Belitsky:2001ns}.
This identification is, however, oversimplified \cite{Belitsky:2008bz,Belitsky:2010jw,Belitsky:2012ch}. We will illustrate
below that in reality all helicity amplitudes contribute to any
given harmonic in the interference term.
Contributions of separate CFFs can be disentangled, generally speaking, by considering linear combinations
of the harmonics $c_{n,{\rm S}}^{\cal I}$, $s_{n,{\rm S}}^{\cal I}$ for various polarizations options.
There exist altogether eight ($2\times 4$) independent linear combinations for $n\in \{1,2\}$, only four, however,
exist for $n=3$ as well as for $n=0$.

The DVCS amplitude squared term, $|{\cal T}^{\rm DVCS}|^2$, can be expanded in contributions of unpolarized, longitudinally
and two transversely polarized parts in complete analogy to Eq.~(\ref{InterferenceTerm}), with each part having
a harmonic expansion
\begin{align}
\label{AmplitudesSquared}
 |{\cal T}^{\rm DVCS}_{\rm S}(\phi,\varphi)|^2
=
\frac{e^6}{y^2 \Q^2}  \biggl\{&\sum_{n=0}^2 c^{\rm DVCS}_{n,{\rm S}}(\varphi)\, \cos(n\phi) \notag
\\
 + &\sum_{n=1}^2 s^{\rm DVCS}_{n,{\rm S}}(\varphi)\, \sin(n \phi) \biggr\}.
\end{align}
The $\phi$-independent $n=0$ term in this expression
is given by an incoherent sum of all contributions with and without photon helicity flip, see Eq.~(\ref{Res-Mom-DVCS-0-imp}) below,
the $n=1$ harmonics originate from the interference of longitudinal-to-transverse helicity-flip amplitudes with the helicity-conserved and
transverse helicity-flip ones, and the $n=2$ terms arise from the interference of the helicity-conserved with the transverse helicity-flip contributions.

Starting from the fully differential cross section in Eq.~(\ref{dsigma}) one can construct various observables.
Availability of both electron and positron beams at HERA experiments allows one to separate the interference term in the cross section.
In an unpolarized experiment, for example, one gains access to the four $n\in\{0,\dots,3\}$  $\cos(n\phi)$-harmonics of the interference term by measuring the
cross section difference for $e^+$ and $e^-$,
\begin{align}
\frac{d\sigma_{\rm odd}}{d \xB d \Q^2 d |t| d \phi}&=\frac{1}{2}\left[\frac{d\sigma_+}{d \xB d \Q^2 d |t| d \phi} -\frac{d\sigma_-}{d \xB d \Q^2 d |t| d \phi} \right]
\nonumber\\
&=
-\frac{\alpha_{\rm em}^3} {8\pi \, y t  \Q^4}
\frac{\sum_{n = 0}^3 c_{n,{\rm unp}}^{\cal I}\, \cos(n \phi)}{\sqrt{1 + \gamma^2}\, {\cal P}_1 (\phi) {\cal P}_2 (\phi)},
\label{dsigma_{odd}}
\end{align}
and to the DVCS squared term from the sum
\begin{align}
\frac{d\sigma_{\rm even}}{d \xB d \Q^2 d |t| d \phi}&=\frac{1}{2}\left[\frac{d\sigma_+}{d \xB d \Q^2 d |t| d \phi} + \frac{d\sigma_-}{d \xB d \Q^2 d |t| d \phi} \right]
\nonumber\\
&=
\frac{\alpha_{\rm em}^3  \xB} {8\pi\,  \Q^6 \sqrt{1 + \gamma^2}}   \sum_{n=0}^2 c^{\rm DVCS}_{n,{\rm unp}}(\varphi)\, \cos(n\phi)
\nonumber\\[1mm]
&\phantom{={}}+ \text{BH~cross~section}\,,
\label{dsigma_{even}}
\end{align}
which, however, contains also the BH cross section that may overwhelm the DVCS contribution in the fixed target kinematics.
The corresponding beam charge asymmetry defined as
\begin{align}
\label{A_C(phi)}
A_C(\phi) = \frac{d\sigma_+(\phi) - d\sigma_-(\phi) }{d\sigma_+(\phi) + d\sigma_-(\phi) }\,,
\end{align}
is easier to measure. A drawback is that it depends non-linearly on the DVCS amplitudes because of the denominator.
One can further project the beam charge asymmetry on the various harmonics,
\begin{align}
A^{\cos(n\phi)}_C =  \frac{2-\delta_{n0}}{2\pi}\int_{-\pi}^\pi d\phi \cos(n \phi) A_C(\phi)\,.
\label{A^{cos(nphi)}_C}
\end{align}
The $A^{\cos(n\phi)}_C $ is governed by $c^{\cal I}_{n, {\rm unp}}$, however, because of the $\phi$-dependent denominator
in (\ref{A_C(phi)}), it is contaminated by all other harmonics as well.

In the case that only an electron beam is available, e.g., in JLAB experiments, one can use single spin flip asymmetries to
access the interference term. First note that the beam spin summed electroproduction cross section differs from the charge even cross section in
Eq.~(\ref{dsigma_{even}}) by the interference term
\begin{align}
\frac{d \Sigma_{\rm BS}\, \sigma}{d \xB d \Q^2 d |t| d \phi} &=\frac{1}{2}\left[\frac{d\sigma_-^{\rightarrow} }{d \xB d \Q^2 d |t| d \phi} + \frac{d\sigma_-^{\leftarrow}}{d \xB d \Q^2 d |t| d \phi} \right]
\nonumber\\
&= \frac{d\sigma_{\rm even}}{d \xB d \Q^2 d |t| d \phi}
\nonumber\\
&\phantom{={}}+\frac{\alpha_{\rm em}^3} {8\pi \, y t  \Q^4}
\frac{\sum_{n = 0}^3 c_{n,{\rm unp}}^{\cal I}\, \cos(n \phi)}{\sqrt{1 + \gamma^2}\, {\cal P}_1 (\phi) {\cal P}_2 (\phi)}\,.
\label{Sigma_{BS}}
\end{align}
The BH cross section, taken in QED LO approximation, drops out in the beam spin difference, however, the interference term (\ref{InterferenceTerm2})
is contaminated by a $\sin(\phi)$ modulation of the DVCS cross section,
\begin{align}
\frac{d \Delta_{\rm BS}\, \sigma}{d \xB d \Q^2 d |t| d \phi} &=\frac{1}{2}\left[\frac{d\sigma_-^{\rightarrow} }{d \xB d \Q^2 d |t| d \phi} -\frac{d\sigma_-^{\leftarrow}}{d \xB d \Q^2 d |t| d \phi} \right]
\nonumber\\
&=
\frac{\alpha_{\rm em}^3} {8\pi \, y t  \Q^4}
\frac{\sum_{n = 1}^2 s_{n,{\rm unp}}^{\cal I}\, \sin(n \phi)}{\sqrt{1 + \gamma^2}\, {\cal P}_1 (\phi) {\cal P}_2 (\phi)}
\nonumber\\
&\phantom{={}}+ \frac{\alpha^3\, \xB\, s_{1,{\rm unp}}^{\rm DVCS}\sin(\phi)}{8\pi \,  \Q^6\sqrt{1+\gamma^2}}\,.
\label{Delta_{BS}}
\end{align}
The latter can at least in principle be distinguished from the interference term by means of the $y$-dependence. The single beam spin asymmetry,
defined as
\begin{align}
A_{{\rm LU},\mp}(\phi) = \frac{d\sigma_{\mp}^{\rightarrow} - d\sigma_{\mp}^{\leftarrow}}{d\sigma_{\mp}^{\rightarrow} + d\sigma_{\mp}^{\leftarrow}}\,,
\label{A_{LU}(phi)}
\end{align}
is dominated by the first harmonic, $n=1$, of the interference term. To get rid of both the odd $n=1$ harmonic in the squared DVCS term (\ref{Delta_{BS}})
and of the interference term in the denominator, one defines the charge-odd  beam spin asymmetry
\begin{align}
A_{{\rm LU},{\cal I}}(\phi) =
\frac{\left[d\sigma_{+}^{\rightarrow} - d\sigma_{+}^{\leftarrow}\right]-\left[d\sigma_{-}^{\rightarrow} - d\sigma_{-}^{\leftarrow}\right]
}{
d\sigma_{+}^{\rightarrow} + d\sigma_{+}^{\leftarrow}+d\sigma_{-}^{\rightarrow} + d\sigma_{-}^{\leftarrow}}\,.
\label{A_{LU,I}(phi)}
\end{align}
Nevertheless, in reality the beam spin asymmetries depend non-linearly on all twelve DVCS amplitudes.
The corresponding odd harmonics,
\begin{align}
A^{\sin(n\phi)}_{{\rm LU},\cdots} = \frac{1}{\pi} \int_{-\pi}^\pi\!d\phi\, \sin(n\phi) A_{{\rm LU},\cdots}(\phi)\,,
\label{A^{sin(n phi)}_{{LU},mp}}
\end{align}
appear to be only in approximate correspondence with the harmonics of the interference term~(\ref{InterferenceTerm1}).

At least in principle, there exist a (over)complete set of observables, measurable in unpolarized, single spin and double spin flip experiments
with both $e^+$ and $e^-$ beams, which is sufficient to disentangle the imaginary and real parts of all twelve DVCS amplitudes \cite{Belitsky:2001ns}.
Such an attempt has been undertaken by the DVCS program of the HERMES collaboration and it has been demonstrated recently that these
asymmetry measurements can indeed be mapped into the space of DVCS amplitudes~\cite{Kumericki:2013br}.

It has been very common in the past to parameterize the DVCS amplitude by the expressions that arise from a partonic calculation
(alias leading-twist QCD calculation at LO accuracy) in terms of GPDs. This procedure is, however, ambiguous and the results depend, e.g.,
on the choice of light-like vectors. In order to overcome this ambiguity one has to perform the analysis using a certain
Lorentz-invariant decomposition of the Compton tensor, not bound to a partonic picture that is necessarily convention dependent.
Such a physically motivated parametrization in terms of CFFs was proposed in Ref.~\cite{Belitsky:2001ns}.
Starting from this parametrization, the electroproduction cross section has been calculated recently
by Belitsky, M{\"u}ller and Ji (BMJ)~\cite{Belitsky:2010jw, Belitsky:2012ch}
for all possible polarization options of the initial electron and nucleon. The corresponding analytic expressions are exact
(for massless electrons) and can also be used in the quasi-real photon regime.
To the best of our knowledge the BMJ framework is presently the only complete, consistent, and published calculational scheme; we will be
using it throughout this paper.

The starting point is the DVCS tensor
\begin{align}
\label{Tmunu1}
&T_{\mu\nu}(q_1,q_2,p_1)= \\
&=
i\!\!\int\!\! d^4 x \, {\rm e}^{i(q_1 + q_2) \cdot x/2}
\vev{p_2,\!s_2 | T \{ j_\mu (x/2) j_\nu (- x/2) \} | p_1,\!s_1}\,,
\nonumber
\end{align}
where $\nu$ ($\mu$) refers to the initial (outgoing) photon.
In the following the BMJ reference frame is taken to be the laboratory frame as specified above,
for details see  App.~\ref{App:BMJ-conventions}.
The BMJ photon helicity amplitudes are defined by the contraction of the DVCS tensor with the polarization vectors, given in
Eqs.~(\ref{eps_1(0)})~--~(\ref{eps_2(1)}),
and are further decomposed in terms of the bilinear spinors~\cite{Belitsky:2012ch} as
\begin{align}
\mathcal{T}^{\rm BMJ}_{a\pm} &= \,
 (-1)^{a-1}\epsilon_2^{\nu\ast}(\pm) T_{\nu\mu} \epsilon_1^{\mu}(a)\,,
\notag\\
   & =  \cffH_{a\pm}\,h +  \cffE_{a\pm}\,e \mp   \cfftH_{a\pm}\,\tilde h \mp  \cfftE_{a\pm}\,\tilde e\,.
\label{CFFs}
\end{align}
Here, $a\in \{-,0,+\}$ labels the helicity of the (initial) virtual photon and the bilinear spinors read
\begin{align}
h =& \frac{1}{P\cdot q}\bar u(p_2)\, \slashed{q}u(p_1)\,,  \;\; \;\;\;
e =\! \frac{1}{P\cdot q}\bar u(p_2)\, \frac{i\sigma_{q\Delta}}{2m} u(p_1)\,,
\notag\\
\tilde h =& \frac{1}{P\cdot q}\, \bar u(p_2)\slashed{q}\gamma_5 u(p_1)\,, \;\;
\tilde e =\! \frac{\Delta\cdot q}{P\cdot q}\,\bar u(p_2)\frac{\gamma_5}{2m}  u(p_1)\,,
\label{BMJstructures}
\end{align}
where
\begin{align}
 P=p_1+p_2\,, && \Delta=p_2-p_1\,, && q=(q_1 + q_2)/2
\end{align}
and we use a shorthand notation $\sigma_{q\Delta}=\sigma_{\alpha\beta}q^\alpha\Delta^\beta$.

The coefficients $\cffH_{ab}\,, \ldots, \cfftE_{ab}$ in the decomposition (\ref{CFFs}) are called photon helicity dependent
CFFs.  The CFFs are functions of the invariant kinematic variables, $\xB$, $t$, and $Q^2$.
We will use a generic notation
\begin{align}
\text{$\cffF_{a+}(\xB,t,Q^2)$ with $\cffF \in \{ \cffH,  \cffE, \cfftH, \cfftE\}$, $a\in \{-,0,+\}$.} \label{cffF}
\end{align}
With the sign convention in Eq.~(\ref{CFFs}) one obtains
$$\cffF_{--}=\cffF_{++}\,,\quad \cffF_{+-}=\cffF_{-+}\,, \quad \cffF_{0-}=\cffF_{0+}.$$
Similar to the photon helicity amplitudes themselves, the photon helicity dependent CFFs are
{\em not} Lorentz-invariant quantities; they depend on the chosen (BMJ) reference frame.

The CFFs $\cffH$ ($\cfftH$) and $\cffE$ ($\cfftE$) can be viewed as nonlocal generalizations of the Dirac (axial-vector) and Pauli (pseudo-scalar) form
factor, respectively. They describe, loosely speaking, the proton helicity-conserved and helicity-flip transitions.
QCD collinear factorization provides the following power counting scheme
\begin{align}
\cffF_{++} &\simeq {\cal O }(1/Q^0)\,,
\notag\\
\cffF_{0+}  &\simeq {\cal O }(1/Q)\,,
\notag\\
\cffF_{-+}  &\simeq {\cal O }(1/Q^2)\,,
\label{cff-counting}
\end{align}
which is not quite accurate as the transverse helicity flip CFFs also contain ${\cal O }(1/Q^0)$ terms in higher orders of perturbation theory
induced by the so-called gluon transversity GPDs \cite{Diehl:1997bu,Hoodbhoy:1998vm,Belitsky:2000jk,Diehl:2001pm}. These contributions are not relevant, however, for the subject of this study.

The BMJ  helicity-flip CFFs satisfy certain kinematical constraints that ensure vanishing of some harmonics in the cross section at the phase space boundaries.
These constraints apply to the `electric' and `magnetic' combinations of the CFFs
\begin{align}
\cffG_{ab}       &\equiv \cffH_{ab}+\frac{t}{4m^2}\cffE_{ab}\,,
\notag\\
\mathcal{M}_{ab} &\equiv \cffH_{ab} + \cffE_{ab}\,,
\label{electric-magnetic}
\end{align}
(and similar for $\cfftH_{ab},\cfftE_{ab}$) that are obvious generalizations of the Sachs form factors
(or axial-vector and pseudo-scalar form factors).
In particular, the `electric' CFFs must have the following behavior for $t\to t_{\rm min}$:
\begin{align}
&\cffG_{0+},\,\cfftG_{0+}  \propto (t_{\rm min} -t)^{1/2}\,,
\notag\\
&\cffG_{-+},\,\cfftG_{-+}  \propto (t_{\rm min} -t)^1\,.
\label{constraint-1}
\end{align}
In contrast, the `magnetic' CFFs $\mathcal{M}_{0+},\widetilde{\mathcal{M}}_{0+}$ may contain a square root
singularity 1/$\sqrt{(t_{\rm min} -t)}$, and $\mathcal{M}_{-+},\widetilde{\mathcal{M}}_{-+}$ do not necessarily
vanish.
In addition, the following  constraints
\begin{align}
&\cffH_{0+}+\frac{\xB(1 +\frac{t}{\Q^2})}{2-\xB+ \frac{\xB t}{\Q^2}}\cfftH_{0+} \propto (t_{\rm min} -t)^{1/2},
\notag\\
&\cffH_{-+}+\frac{\xB(1 +\frac{t}{\Q^2})}{2-\xB+ \frac{\xB t}{\Q^2}}\cfftH_{-+} \propto (t_{\rm min} -t)^{1},
\label{constraint-2}
\end{align}
and the similar ones for  $\cffH, \cfftH  \to \cffE,\cfftE $ have to be satisfied~\cite{Belitsky:2012ch}.
From these four combinations for longitudinal (or transverse helicity) flip, three are independent. A forth independent
combination, suggested by the BMP result, is quoted in App.~\ref{App:analyticity}.

The harmonic coefficients of the interference (\ref{InterferenceTerm1}) and DVCS amplitude squared (\ref{AmplitudesSquared}) term
that are directly related to experimental observables,
e.g., Eqs.~(\ref{dsigma_{odd}})--(\ref{A^{sin(n phi)}_{{LU},mp}}), can be calculated in terms of linear and bilinear combinations of CFFs (\ref{cffF}).
The power counting  scheme, given in Eq.~(\ref{cff-counting}), implies that the $n=1$ harmonics $c^{\cal I}_{1, \rm S}$ and $s^{\cal I}_{1, \rm S}$
of the interference term~(\ref{InterferenceTerm1}) provide the dominant contributions  in the DVCS regime.
For an unpolarized nucleon these harmonics are given to the leading twist-two accuracy by the following linear combinations
\begin{widetext}
\begin{align}
\label{Res-IntTer-unp}
\begin{Bmatrix}c^{\cal I}_{1, \rm unp} \\ s^{\cal I}_{1, \rm unp}\end{Bmatrix}
&=
\frac{8\tK\sqrt{1-y-\frac{y^2 \gamma^2}{4}}}{\Q(1+\gamma^2)^2}
\begin{Bmatrix}  -\left[2-2 y+y^2\big(1+\frac{\gamma^2}{2}\big)\right] \\ \lambda y (2-y) (1+\gamma^2) \end{Bmatrix}
\begin{Bmatrix}\Re{\rm e} \\ \Im{\rm m}  \end{Bmatrix} {\cal C}^{\cal I}_{\rm unp}(\cffF_{++}) + {\cal O}(1/\Q^2)\,,
\end{align}
\end{widetext}
where $\lambda = \pm 1$ is the electron polarization (helicity),
\begin{align}
{\cal C}^{\cal I}_{\rm unp}(\cffF)=
 F_1 \cffG +(F_1 + F_2)\left[\frac{\xB(1+\frac{t}{\Q^2})}{2 - \xB+\frac{\xB t}{\Q^2}} \cfftH-\frac{t}{4\M^2}  \cffE \right],
\label{{cal C}^{I}_{unp}}
\end{align}
$F_1(t)$ and $F_2(t)$ are the Dirac and Pauli proton form factors, and $\tK = \mathcal{O}(Q^0)$ is a kinematical factor
which has mass dimension one. This factor, defined in Eq.~(\ref{tK}), vanishes at the momentum transfer boundaries $t=t_{\rm min}$ and $t=t_{\rm max}$,
\begin{align}
t_{\rm min/max} \equiv -\Q^2 \frac{2 (1 - \xB) \left( 1 \mp \sqrt{1 + \gamma^2}\right)+ \gamma^2}{4 \xB (1 - \xB) + \gamma^2}\,,
\label{t_{min/max}}
\end{align}
{}[upper (lower) sign correspond to the minimal (maximal) allowed value $-t_{\rm min}$ ($-t_{\rm max}$)] as well as at the maximal allowed value of
Bjorken variable $x_{\rm B\, max}(t,\Q^2)$, see discussion  of Eq.~(10) in Ref.~\cite{Belitsky:2012ch}.

The linear combination (\ref{{cal C}^{I}_{unp}}) of CFFs  is written
in such a manner that the kinematical constraints (\ref{constraint-1}) and (\ref{constraint-2}) are implemented.
The omitted terms $\mathcal{O}(1/Q^2)$ in Eq.~(\ref{Res-IntTer-unp}) contain
the contributions of the helicity-flip CFFs and some further kinematical corrections in which it is also  ensured that the
kinematical singularities in $\cffF_{0+}$ are explicitly canceled.
The complete formula for the unpolarized $n=1$ odd harmonic (\ref{Res-IntTer-unp}) is provided below in Eq.~(\ref{s^{I}_{1,{unp}}-exact}).
Note that for typical DVCS kinematics ($\xB \ll 1$, $-t \ll 4\M^2$) the expression for
${\cal C}^{\cal I}_{\rm unp}(\cffF)$ in Eq.~(\ref{{cal C}^{I}_{unp}}) is dominated by the
first term which involves the `electric' combination $\cffG$ of the CFFs (\ref{electric-magnetic}).
Similar expressions can be derived for a polarized target;
they can be found in Sec.~2.3 of Ref.~\cite{Belitsky:2012ch}.
However, only the unpolarized result (${\cal C}^{\prime \cal I}_{\rm unp}$ in the notations of~\cite{Belitsky:2012ch})
is presently available in a compact and explicitly kinematical singularity-free form.

The main contribution to the cross section of the DVCS amplitude squared term (\ref{AmplitudesSquared}) comes from the constant $n=0$ harmonics,  e.g., for an unpolarized target one obtains the expression
\begin{widetext}
\begin{align}
c_{0,\rm unp}^{\rm DVCS}
=
2 \frac{2-2 y + y^2+\frac{\gamma^2}{2} y^2}{1+\gamma^2}\left\{
\biggl[{\cal C}_{\rm unp}^{\rm DVCS} (\cffF_{++},\cffF_{++}^\ast)
+ {\cal C}_{\rm unp}^{\rm DVCS} (\cffF_{-+},\cffF_{-+}^\ast)\biggr]
+ 2\varepsilon(y)\, {\cal C}_{\rm unp}^{\rm DVCS}({\cal F}_{0+}, {\cal F}_{0+}^\ast)\right\},
\label{Res-Mom-DVCS-0-imp}
\end{align}
where ${\cal C}_{\rm unp}^{\rm DVCS}$ stand for the bilinear combinations of CFFs
\begin{align}
\label{Def-CDVCSunp}
{\cal C}_{\rm unp}^{\rm DVCS}(\cffF,\cffF^\ast) =\frac{4}{\left(2-\xB + \frac{\xB t}{\Q^2}\right)^2}\Biggl[&
(1-\xB)\left(1+\frac{\xB t}{\Q^2}\right) \left[\cffG \cffG^\ast + \cfftG \cfftG^\ast\right]+ \left(2+\frac{t}{\Q^2}\right) \frac{\xB^2\,\M^2}{\Q^2}\cfftG \cfftG^\ast
 \nonumber\\
&+\frac{\tK^2}{4 \M^2}
\left\{\cffG \cffE^\ast + \cffE \cffG^\ast +\cfftG \cfftE^\ast + \cfftE \cfftG^\ast
+\frac{4 \M^2-t}{4 \M^2} \cffE \cffE^\ast- \frac{t}{4 \M^2} \cfftE \cfftE^\ast
\right\} \Biggr]
\end{align}
\end{widetext}
and the ratio of longitudinal to transversal photon flux is
\begin{align}
\varepsilon(y)= \frac{1-y - \frac{\gamma^2}{4} y^2}{1-y + \frac{1}{2} y^2+\frac{\gamma^2}{4} y^2}.
\label{varepsilon}
\end{align}
For a typical DVCS experiment  $\tK^2 \ll 4 m^2$. In this case, taking into account the power counting rules (\ref{cff-counting}),
$c_{0,\rm unp}^{\rm DVCS}$ is dominated at large $Q^2$ by the helicity conserving `electric' CFFs $\cffG_{++}$ and $\cfftG_{++}$.

The $n=0$ harmonic (\ref{Res-Mom-DVCS-0-imp}) is
formally suppressed by an additional factor $\mathcal{O}(1/Q)$ as compared to the interference term, e.g.,
for the unpolarized case one infers from  Eqs.~(\ref{dsigma_{odd}}), (\ref{dsigma_{even}}), (\ref{Res-IntTer-unp}),
and (\ref{Res-Mom-DVCS-0-imp}) the relative factor $y t {\cal P}_1(\phi){\cal P}_2(\phi)/\tK \Q \sim \mathcal{O}(1/Q)$.
Note that the interference term can get weakened by integration over $\phi$  and that there is no
$1/\Q$-suppression if we compare the $n=0$ harmonic (\ref{Res-Mom-DVCS-0-imp}) with those of the interference term.

The $n=1$ harmonics in (\ref{AmplitudesSquared}) originate from the interference of longitudinal helicity flip CFFs $\cffF_{0+}$ with the transverse ones
and the $n=2$ harmonics arise from the interference of $\cffF_{-+}$ with $\cffF_{++}$. All of these harmonics  can be
expressed in terms of bilinear combinations of the CFFs, similar to Eq.~(\ref{Def-CDVCSunp}), and are listed in Sec.~2.2 of Ref.~\cite{Belitsky:2012ch}.
The power counting scheme (\ref{cff-counting}) implies that these harmonics are formally suppressed by $1/\Q^2$ as
compared to the corresponding ones of the interference term.

To summarize, although the power counting in Eq.~(\ref{cff-counting}) suggests that the properly chosen experimental observables
are dominated by one particular CFF (e.g.~the $n=1$ harmonics of the interference term by photon helicity conserved and the $n=2$  harmonics
by the longitudinal-to-transverse helicity flip CFFs), exact expressions are rather intricate and contain contributions of all remaining CFFs as well.
In the data analysis that is not restricted to the formal large $Q^2$ limit that, we believe, is not appropriate for both the existing and the expected
future data, all such subleading contributions have to be taken into account. The point that we want to stress here is that the \emph{definition}
of the CFFs themselves is ambiguous to the $1/Q$ accuracy; this ambiguity is resolved at the level of physical observables only, in the sum of
all contributions. Similarly, kinematical singularities in the helicity dependent CFFs cancel each other in the exact expressions for the amplitudes
which can be rather lengthy.

Last but not least, we want to note that
in present DVCS phenomenology only the non-flip CFF $\cffH_{++}$ can be accessed from the $n=1$ even and odd harmonics in
unpolarized experiments \cite{Kumericki:2009uq} and its parity-odd analog $\cfftH_{++}$ is constrained by measurements on longitudinal
polarized target \cite{Guidal:2010de,Kumericki:2013br}.
The nucleon helicity flip contributions, $\cffE_{++}$ or $\cfftE_{++}$, are essentially not constrained at all \cite{Kumericki:2013br}.
Furthermore, it is generally accepted that the photon helicity flip contributions,
which are suppressed, are compatible with zero within the present day experimental errors.

\section{Power corrections to Compton form factors}
\label{Sec:Compton}
\subsection{Partonic description of DVCS and beyond}
\label{Sec:description}

The parton model corresponds to the LO QCD perturbative calculation to leading twist-two accuracy.
At this level there are four CFFs $\cffF \in \{\cffH,\cffE,\cfftH,\cfftE\}$ that are given by
convolution integrals of GPDs $\gpdF\in \{\gpdH,\gpdE,\gpdtH,\gpdtE\}$ over the momentum fraction $x$ with simple coefficient functions,
\begin{align}
\cffF & \stackrel{\rm LO}{=}  \sum_q e_q^2 \int_{-1}^1\!\! dx\! \left[\frac{1}{\xi-x -i\epsilon} - \frac{\sigma(F)}{\xi+x- i\epsilon} \right]\! \gpdF^q(x,\xi,t)\,,
\label{DVCS-LO}
\end{align}
with an obvious correspondence
$$\cffH \leftrightarrow \gpdH\,,\;\;\cffE \leftrightarrow \gpdE\,,\;\; \cfftH \leftrightarrow \gpdtH\,,\;\;\cfftE \leftrightarrow \gpdtE\,.$$
Here and below we assume that the GPDs are defined with the established conventions, e.g., given in \cite{Diehl:2003ny},
\begin{align}
\sigma(\gpdH)=\sigma(\gpdE)=1 \quad  \mbox{and}\quad \sigma(\gpdtH)=\sigma(\gpdtE)=-1,
\label{sigma}
\end{align}
is a signature factor, $\xi\simeq \xB/(2-\xB) $ is the skewedness variable,
and  $e_q$ are the fractional quark charges. The scale dependence of the GPDs is not shown for brevity.
To the NLO accuracy the coefficient functions are modified by $\mathcal{O}(\alpha_s)$ corrections and become
more complicated. Such corrections are not relevant for the present study, we ignore them in what follows.

Note that only charge conjugation even $C = +1$ combinations of the GPDs
\begin{align}
\gpdF^{q^{(+)}}(x,\xi,t) = \gpdF^{q}(x,\xi,t) -\sigma(F)\, \gpdF^{q}(-x,\xi,t)\,
\label{gpd-Ceven}
\end{align}
can contribute to the DVCS,  which is reflected in Eq.~(\ref{DVCS-LO}) by the (anti)symmetrization of the coefficient function in $x$.
Using this symmetry we can rewrite (\ref{DVCS-LO}) as
\begin{align}
\cffF & \stackrel{\rm LO}{=} \sum_q e_q^2 \int_{-1}^1\!\! \frac{dx}{2\xi}\,
\C_{0}\biggl(\!\frac{\xi+x- i\epsilon}{2(\xi-i\epsilon\!)}\biggr)
 \gpdF^{q^{(+)}}(x,\xi,t)
\notag\\
  & \stackrel{\rm LO}{\equiv}  \C_0\!\circledast\!\gpdF\,,
\label{T-convolution}
\end{align}
where the (anti)symmetrized kernel is replaced by
\begin{align}
 \C_0(u) &=\, \frac{1}{1-u}
\end{align}
and in the second line we have introduced
a notation `$\circledast$' for the (normalized) convolution integral, including the sum over the quark flavors.

If the QCD calculation is done to the ${1/Q^2}$ accuracy, the following complications occur and must be taken into account:
\begin{itemize}
\item{} The skewedness parameter $\xi$ must be \emph{defined} with a power accuracy
\begin{align}
   \xi \to \xi(x_B,t,Q^2) = \frac{x_B}{2-x_B} + \mathcal{O}(1/Q^2)\,,
\end{align}
\item{} The CFFs must be \emph{defined} through a certain decomposition of the DVCS tensor (\ref{Tmunu1}). The BMJ decomposition (\ref{CFFs})
        is one possibility; the BMP decomposition discussed below is another valid option. In both cases the LO CFFs (\ref{DVCS-LO}) are
        recovered as the scaling limit of the helicity-conserving CFFs, that is
\begin{align}
 {}\hspace*{0.6cm}\mathcal{F}_{++} &= \C_0\!\circledast\!\gpdF\Big|_{\xi\to \xi(x_B,t,Q^2)}  + \mathcal{O}(1/Q, 1/Q^2),
\end{align}
 where the expression for the $\mathcal{O}(1/Q, 1/Q^2)$ addenda depends both on the chosen form factor decomposition
(e.g.~BMJ vs.~BMP) and on the convention used for the skewedness parameter.
\item{} There are eight more CFFs $\mathcal{F}_{0+},\mathcal{F}_{-+} $ corresponding to photon helicity flip transitions that must be taken into account
        in the same approximation.
\end{itemize}
In what follows we discuss the convention dependence of various elements in this setup in some detail. It is important to realize
that the corresponding ambiguities only cancel at the level of physical observables.

In the literature the skewedness variable $\xi(\xB,t,Q^2)$ is defined in various manners. This ambiguity is related to the choice of the
reference frame in which one performs the calculation, see a discussion in Ref.~\cite{Belitsky:2005qn}.
The KM convention, used by Kumeri{\v c}ki and M\"uller in global DVCS fits, is
\begin{align}
  \xi_{\rm KM} = \frac{\xB}{2-\xB}\,.
\end{align}
It is known that the Vanderhaeghen-Guichon-Guidal (VGG) convention, used by Guidal, for local CFF fits is practically not
very different from the KM one, a discussion for scalar target can be found in \cite{Belitsky:2008bz}, and those used by
Kroll, Moutarde, and Sabatie in \cite{Kroll:2012sm}. All these definitions are motivated by using a certain generalization
of the standard DIS reference frame where the initial \emph{photon} and \emph{proton}  momenta form the longitudinal plane.
In contrast to this traditional approach,
BMP~\cite{Braun:2012bg,Braun:2012hq} define the longitudinal plane as spanned by the \emph{two photon} momenta
$q_1$ and $q_2$, see  App.~\ref{BMP-notation}.
For this choice the momentum transfer to the target $\Delta = q_1-q_2$ is
purely longitudinal and both --- initial and final state --- protons have the same nonvanishing transverse momentum $P_{\perp}$,
\begin{align}
  |\xi P_\perp|^2 = \frac{1-\xi^2}{4} (t_{\rm min}-t)\,,\quad  t_{\rm min} = -\frac{4 m^2 \xi^2}{1-\xi^2}\,,
\label{xiPperp}
\end{align}
where $\xi = \xi_{\rm BMP}$ is the BMP skewedness parameter defined
with respect to the real (final state) photon momentum $q_2^2=0$:
\begin{align}
  \xi_{\rm BMP} = \frac{p_1\cdot q_2 - p_2\cdot q_2}{p_1\cdot q_2 + p_2\cdot q_2} = \frac{\xB(1+t/Q^2)}{2-\xB(1-t/Q^2)}
\label{xB2xiBMP}
\end{align}
and $t_{\rm min}$ is exactly equivalent to the expression (\ref{t_{min/max}}).
Consequently, the condition $|P_\perp |^2 \ge 0$ translates to the lower bound for the negative momentum transfer square, $ -t \ge -t_{\rm min}.$

The BMP choice is advantageous in two respects. First, it is easy to convince oneself that most contributions to the
longitudinal-to-transverse helicity flip amplitudes (\ref{A0}) and the transverse flip amplitudes (\ref{Aflip})
are proportional to the first and the second power of $|\xi P_\perp| \propto \sqrt{t_{\rm min}-t}$, respectively, and also the remaining terms are
compatible with the expected threshold behavior (\ref{constraint-1}) and (\ref{constraint-2}).
Second, as shown in Ref.~\cite{Braun:2012bg}, the DVCS amplitudes on a
scalar target have an expansion in $t/Q^2$ and $|\xi P_\perp|^2$ and do not contain \emph{any} target mass corrections $m^2/Q^2$ apart from those
absorbed in $|\xi P_\perp|$ through the expression for
$$
 t_{\rm min} = -4 m^2 \xi^2/(1-\xi^2) \sim -m^2 x_B^2 \quad \text{for}\quad x_B \ll 1\,.
$$
This property can be viewed as the generalization of the well-known result that target mass corrections in DIS are organized in terms of the
Nachtmann variable and involve the expansion in powers of $m^2 x_B^2$ rather than $m^2$. An interesting feature of DVCS is that all such corrections
contribute through the combination $|\xi P_\perp|^2 \propto (t_{\min}-t )$ so that in the physical region $-t \ge -t_{\rm min}$ Nachtmann-type target-mass
corrections are always overcompensated by the finite-$t$ effects, i.e., the sign of the overall kinematic correction is opposite.
For spin-1/2 targets there are some additional mass
corrections~\cite{Braun:2012hq} that have a simple structure, however. They arise entirely from the algebra of spinor bilinears.

Another difference of the BMP and BMJ conventions is that the photon helicity amplitudes are defined in Ref.~\cite{Braun:2012bg} with respect
to a different set of polarization vectors $\varepsilon^{\pm,0}_\mu$ (\ref{varepsilon^a_mu})
\begin{align}
\mathcal{T}^{\rm BMP}_{a\pm} &=
 (-1)^{a-1}\varepsilon_\nu^{\pm} T^{\nu\mu} \varepsilon_\mu^{a,\ast}
\notag\\
&=\cffHbmp^q_{a\pm}h+\cffEbmp_{a\pm}e
\mp \cfftHbmp_{a\pm}\tilde h\mp \cfftEbmp_{a\pm}\tilde e\,,
\label{CFFs-BMP}
\end{align}
cf.~Eq.~(\ref{CFFs}). The relation between the BMP CFFs (\ref{CFFs-BMP})
$\cffFbmp \in\{\cffHbmp,\cffEbmp, \cfftHbmp, \cfftEbmp\}$
 and the BMJ CFFs (\ref{CFFs}) $\cffF\in \{\cffH,\cffE,\cfftH,\cfftE\}$ is purely kinematical and can easily be worked out,
see App.~\ref{BMP-notation}:
\begin{align}
\cffF_{\pm+} &=  \cffFbmp_{\pm+} + \frac{\varkappa}{2}\Big[\cffFbmp_{++}+\cffFbmp_{-+}\Big]- \varkappa_0\, \cffFbmp_{0+},
\notag\\
\cffF_{0+} &=  - \left(1+\varkappa\right) \cffFbmp_{0+} + \varkappa_0\Big[\cffFbmp_{++} + \cffFbmp_{-+}\Big]
\label{CFF-F2F}
\end{align}
with an obvious correspondence $\cffH \leftrightarrow \cffHbmp$, etc.
Here
\begin{align}
\varkappa_0 &= \frac{\sqrt{2} Q \widetilde K}{\sqrt{1 + \gamma^2}(Q^2+t)}  = \mathcal{O}(1/Q)\,,
\notag\\
\varkappa &=\frac{{ Q}^2  - t + 2 \xB t}{\sqrt{1 + \gamma^2}(Q^2+t)}-1   = \mathcal{O}(1/Q^2)\,.
\label{varkappa}
\end{align}
Since $\cffFbmp_{++}=\mathcal{O}(1/Q^0)$, $\cffFbmp_{0+}= \mathcal{O}(1/Q)$, and $\cffFbmp_{-+} = \mathcal{O}(1/Q^{2})$,
the relations (\ref{CFF-F2F}), strictly speaking,
are beyond the accuracy of the BMP calculation for the helicity amplitudes.
For consistency one may use approximate relations
$$ \mathcal{F}_{\pm+} \simeq  \cffFbmp_{\pm+}\! + \frac{\varkappa}{2} \cffFbmp_{++} \! - \varkappa_0\, \cffFbmp_{0+}
\mbox{ and }
 \mathcal{F}_{0+} \simeq  -\cffFbmp_{0+}\! + \varkappa_0\, \cffFbmp_{++}$$
that differ from (\ref{CFF-F2F}) by terms proportional to $1/Q^3$ and $1/Q^4$.  However, using the exact transformation
formulas from the BMP to the BMJ basis, Eq.~(\ref{CFF-F2F}),
has the advantage that the results for physical observables expressed in terms of the BMJ CFFs coincide
with the corresponding results which one would obtain by a direct calculation by means of the original BMP parametrization.
We will stick to this `exact' transformation in the following.

Explicit expressions for the BMP CFFs $\cffFbmp \in \{\cffHbmp, \cffEbmp, \cfftHbmp, \cfftEbmp\}$ are collected in Eqs.~(\ref{mathfrak++})~--~(\ref{mathfrak-+}).
They include also some ${\cal O}(1/Q^3)$ and ${\cal O}(1/Q^4)$ corrections that are due to the transformation of the original BMP
expressions (\ref{Anonflip})--(\ref {Aflip}) to the basis of spinor bilinears in Eq.~(\ref{BMJstructures}).
The resulting ambiguity --- to include such terms or leave them out --- signals the uncertainty which is left.
For example, the BMP result for the helicity conserved CFF $\cffHbmp_{++}$ reads
\begin{align}
\cffHbmp_{++} &=
  {T}_0\circledast H
+ \frac{-t}{Q^2}\Big[\frac{1}{2}{T}_0 - {T}_1
- 2 \xi  \mathbb{D}_\xi\,{T}_2 \Big]\circledast H
\notag\\
&\phantom{={}}+ \frac{2t}{Q^2} \xi^2 \partial_\xi \xi  {T}_2 \circledast (H+E)\,.
\label{H++example}
\end{align}
The first convolution integral on the r.h.s.~of this equation corresponds to the
leading-order parton model result (\ref{T-convolution}) calculated using the BMP convention with the skewedness parameter
$\xi = \xi_{\rm BMP}$ (\ref{xB2xiBMP}). The remaining terms are the kinematical twist-four corrections of order $\mathcal{O}(1/Q^2)$.
They are given by similar convolution integrals that involve new coefficient functions $T_1(u), T_2(u),\ldots $ and, in general, other GPDs.
These convolutions are also decorated by powers of the skewedness parameter and the derivatives $\partial_\xi = \partial/\partial \xi$.
The differential operator $\mathbb{D}_\xi$ is defined as
\begin{align}
   \mathbb{D}_\xi &= \partial_\xi + 2 \frac{|\xi P_\perp|^2}{t}\partial^2 _\xi \xi
         \notag\\&= \partial_\xi - \frac{t-t_{\rm min}}{2t} (1-\xi^2) \partial^2 _\xi \xi\,.
\label{mathbbDxi}
\end{align}
The expressions for the other CFFs (\ref{mathfrak++})~--~(\ref{mathfrak-+})  have similar structure.
The full list of the coefficient functions appearing in the BMP results is
\begin{subequations}
\label{T-coef}
\begin{align}
\C_0(u) =&\, \frac{1}{1-u},
\label{T0-coef}\\
\C^{(+)}_{1}(u) =&\,\frac{(1-2u)\ln(1-u)}{u},
\label{T1p-coef}\\
\C_{1}^{(-)}(u)\equiv &\, \C_{1}(u) = -\frac{\ln(1-u)}{u},
\label{T1m-coef}\\
\C_2(u) =&\, \frac{{\rm Li}_2(1)-{\rm Li}_2(u)}{1-u} + \frac{\ln(1-u)}{2u},
\label{T2-coef}
\end{align}
\end{subequations}
where the notation follows Ref.~\cite{Mueller:2013caa}.
These functions are holomorphic in the complex $u$-plane except for a pole at $u=1$ in the LO kernel (\ref{T0-coef})
or rather harmless, logarithmic $[1,\infty]$-cuts for the kernels (\ref{T1p-coef})--(\ref{T2-coef}) which contribute to the higher twist corrections.
All of them enter the convolution integrals
with Feynman`s causality prescription, as exemplified in (\ref{T-convolution}), that gives rise to a positive imaginary part.
Hence, for a positive GPD the resulting imaginary part from the convolution is positive, too.
Note that in contrast to all other kernels $\C^{(+)}_{1}(u)$, defined in (\ref{T1p-coef}), does not vanish  in the limit $u\to \infty$, however,
this peculiarity will be cured by applying the differential operator $\partial_\xi \xi$ to the corresponding convolution integral.

\subsection{GPD model}
\label{Sec:GPDModel}

To gain some generic insights in the structure of power corrections in this Section
we use a $t$-independent toy GPD model that is  based on Radyushkin's double distribution ansatz (RDDA) \cite{Radyushkin:1998es,Musatov:1999xp},
\begin{align}
\label{GPD-toy}
F^{q^{(+)}}\!\!(x,\xi) &= \int\limits_{0}^1\!dy\!\!\int\limits_{-1+y}^{1-y}\!dz\, \delta(x\!-\!y\!-\!\xi z)\,
\frac{105}{128} \frac{(1\!-\!y)^2-z^2}{\sqrt{y}}
\nonumber\\
&\phantom{={}}  -\sigma(F) \{x\to -x\}\,.
\end{align}
This model corresponds to the generically correct valence-like quark density $q(x)=(35/32)x^{-\alpha}(1-x)^\beta$ with $\alpha=1/2$ and $\beta=3$, normalized to one, and the so-called profile function $(3/4)(1-w^2)$ with $w=z/(1-y)$. The convolution of this GPD with the leading-order kernel
$T_0$ provides a signature-independent imaginary part ${\Im\text{m}}\cffF^{\rm LO}  = \pi F^{q}(\xi,\xi)$, where
\begin{align}
F^q(\xi,\xi) = \frac{7}{4(1+\xi)} \left(\!\frac{2 \xi}{1+\xi}\!\right)^{-1/2}\frac{1-\xi}{1+\xi }.
\label{F^q(xi,xi)}
\end{align}
In such a model the $\xi\to 1$ behavior  of $F^q(\xi,\xi)$ is determined by the profile function rather than the $x\to 1$ behavior of the
parton distribution function (PDF),
in our case $F^q(\xi,\xi) \sim (1-\xi)^1$. The small $\xi$-asymptotics is the same as for the PDF, corresponding to
a `Reggeon intercept' $\alpha=1/2$.
Skewedness changes, however, the value of the residue.

We note in passing that it is possible to rewrite the BMP results~\cite{Braun:2012bg,Braun:2012hq} directly in terms of the double distributions.
This can be useful in the applications. The corresponding expressions are given in App.~\ref{App:DD}.

The GPD on the cross-over line  $x=\xi$ (\ref{F^q(xi,xi)}) can be used to evaluate the real part of the convolution integral
$T_0\!\circledast \!F$ via signature-even or -odd dispersion relations \cite{Teryaev:2005uj},
which presents an alternative to a direct numerical calculation of the LO convolution integral (our notation will be consistent with
those of Sec.~3.2 in Ref.~\cite{Mueller:2013caa}).

A dissipative framework can be also used for the evaluation of power corrections. As the first step one calculates the imaginary parts,
\begin{align}
\Im{\rm m}\, \C\!\circledast \!\gpdF = \pi \sum_q e_q^2 \int_\xi^1\!\frac{dx}{x}\,t(x) F^{q^{(+)}}(\xi/x,\xi)\,,
\label{Im-conv}
\end{align}
that arise from the convolution with the imaginary parts of the kernels
$\C \in \{T_0, T_1^{(\pm)},T_2\}$ defined in Eq.~(\ref{T-coef}),
\begin{subequations}
\label{t-coef}
\begin{align}
t_0(x) &= \delta(1-x),
\label{t0-coef}\\
t^{(+)}_{1}(x) &=\frac{1}{x(1+x)},
\label{t1p-coef}\\
t_{1}^{(-)}(x)&\equiv  t_{1}(x) = \frac{1}{1+x},
\label{t1m-coef}\\
t_2(x) &= \frac{\ln\frac{1+x}{2 x}}{1-x}-\frac{1}{2 (1+x)}\,.
\label{t2-coef}
\end{align}
\end{subequations}
For technical details and notation see Sec.~3.2 in \cite{Mueller:2013caa}.
The real parts of the photon helicity conserved CFFs $\cffFbmp_{++}$ can be recovered from dispersion relations,
unsubtracted for the signature-odd CFFs and involving
the $D$-term related subtraction constant for signature-even CFFs
$\cffHbmp_{++}$ and $\cffEbmp_{++}$, modified as compared to the leading-order
leading twist result.

The BMP results for helicity flip CFFs can be treated in the same
framework, however, it is desirable to remove first the kinematical
constraints by suitable prefactors.

We add that the applicability of the dissipative framework was established
for NLO  corrections at  leading twist-two and also for the LO result at twist-three level for a scalar target in
Ref.~\cite{Diehl:2007jb} and Ref.~\cite{Moiseeva:2008qd}, respectively.
In App.~\ref{App:analyticity} we show that it holds for twist-four kinematical corrections as well.

The imaginary parts (\ref{Im-conv}) only involve the GPD in the outer region with the argument $\xi \le \xi/x \le 1$.
The corresponding expression is readily obtained from (\ref{GPD-toy}) and reads
\begin{align}
F^q(\xi/x,\xi) &=
\frac{7 \left(1-\xi ^2\right)}{32 \sqrt{\xi}\, x^{\frac{5}{2}}}
\Biggl[\!
\left(\!\frac{1+x}{1+\xi }-5\frac{1-x}{1-\xi }\!\right)\!\left(\!\frac{1+x}{1+\xi}\!\right)^{\frac{3}{2}}
\notag\\
&\phantom{={}}- \left(\frac{1-x}{1-\xi}-5\frac{1+x}{1+\xi}\!\right)\!\left(\!\frac{1-x}{1-\xi}\!\right)^{\frac{3}{2}}\!\Biggr].
\label{F^q(xi/x,xi)}
\end{align}
For $x=1$ this function is given by the GPD on the cross-over line, see (\ref{F^q(xi,xi)}), while for $x\to \xi$ it
has a PDF-like behavior,
$$F^{q}(\xi/x,\xi) \stackrel{\xi/x\to 1}{=} \frac{35 x^3 \left(1-\xi/x\right)^3}{32 \xi^3 \left(1-\xi ^2\right)^2}$$
characterized by a generic $(1-\xi/x)^3$ falloff.
Since all kernels in Eq.~(\ref{t-coef}) except for the LO $t_0(x)$  have a constant behavior for
$x\to 1$, the convolution (\ref{Im-conv}) weakens the $\xi\to 1$ asymptotics compared to the GPD at the cross-over line by one power, i.e.,
in our model we obtain $\sim (1-\xi)^2$.
The derivatives over skewedness in the expressions
for the power corrections, $\partial_\xi$ or $(1-\xi^2)\partial_\xi^2$, cf.~(\ref{H++example}), reduce the power again
and restore the original $\sim (1-\xi)$ behavior.
Thus the higher-twist corrections have, generically, the same behavior at $\xi\to 1$ as the LO term.

In the small $\xi$-region we read off from Eq.~(\ref{F^q(xi/x,xi)}) the expected Regge behavior $\sim\xi^{-1/2}$,
\begin{align}
F^{q}(\xi/x,\xi) &\!\stackrel{\xi\to 0}{=}\! \frac{7(1+x)^{3/2} }{16 x^{3/2} \sqrt{\xi }}
\Biggl\{6+\frac{2\!+\!3x}{x}\Biggl[\biggl(\frac{1\!-\!x}{1\!+\!x}\biggr)^{\frac{3}{2}}\!\!-1\Biggr]\Biggr\}.
\label{F^q(xi/x,xi)-smallxi}
\end{align}
Note that this function vanishes for $x\to 0$ as $\sqrt{x}$ and approaches a constant for $x\to 1$.
With an exception of
$$t_1^{(+)}(x) = \frac{1}{x} - t_1^{(-)}(x),$$
which possesses a $1/x$-singularity, the remaining kernels in Eq.~(\ref{t-coef}) are regular at $x=0$.
Thus, apart from this singular case, one can safely set the lower limit of the integration in (\ref{Im-conv}) to zero
which reveals that the convolution integral behaves as $1/\sqrt{\xi}$ as well.
The additional $1/x$-singularity in  $t_1^{(+)}(x)$ yields an extra $1/\xi$-pole, however, it is annihilated in the final expressions
by the application of the differential operator $\partial_\xi \xi$.

\begin{figure}
\includegraphics[width=8.6cm]{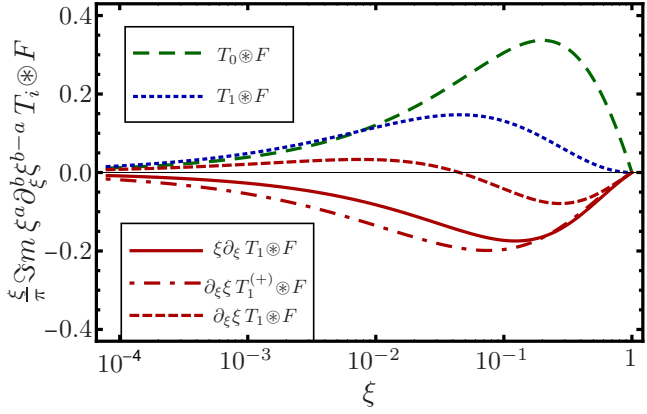}
\caption{
Imaginary parts of typical contributions to the CFFs, multiplied with $\xi/\pi$, from the toy GPD model (\ref{GPD-toy}):
LO contribution $F(\xi,\xi)$ (dashed), convolution integral $T_1\!\circledast \!\gpdF$ (dotted), acting on it with the differential operator
$\xi \partial_\xi$ (solid) and  $\partial_\xi \xi$ (short-dashed), as well as  $\partial_\xi \xi T^{(-)}_1\!\circledast \!\gpdF$ (dash-dotted).
Normalization of the u(d)-quark PDF is set to 2 (1).
\label{Fig:ImT}}
\end{figure}

The small-$\xi$ and large-$\xi$ behavior of various contributions to the power corrections can be studied in the similar manner
for a more general RDDA  such that the GPD on the cross-over line reads as
$$
F(\xi,\xi)\sim\frac{1}{1+\xi} \left(\!\frac{2\xi}{1+\xi}\!\right)^{-\alpha}\left(\!\frac{1-\xi}{1+\xi}\!\right)^{b}
$$
with parameters $\alpha >0$ and  $b >0$ governing the $\xi\to 0$ and $\xi\to1$ asymptotics, respectively.
We find that also in this case the small-$\xi$ and large-$\xi$ asymptotics of the twist-three and twist-four corrections
will follow the LO behavior. This conclusion seems to be rather generic.
For a large class of GPDs the convolutions (\ref{Im-conv}) yield  functions that monotonously decrease with $\xi$.
The consequent application of the homogeneous differential operator $\xi \partial_\xi$ on a convolution integral changes the sign and leaves the
functional form of $F(\xi,\xi)$ roughly intact. Another possibility, the application of the differential operator $\partial_\xi \xi  = 1+  \xi \partial_\xi  $
yields a sum of positive and negative contributions such that the negative one overwhelms at large-$\xi$ whereas for small-$\xi$ the positive contribution
dominates if $0<\alpha<1$. Some selected examples which illustrate this discussion are displayed for our toy model in Fig.~\ref{Fig:ImT}.

Closing this general discussion, we mention that for integer $b$ (profile parameter) and $\beta$ (PDF parameter),
e.g., for our toy model with $\alpha=1/2$,
all convolution integrals with the kernels in Eq.~(\ref{t-coef}) can be calculated analytically in  terms of elementary,
logarithmic, and dilogarithmic functions. Starting from these expressions one can
calculate the corresponding dispersion
integrals, again in an analytic manner, and finally apply the corresponding differential operators.
For the GPD model of Goloskokov and Kroll, which we will utilize below, the imaginary part can be analytically evaluated in
terms of hypergeometric functions $_2F_1$.  As shown in App.~\ref{App:analyticity}, one can then utilize
dispersion relations to calculate the real part in a direct manner, i.e.~no differentiation of the real part is needed.
%
%
We are now in a position to consider higher-twist power corrections to various (BMP) CFFs in some detail.

\subsection{Helicity conserved CFFs $\cffFbmp_{++}$}

The original BMP results~\cite{Braun:2012hq}
for the photon helicity-conserved CFFs, exactly transformed to the basis (\ref{CFFs-BMP}), are collected in Eq.~(\ref{mathfrak++}). They can be written in a compact form as follows,
\begin{align}
\cffFbmp_{++}\!= & \C_0\!\circledast \!\gpdF
+ \frac{-t}{Q^2}\Big[\frac{1}{2} \C_0 -\C_1 - 2 \xi^{\frac{1+\sigma}{2}}     \mathbb{D}_\xi  \xi^{\frac{1-\sigma}{2}}\, \C_2
\Big]\!\circledast \!\gpdF
\notag\\
&\, +\delta_{\cfftEbmp\cffFbmp} \frac{4m^2}{\Q^2}\left[
\C_0 + \frac{-t}{Q^2}\Big(\frac{1}{2} \C_0 -\C_1 - 2 \mathbb{D}_\xi  \xi\, \C_2\Big)
\right]\!\circledast \!\gpdtG
\notag\\
&\,- \frac{4 m^2\delta_{\cffEbmp\cffFbmp}-t\, \delta_{\cffHbmp\cffFbmp}}{Q^2}\, 2 \xi^2 \partial_\xi \xi\, \C_2\!\circledast \![H+E]
\notag\\
&\, - \frac{4 m^2\delta_{\cfftEbmp\cffFbmp}-t\, \delta_{\cfftHbmp\cffFbmp}}{Q^2}\,
2\xi \partial_\xi\, \C_2\!\circledast \!\gpdtH\,,
\label{cffFbmp_{++}}
\end{align}
where $\cffFbmp_{++} \in \{\cffHbmp_{++},\cffEbmp_{++},\cfftHbmp_{++},\cfftEbmp_{++}\}$,
$\delta_{\cffFbmp^\prime\cffFbmp}$ is the Kronecker symbol (equal to one if the CFFs $\cffFbmp^\prime$ and $\cffFbmp$ coincide and zero otherwise),
$\sigma\equiv \sigma(F)$ is the signature factor (\ref{sigma}), and the `electric' GPD $\gpdtG = \gpdtH + (t/4\M^2) \gpdtE$ is defined in analogy
to the `electric' CFFs (\ref{electric-magnetic}).
The differential operator $\mathbb{D}_\xi$ is defined in Eq.~(\ref{mathbbDxi}).
Note that $\mathbb{D}_\xi =\partial_\xi$ for $t=t_{\rm min}$, and
$\mathbb{D}_\xi= \partial_\xi - (1/2)(1-\xi^2) \partial^2 _\xi \xi$ for $-t \gg -t_{\rm min}$,
i.e.~in the both limiting cases $t$-dependence drops out.
The extra term $\sim \M^2/\Q^2$  in the second line in Eq.~(\ref{cffFbmp_{++}}) has the same
combination of coefficient functions as shown in the first line for $\sigma=-1$ and it contributes only to $\cfftEbmp_{++}$, however, is determined
by the `electric' GPD $\gpdtG$.  It arises from the rewriting of BMP bilinear spinors in the BMJ basis, clearly visible in Eq.~(\ref{BMJ2BMPspinors1}) of App.~\ref{App:BMP2BMJ-spinors}. Note that this rewriting is also associated with an additional $t/Q^2$ correction, which is hidden here
in $(4\M^2/\Q^2) \gpdtG = (4\M^2/\Q^2) \gpdtH +(t/Q^2) \gpdtE $.
Strictly speaking the  twist-six terms $\sim m^2 t/\Q^4$ and $\sim t^2/\Q^4$ are beyond our accuracy, however, keeping them ensures
that we discuss the original BMP result in another representation.

As can be expected on general grounds, signature-even (i.e.~parity-even) and signature-odd (i.e.~parity-odd) CFFs, $\cffHbmp_{++},\cffEbmp_{++}$ and $\cfftHbmp_{++},\cfftEbmp_{++}$,
arise only from the GPDs with the same signature (parity), $H,E$ and $\widetilde{H},\widetilde{E}$, respectively.
The $4m^2/Q^2$ terms are absent in the target helicity conserved CFFs $\cffHbmp_{++}$ and $\cfftHbmp_{++}$ so that their twist-four corrections
are entirely proportional to $-t/\Q^2$ (apart from the term in $t_{\rm min}$ in $\mathbb{D}_\xi$ which is numerically insignificant),
whereas they do contribute to the target helicity flip CFFs $\cffEbmp_{++}$ and $\cfftEbmp_{++}$.
Although there is no kinematical necessity, we observe that
the terms in the third and forth line of Eq.~(\ref{cffFbmp_{++}})
drop out in the `electric' CFFs
$$
\cffGbmp_{++}=\cffHbmp_{++}+\frac{t}{4m^2}\cffEbmp_{++}\,, \;\;
\cfftGbmp_{++}=\cfftHbmp_{++}+\frac{t}{4m^2}\cfftEbmp_{++}\,,
$$
that are expressed in terms of the `electric' GPDs of the same signature (or parity)
$$
   G = H +\frac{t}{4m^2} E\,,\;\; \widetilde{G} = \widetilde{H} +\frac{t}{4m^2} \widetilde{H}\,,
$$
so that for these combinations the whole twist-four contributions are proportional to $-t/\Q^2$ as well.

In order to quantify these corrections, we define the (relative) coefficients $k_{{++}}^{\cffFbmp}$ as
\begin{subequations}
\label{k_{{++}}^{cffFbmp}}
\begin{align}
\frac{\Im{\rm m}\, \cffFbmp_{++}(\xi,t,\Q^2)}{\Im{\rm m}\, \cffFbmp^{\rm LT}_{++}(\xi,t,\Q^2)} = \left[1-\frac{t}{\Q^2} k_{++}^\cffF(\xi,t_{\rm min}/t) \right],
\end{align}
where the value $k_{{++}}^{\cffFbmp} = 1$ corresponds to a (enhanced) higher-twist multiplicative correction factor $(1-t/Q^2)$ to the imaginary part of
a given CFF $\cffFbmp_{++}$ with respect to the LO leading-twist expression. As reference we take the original BMP result to leading twist
accuracy, which is obtained from Eq.~(\ref{cffFbmp_{++}}) by dropping the last two lines and  all explicit $t/\Q^2$ corrections in the first two lines,
\begin{align}
\cffFbmp^{\rm LT}_{++} = \left\{ {
\C_0\!\circledast\!F \quad \mbox{for}\quad \cffFbmp \leftrightarrow \gpdF \in \{\gpdH,\gpdE, \gpdtH\}
\atop
\C_0\!\circledast\! \gpdtE + \frac{m^2}{\Q^2} \C_0\circledast\left[\gpdtH + \frac{t}{4\M^2} \gpdtE\right] \quad\mbox{for}\quad \cffFbmp =\cfftEbmp
}\right..
\end{align}
\end{subequations}
The twist-four term $\sim m^2/Q^2$ in the CFF $\cfftE_{++}$ arises again from the transformation of bilinear spinors and is discussed in more detail in Sec.~\ref{Sec:map}, see Eq.~(\ref{BMP-convention1}). A very important point here is also  that
the LO expression $T_0\circledast F$ is calculated using BMP convention (\ref{xB2xiBMP}) for the skewedness parameter $\xi=\xi_{\rm BMP}$.
The expansion of $\xi_{\rm BMP}(x_B,t/Q^2)$ in powers of $t/Q^2$ yields additional corrections that will be discussed separately.

We choose to begin with the `electric' combinations of the CFFs where the corrections  have simpler structure.
From Eqs.~(\ref{cffFbmp_{++}}) and (\ref{k_{{++}}^{cffFbmp}}) one easily obtains
\begin{align}
k_{{++}}^{\cffGbmp}=&\;\frac{
\Im{\rm m}\Big\{
\frac{1}{2}\C_0\!\circledast \!\gpdG  - \C_1\!\circledast \!\gpdG - 2 \xi \partial_\xi\, \C_2\!\circledast \!\gpdG\Big\}
}{
\Im{\rm m}\C_0 \!\circledast \!\gpdG
}
\notag\\
&\;+ \frac{t-t_{\rm min}}{t} \frac{
\Im{\rm m} (1-\xi^2)  \xi \partial^2_\xi \xi\, \C_2 \!\circledast \!\gpdG
}{
\Im{\rm m}\C_0 \!\circledast \!\gpdG
},
\label{k_{{++}}^{cffGbmp}}
\end{align}
and
\begin{align}
k_{{++}}^{\cfftGbmp}=&\;\frac{
\Im{\rm m}\left\{
\frac{1}{2}\C_0\!\circledast \!\gpdtG  - \C_1\!\circledast \!\gpdtG - 2 \partial_\xi\xi \, \C_2\!\circledast \!\gpdtG
\right\}
}{
\Im{\rm m}\C_0\!\circledast \!\gpdtG
}
\notag\\
&\;+\frac{t-t_{\rm min}}{t} \frac{
\Im{\rm m}(1-\xi^2) \partial^2_\xi \xi^2\, \C_2\!\circledast \!\gpdtG
}{
\Im{\rm m}\C_0\!\circledast \!\gpdtG
}.
\label{k_{{++}}^{cfftGbmp}}
\end{align}
These two factors
are displayed in Fig.~\ref{Fig:cffF_{++}} as functions of BMP skewedness parameter
for the GPD model specified in Eq.~(\ref{GPD-toy}) and two choices of the momentum transfer:
$-t\gg -t_{\rm min}$ (solid curves) and  $t=t_{\rm min}$ (dashed curves).
%
\begin{figure}
\includegraphics[width=8.6cm]{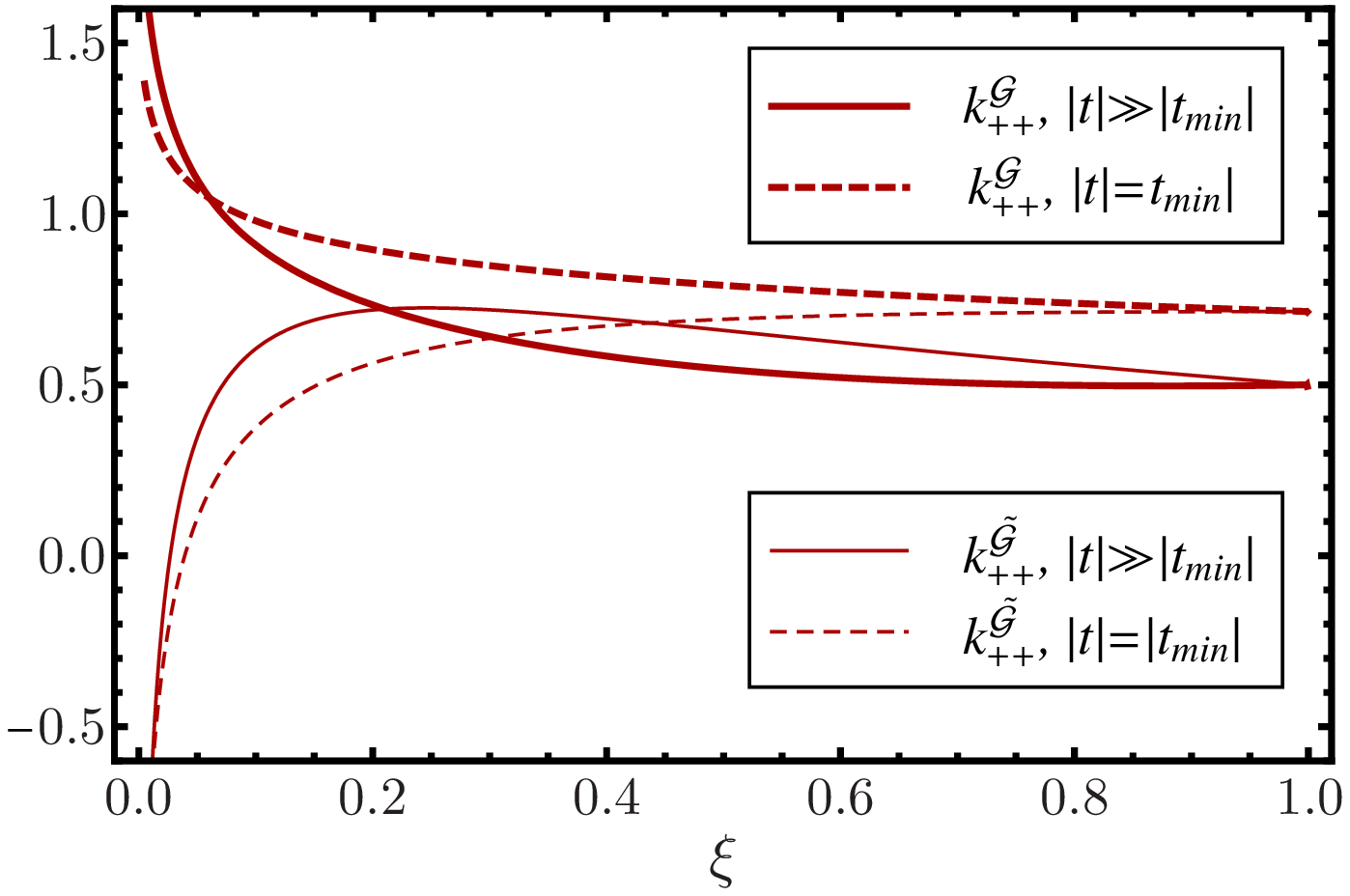}
\vspace{-4mm}
\caption{
Effective coefficients $k^\cffFbmp_{++}$ of $-t/\Q^2$ corrections (\ref{k_{{++}}^{cffFbmp}})
for the `electric' signature-even $\cffFbmp = \cffGbmp_{++}$ (thick) and signature-odd  $\cffFbmp = \cfftGbmp_{++}$ (thin)
CFFs evaluated for the GPD (\ref{GPD-toy}). The solid and dashed curves are calculated for
$-t\gg -t_{\rm min}$ and $t=t_{\rm min}$, respectively.
}
\label{Fig:cffF_{++}}
\end{figure}
%
The difference between solid and dashed curves is marginal, which signals that the
$(t-t_{\rm min})/t$ terms are numerically less important. We observe also that for $\xi \gtrsim 0.1$ the $k_{++}$ factors
in the signature-even (thick curves) and -odd (thin curves) sector are rather similar and  that all
curves are rather flat and $k^\cffGbmp_{++} \simeq k^{\cfftGbmp}_{++} \simeq 0.5-1$.
Approaching the  small-$\xi$ region $k^\cffGbmp_{++}$ increases while $k^{\cfftGbmp}_{++}$ decreases.
The limiting values at $\xi\to 0$, which are not displayed, remain finite. They depend on model details and can be calculated analytically, see Sec.~\ref{sec:ColliderKinematics}.

Next, we consider the signature-even `magnetic' combination, $\cffHbmp+\cffEbmp$.
In this case an additional contribution proportional to  $2\xi^2(4m^2-t)/Q^2$ appears that involves a convolution
with `magnetic' GPD $H+E$. In a typical DVCS kinematics ($\Q^2 \gtrsim 2m^2$) this factor is roughly $2m^2\,\xB^2/Q^2 \lesssim   \xB^2$ and can be considered as small
apart from the region of very large  $\xB$. Hence this extra contribution is numerically not very important (at least in the valence region) and
therefore $k^{\cffHbmp+\cffEbmp}_{++} \approx k^\cffGbmp_{++}$. It follows that the twist-four corrections to the CFFs $\cffHbmp$ and $\cffEbmp$ themselves are of the same
order as for the `magnetic' combination, $k_{{++}}^{\cffHbmp} \approx k_{{++}}^{\cffEbmp} \approx k_{{++}}^{\cffGbmp}$, displayed in Fig.~\ref{Fig:cffF_{++}}.

Finally, we consider the signature-odd CFFs. The coefficient $k_{++}^{\cfftHbmp}$ of the $-t/Q^2$ proportional correction to $\cfftHbmp$
has the same structure as the corresponding coefficient for the signature-odd `electric' CFF $\cfftGbmp$
(\ref{k_{{++}}^{cfftGbmp}}), with an extra term
$$
\frac{t}{Q^2} \frac{\Im{\rm m}\, \partial_\xi\xi \, \C_2\!\circledast \!\gpdtH }{\Im{\rm m}\, \C_0\!\circledast \!\gpdtH}\,.
$$
The ratio of imaginary parts in this expression is rather small because of the differential operator $\partial_\xi\xi$
in the numerator, cf.~analogous convolutions shown by short (for $\partial_\xi \xi T_1$) and long dashes (for $T_0$) in Fig.~\ref{Fig:ImT}. Thus this
extra contribution is not very significant. It follows that the $-t/Q^2$ corrections to $\cfftHbmp$  are positive
and roughly of the same magnitude as for $\cfftGbmp$ shown in  Fig.~\ref{Fig:cffF_{++}}.
The $-t/Q^2$  corrections to $\Im{\rm m}\cfftEbmp_{++}$ are entirely determined by $k_{++}^{\cfftGbmp}\sim 0.5$, however, for this CFF the extra term
$$
- \frac{
4\M^2\, \Im{\rm m}\partial_\xi\xi\,\C_2\!\circledast \!\gpdtH}{
\Im{\rm m}\, \C_0\!\circledast\!\left[(\Q^2+t)\,\gpdtE   + 4\M^2\,\gpdtH\right] }
$$
appears. For vanishing GPD $\gpdtE$  this term simplifies to
$-\Im{\rm m}\, \partial_\xi\xi \, \C_2\!\circledast \!\gpdtH /\Im{\rm m}\, \C_0\!\circledast \!\gpdtH$, which as we have discussed is a smaller (positive) modification, which will decrease further for a positive $\Im{\rm m} \C_0\!\circledast \!\gpdtE $.  Note also
that the corresponding spinor bilinear $\tilde{e}$
contains also a small prefactor $\Delta\cdot q/P\cdot q = -\xi/(1+t/Q^2)$, see Eq.~(\ref{BMJstructures}),
so that the full $\cfftEbmp_{++}$ contribution is suppressed in the experimental observables by an additional factor $\xi$, which makes
the effect of the $4m^2/Q^2$ correction even milder.
Furthermore, this CFF drops out entirely  in the unpolarized interference term in the cross section, $\cal{I}_{\rm unp}$
in Eq.~(\ref{InterferenceTerm}).

\subsection{Longitudinal-to-transverse helicity flip  CFFs $\cffFbmp_{0+}$}

The longitudinal-to-transverse helicity flip CFFs $\cffFbmp_{0+}$ are twist-three, i.e.~suppressed by $1/Q$ compared to the
helicity-conserving contributions, and the power corrections to them are twist-five, of order $1/Q^3$ which is beyond our accuracy.
The leading, twist-three, expressions are known since a decade, and have been confirmed once more in Ref.~\cite{Braun:2012hq}.
The BMP results for $\cffFbmp_{0+}$ in the  representation (\ref{CFFs-BMP}) are collected in Eq.~(\ref{mathfrak0+}) and can be cast in the
following form
\begin{align}
\cffFbmp_{0+}\!\simeq & -\frac{4 |\xi P_\perp|}
{\sqrt{2}Q}
\left(\!1+\frac{(1-\sigma)t}{2\Q^2}\!\right)
\xi^{\frac{1+\sigma}{2}}  \partial_\xi \xi^{\frac{1-\sigma}{2}}\, \C_1\!\!\circledast \!\gpdF
\notag\\
&+ \frac{4m^2(\delta_{\cffEbmp\cffFbmp} -\delta_{\cfftEbmp\cffFbmp} )-t (\delta_{\cffHbmp\cffFbmp}- \delta_{\cfftHbmp\cffFbmp})}{\sqrt{2}Q |\xi P_\perp|}
\xi^{\frac{1+\sigma}{2}}
\notag\\
& \times  \left\{\xi \C_1\!\circledast \![H+E] - \C_1\!\circledast \!\gpdtH  \right\},
\label{cffFbmp_{0+}}
\end{align}
where the notation is similar to Eq.~(\ref{cffFbmp_{++}}) and we neglected twist-five terms proportional to
$(|\xi P_\perp|t/Q^3) \partial_\xi \xi\,{T}_1 \circledast (H+E)$ and
$(|\xi P_\perp|4\M^2/Q^3) \partial_\xi \xi\, \C_1\!\!\circledast \!\gpdtH$ which are present
in the exactly transformed expressions for $\cffHbmp_{0+}$ and $\cfftEbmp_{0+}$, cf.~Eqs.~(\ref{mathfrakH0+}) and (\ref{mathfraktE0+}).

The contribution in the first line in Eq.~(\ref{cffFbmp_{0+}}) involves
the kinematical factor
$$
\frac{4 |\xi P_\perp|}{\sqrt{2}Q}= \frac{2\sqrt{2}\widetilde{K}}{Q\bigl(2-\xB+ \frac{\xB\,t}{Q^2}\bigr)}=\frac{\sqrt{2(t_{\rm min}-t)(1-\xi^2)}}{Q}\,,
$$
which vanishes at the phase space boundaries.
Note that it can be expressed in terms of the kinematical factor $\widetilde{K}$ which is used in Ref.~\cite{Belitsky:2012ch}.

The contributions in the second and third lines in Eq.~(\ref{cffFbmp_{0+}}) are shown exactly  as they arise
from the BMP calculation (no approximation are done here).
These terms have a kinematical $1/|\xi P_\perp|$ singularity  which drops out in `electric'
combinations
\begin{align}
\cffGbmp_{0+}=\cffHbmp_{0+}+\frac{t}{4\M^2}\cffEbmp_{0+}, &&  \cfftGbmp_{0+}=\cfftHbmp_{0+}+\frac{t}{4\M^2}\cfftEbmp_{0+}
\notag
\end{align}
as well as in $\cffHbmp_{0+}+\xi \cfftHbmp_{0+}$ and $\cffEbmp_{0+}+\xi \cfftEbmp_{0+}$. These cancelations ensure that
all angular harmonics in the cross section have the correct behavior at $t\to t_{\rm min}$ as discussed in Sec.~\ref{Sec:electroproduction},
cf.~Eqs.~(\ref{constraint-1}) and (\ref{constraint-2}).

The size of these kinematical singularity free combinations of the twist-three CFFs is governed by the convolution of the corresponding combinations of
GPDs with the kernel $T_1$ (\ref{T1m-coef}), and applying a homogeneous differential operator $\xi\partial_\xi$ (signature-even) or $\partial_\xi\xi$ (signature-odd).
As we have seen already, such convolution integrals are rather mild. The corresponding imaginary parts normalized to the leading-twist helicity conserving
contributions,
\begin{align}
 k_{0+}^{(+)}=-\frac{\Im{\rm m}\,\xi\partial_\xi\, \C_1\!\circledast \!\gpdF}{\Im{\rm m}\, \C_0\!\circledast\! \gpdF}, &&
k^{(-)}_{0+}=-\frac{\Im{\rm m}\,\partial_\xi\xi\, \C_1\!\circledast \!\gpdF} {\Im{\rm m}\, \C_0\!\circledast\! \gpdF},
\label{k0+}
\end{align}
are shown by the thick and thin short-dash-dotted
curves in Fig.~\ref{Fig:cffF_{0+}}, respectively. We see that these ratios are at most $\sim 1/2$. Thus the magnitude of
the singularity free combinations of the twist-three CFFs can be estimated as
$\lesssim \sqrt{(t_{\rm min}-t)(1-\xi^2)/2Q^2}\,\Im{\rm m} \C_0\!\circledast\! \gpdF$, which for DVCS kinematics, say $-4t/\Q^2 \lesssim 1$, is a reasonably small number.

%
\begin{figure}[t]
\includegraphics[width=8.6cm]{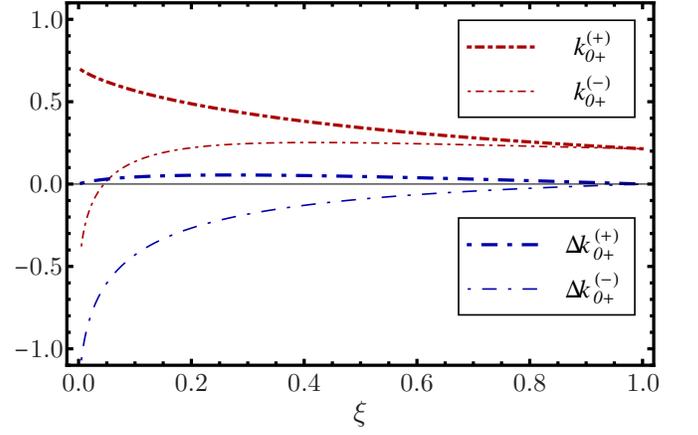}
\vspace{-4mm}
\caption{The ratios $k_{0+}^{(+)}$ (thick dash-dotted curves) and
$k^{(-)}_{0+}$ (thin dash-dotted curves), cf.~Eq.~(\ref{k0+}),
characterizing the magnitude of the contributions in the first line in Eq.~(\ref{cffFbmp_{0+}}) to the
longitudinal-to-transverse helicity flip  CFFs $\cffFbmp_{0+}$, evaluated for the GPD model in Eq.~(\ref{GPD-toy}).
The thick and thin long dash-dotted curves show the ratios $\Delta k_{0+}^{(+)}$ and $\Delta k_{0+}^{(-)}$, respectively, which
are defined in Eq.~(\ref{Delta k_{0+}}) and determine the numerical size of the addenda in
the two last lines in Eq.~(\ref{cffFbmp_{0+}}).}
\label{Fig:cffF_{0+}}
\end{figure}
%
The numerical size of the addenda in the two last lines in Eq.~(\ref{cffFbmp_{0+}}) is determined
by the convolution integral $$\xi \C_1\!\circledast \![H+E] - \C_1\!\circledast \!\gpdtH,$$
where the $H+E$ combination enters with an additional factor $\xi$. To exemplify the numerical size of the addenda we show
in Fig.~\ref{Fig:cffF_{0+}} the quantities
\begin{align}
\Delta k_{0+}^{(+)}= - \xi k_{0+}^{(-)}\,, \quad
\Delta k_{0+}^{(-)}=-\frac{\Im{\rm m}\, \C_1\!\circledast \!\gpdF}{\Im{\rm m}\, \C_0\!\circledast\! \gpdF}
\label{Delta k_{0+}}
\end{align}
as thick and thin long dash-dotted curves, respectively.
Note that for the signature-even combination $\cffHbmp_{0+}+\cffEbmp_{0+}$ there is one more factor $\xi$ in front.
These terms  will either disappear in physical observables or their kinematical singularities will be softened  and they will be dressed with additional suppression factors, e.g.~$\xi t/\Q^2$.

\subsection{Transverse helicity flip CFFs $\cffFbmp_{-+}$}

The CFFs $\cffFbmp_{-+}$, involving photon helicity flip by two units, are suppressed by two powers of the
large momentum, i.e.~they are twist-four (and include twist-six etc. corrections). They are interesting in their own right
as a background to possible leading-twist gluon transversity GPD contributions
to the same amplitudes and can be of phenomenological importance in this context~\cite{Kivel:2001rw}.
The leading twist-four quark contribution to $\cffFbmp_{-+}$ was calculated in Ref.~\cite{Braun:2012hq}.
The result is given in  Eq.~(\ref{mathfrak-+}) and can be cast in the following form
\begin{align}
\hspace*{-0.1cm}
\cffFbmp_{-+}\!\simeq & (-1)^{\delta_{\cfftHbmp\cffFbmp}}\frac{4|\xi P_\perp|^2}{Q^2}
\left(\!1\!+\!\frac{(1\!-\!\sigma)t}{2\Q^2}\!\right)
\xi^{\frac{1+\sigma}{2}}\partial_\xi^2 \xi^{\frac{3-\sigma}{2}}\, \C_{1}^{(\sigma)}\!\!\circledast \!\gpdF
\notag\\
& -\frac{4m^2(\delta_{\cffEbmp\cffFbmp}-\delta_{\cfftEbmp\cffFbmp})-t (\delta_{\cffHbmp\cffFbmp}-\delta_{\cfftHbmp\cffFbmp})}{Q^2} \xi^{\frac{1+\sigma}{2}}\,
\notag\\
&\times 2\!
\left\{\xi\partial_\xi \xi\, \C_{1}^{(+)}\!\circledast \![H+E] + \partial_\xi \xi\,\C_1^{(-)}\!\circledast \!\gpdtH\right\},
\label{cffFbmp_{-+}}
\end{align}
where we now neglected additional twist-six contributions to $\cffHbmp_{-+}$ and $\cfftEbmp_{-+}$,
proportional to $(t/Q^2) \partial^2_\xi \xi^2 \C_{1}^{(+)}\!\circledast \![H+E]$ and $-(4\M^2/Q^2) \partial_\xi^2 \xi^2\, \C_{1}^{(-)}\!\circledast \!\gpdtH$,
respectively, see Eqs.~(\ref{mathfrakH-+}) and (\ref{mathfraktE-+}).

The general structure of the expression (\ref{cffFbmp_{-+}}) resembles what we observed already for the longitudinal-to-transverse CFFs.
The contributions in the first line vanish at the kinematic boundaries thanks to the prefactor
$$
\frac{4|\xi P_\perp|^2}{Q^2} =\frac{4\widetilde{K}^2}{Q^2\bigl(2-\xB+ \frac{\xB\,t}{Q^2}\bigr)^2}= \frac{t_{\rm min} -t}{Q^2}(1-\xi^2),
$$
whereas the addenda in the second and the third lines drops out in  `electric' CFFs
\begin{align}
\cffGbmp_{-+}=\cffHbmp_{-+}+\frac{t}{4\M^2}\cffEbmp_{-+}\,, && \cfftGbmp_{-+}=\cfftHbmp_{-+}+\frac{t}{4\M^2}\cfftEbmp_{-+}\,,
\notag
\end{align}
as well in the $\cffHbmp_{-+}+\xi \cfftHbmp_{-+}$ and  $\cffEbmp_{-+}+\xi \cfftEbmp_{-+}$ combinations.
Hence these combinations vanish linearly as $t\to t_{\rm min}$, in agreement with  Eqs.~(\ref{constraint-1}) and (\ref{constraint-2}).
The magnitude of these, kinematical singularity free, combinations of CFFs, in units of $(t_{\rm min}-t)/Q^2$, is
governed by the convolution of the corresponding GPDs with the kernels  $\C_{1}^{(+)}$ (\ref{T1p-coef}) and $\C_{1} \equiv \C_{1}^{(-)}$ (\ref{T1m-coef})
decorated by the  second order differential operators  $(1-\xi^2)\xi \partial_\xi^2 \xi $ (signature-even) or $(1-\xi^2) \partial_\xi^2 \xi^2 $ (signature-odd).

%
\begin{figure}
\includegraphics[width=8.6cm]{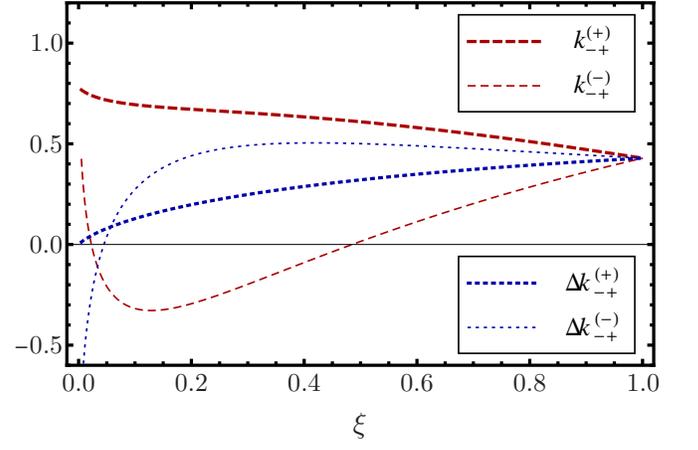}
\vspace{-4mm}
\caption{
The ratios $k_{-+}^{(+)}$ (thick dashed curves) and $k_{-+}^{(-)}$ (thin dashed curves), defined in Eq.~(\ref{k-+}),
characterizing the magnitude of the contributions in the first line in Eq.~(\ref{cffFbmp_{-+}}) to the
transverse-to-transverse helicity flip  CFFs $\cffFbmp_{0+}$, evaluated for the GPD model (\ref{GPD-toy}).
The dashed and dotted curves show the ratios $\Delta k_{-+}^{(+)}$ and
$\Delta k_{-+}^{(-)}$, respectively, which characterize the numerical size of
the addenda in the two last lines in Eq.~(\ref{cffFbmp_{-+}}), as defined in Eq.~(\ref{deltak-+}).
}
\label{Fig:cffF_{-+}}
\end{figure}
%

According to our discussion in Sec.~\ref{Sec:GPDModel} one should expect that the net results for the imaginary parts behave in the
$\xi\to 0$ and $\xi\to 1$  limits similarly to the LO convolution integrals.
In Fig.~\ref{Fig:cffF_{-+}} we plot the corresponding ratios
\begin{align}
 k_{-+}^{(+)}&=\frac{\Im{\rm m}(1-\xi^2)\xi\partial^2_\xi \xi \C_1^{(+)}\!\!\circledast\!\gpdF}{\Im{\rm m}\C_0\!\circledast\!\gpdF}\,,
\notag\\
 k_{-+}^{(-)}&=\frac{\Im{\rm m}(1-\xi^2)\partial_\xi^2 \xi^2 \C_1^{(-)}\!\!\circledast \!\gpdF}{ \Im{\rm m} \C_0\!\circledast\! \gpdF}
\label{k-+}
\end{align}
by the thick and thin dashed curves, respectively.
One sees that $k_{-+}^{(+)} \simeq +0.5$ whereas $k_{-+}^{(-)}$ changes sign at $\xi\sim 0.5$ but becomes positive again at $\xi \to 0$.

The addenda in the second and the third line in Eq.~(\ref{cffFbmp_{-+}}) has the same structure as for
the longitudinal-to-transverse helicity flip
CFFs $\cffFbmp_{0+}$ considered in the previous section, cf. Eq.~(\ref{cffFbmp_{0+}}).
Hence, it will disappear in physical observables or will be dressed with additional suppression factors like $(t-t_{\min})/\Q^2$.
The size of these contributions is governed by the convolution integral
$$-2\!\left\{\xi\partial_\xi \xi\, \C_{1}^{(+)}\!\circledast \![H+E] + \partial_\xi \xi\,\C_1\!\circledast \!\gpdtH\right\}.$$
The contribution of the `magnetic' GPD combination $H+E$ involves an extra factor $\xi$ as compared to the second term so that
its contribution is suppressed and less important for smaller $\xi$ values, whereas the contribution of $\gpdtH$ possesses a node
because of the differential operator $\partial_\xi \xi$.
For illustration we show in  Fig.~\ref{Fig:cffF_{-+}} the ratios
\begin{align}
  \Delta k_{-+}^{(+)}&=-2\frac{\Im{\rm m}\, \xi\partial_\xi \xi\C_1^{(+)}\!\circledast \!\gpdF}{ \Im{\rm m}\, \C_0\!\circledast\! \gpdF}\,,
\notag\\
  \Delta k_{-+}^{(-)}&=-2\frac{\Im{\rm m}\, \partial_\xi \xi\,\C_1^{(-)}\!\circledast \!\gpdF} {\Im{\rm m}\, \C_0\!\circledast\! \gpdF}\,
\label{deltak-+}
\end{align}
as thick and thin  short--dashed curves, respectively.

\section{Power corrections to DVCS observables}
\label{Sec:corrections}

\subsection{Mapping the BMP and BMJ Compton form factors}
\label{Sec:map}

To evaluate observables, we need to express the electroproduction cross section (\ref{dsigma})
in terms of the BMP helicity dependent CFFs $\cffFbmp_{ab}$.
Instead of a new calculation one can overtake the results from Ref.~\cite{Belitsky:2012ch}
making use of the transformation (\ref{CFF-F2F}) of the BMP CFFs to the BMJ basis, $\cffFbmp_{ab} \to \cffF_{ab}$.
As we have already mentioned, these relations are purely kinematic and can be thought of as, loosely speaking,
a Lorentz transformation to a different reference frame. The relations in Eq.~(\ref{CFF-F2F}) are exact (no approximation has been made)
and contain terms proportional to $1/Q^3$ and $1/Q^4$ that are beyond the twist-four accuracy of the BMP amplitudes~\cite{Braun:2012hq}.
The corresponding ambiguity --- use the exact relations or truncate them to $1/Q^2$ accuracy ---
is part of the remaining uncertainty $1/Q^3$ of our calculation.
We have chosen to use exact transformations because in this way
the results for physical observables expressed in terms of the BMJ CFFs coincide identically
with the corresponding results which one would obtain by a direct calculation by means of the original BMP parametrization.

Since the BMJ CFF basis is designed to make absence of kinematic singularities explicit,
using it at the intermediate step offers a useful insight in the threshold behavior of the results near kinematic
boundaries, e.g.~$t\to t_{\rm min}$.
It is easy to check that the coefficients $\varkappa_0$,\, $\varkappa$,
appearing in the relations between BMP and BMJ CFFs (\ref{CFF-F2F}) and defined in Eq.~(\ref{varkappa}), have the following behavior in this limit:
$$
 \varkappa_0 \sim (t_{\rm min}-t)^{1/2}\,, \quad \varkappa \sim (t_{\rm min}-t)^{1}\,.
$$
Thus, the admixture of the longitudinal-to-transverse helicity-flip BMP CFFs $\cffFbmp_{0+}$ to the helicity-conserved $\cffF_{++}$
or transverse helicity flip  $\cffFbmp_{-+}$ BMJ CFFs in the first line in Eq.~(\ref{CFF-F2F}) is proportional to $(t_{\rm min}-t)^{1/2}$
and in this way the kinematical singularities of $\cffFbmp_{0+}$, see Eq.~(\ref{cffFbmp_{0+}}),
(or the original BMP result in Eq.~(\ref{mathfrak0+})) are removed.
The $\cffFbmp_{++}+\cffFbmp_{-+}$ admixture is multiplied with $\varkappa \sim (t_{\rm min}-t)^1$
and vanishes at the threshold. For the case of $\cffFbmp_{-+}$ the contributions of the addenda in the last two lines
in Eqs.~(\ref{cffFbmp_{0+}}) and (\ref{cffFbmp_{-+}}) do not vanish at threshold, however,
in physically observables they will be dressed with extra kinematical factors $\sim (t_{\rm min}-t)$.
The expression for $\cffF_{0+}$ in the second line of Eq.~(\ref{CFF-F2F}) is consistent with the threshold behavior as well.

An important issue that we want to discuss in detail is the ambiguity of the leading-twist (LT) calculations.
Starting from the BMJ conventions, the LT approximation to LO accuracy
can be summarized as follows:
\begin{align}
  \text{LT} \equiv \text{LT}_{\rm KM} &: ~\begin{cases}
         \cffF_{++}=\C_0\!\circledast\!F, & \cffF_{0+}= 0,
          \\
         \cffF_{-+}=0,  & \xi = \xi_{\rm KM}
        \end{cases}
\label{KM-convention}
\end{align}
i.e.~the BMJ helicity-conserving CFF is calculated in the LO approximation using
$\xi_{\rm KM} = x_B/(2-x_B)$ for the skewedness parameter and the other CFFs are put to zero.
This ansatz is used by Kumeri{\v c}ki and M\"uller \cite{Kumericki:2009uq,Kumericki:2010fr,Kumericki:2011zc,Kumericki:2013br} in global DVCS fits,
and in practical terms it is not very different from the VGG convention, used by Guidal, (see a discussion in \cite{Belitsky:2008bz})
and also the convention used by Kroll, Moutarde, and Sabatie in \cite{Kroll:2012sm}. We will, therefore, refer to Eq.~(\ref{KM-convention})
as the `standard' LO approximation in what follows.

Starting instead from the BMP framework, the analogous LT LO approximation, derived from (\ref{AnonflipLT}), reads
\begin{subequations}
\label{BMP-convention}
\begin{align}
  \text{LT}_{\rm BMP} &: ~\begin{cases}
         \cffFbmp_{++}=\C_0\!\circledast\!F, & \mbox{for} \quad \cffFbmp\in \{ \cffHbmp, \cffEbmp, \cfftHbmp\}
         \\[1mm]
         \cfftEbmp_{++} = \big(1+\frac{t}{Q^2}\big)\!\!\!\!\! &\!\!\!\!\! \C_0\!\circledast\!\gpdtE + \frac{4m^2}{Q^2}\C_0\!\circledast\!\gpdtH
          \\[1mm]
        \cffFbmp_{0+}=\cffFbmp_{-+}=0, & \xi = \xi_{\rm BMP}.
        \end{cases}
\label{BMP-convention1}
\end{align}
As already said above, the more complicated expression for $\cfftEbmp_{++}$ as compared to $\cffHbmp_{++}, \cffEbmp_{++}, \cfftHbmp_{++}$
is due to the rewriting of the original BMP amplitudes in terms of the BMJ spinor bilinears.
The difference with the `naive' choice $\cfftEbmp_{++}^{{\rm LT}_{BMP}} =  \C_0\!\circledast\!\gpdtE$ is a twist-four correction $\mathcal{O}(t/Q^2,m^2/Q^2)$.
Including this correction in the LT approximation or adding it to the addenda of higher-twist contributions is mostly a matter of taste as
only the sum is defined to the $\mathcal{O}(1/Q^2)$ accuracy, and is just another facet of the ambiguity of the twist separation.
We include this correction in (\ref{BMP-convention1}) so that this ansatz corresponds literally to the leading-twist BMP amplitudes.
Numerically, the difference is rather large for the CFF $\cfftEbmp_{++}$ but appears to be very small for all observables that we consider below
for unpolarized and longitudinally polarized targets. We stress that the full result including power suppressed contributions to the BMP
amplitudes is well defined  to this accuracy, only the separation of the LT part involves some freedom and is prescription dependent.

Finally, using the transformation rules (\ref{CFF-F2F}), the approximation in Eq.~(\ref{BMP-convention1}) is equivalent to
\begin{align}
  \text{LT}_{\rm BMP} &: ~\begin{cases}
         \cffF_{++}=\left(\!1+\frac{\varkappa}{2}\!\right) \cffFbmp_{++}
         & \cffF_{0+}= \varkappa_0\, \cffFbmp_{++}, 
          \\[1mm]
         \cffF_{-+}=\frac{\varkappa}{2}\cffFbmp_{++}, 
         & \xi = \xi_{\rm BMP},
        \end{cases}
\label{BMP-convention2}
\end{align}
\end{subequations}
where the LT CFFs $\cffFbmp_{++}$ are specified in (\ref{BMP-convention1}) and $\xi_{\rm BMP} = \xi_{\rm BPM}(\xB,t,Q^2)$ is defined in Eq.~(\ref{xB2xiBMP}).

It is important to realize that the two LT ans\"atze in Eq.~(\ref{KM-convention}) and Eq.~(\ref{BMP-convention1}) are perfectly
legitimate. Their difference reveals that both the distinction between helicity-conserving and
helicity-flip CFFs, and the expression for skewedness parameter in terms of kinematic invariants,
depend to power $1/Q$ accuracy on the reference frame.

\begin{figure}[t]
\includegraphics[width=8.6cm]{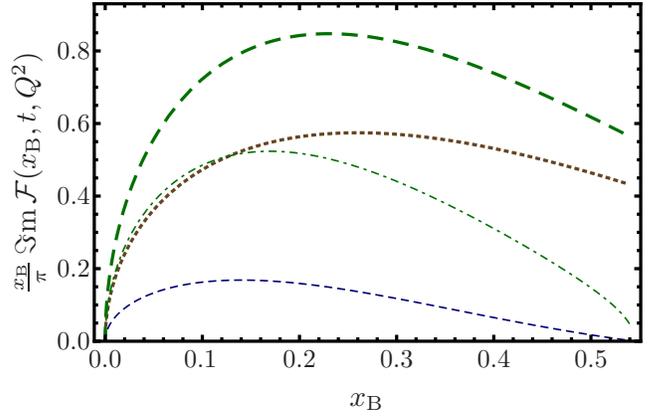}
\vspace{-4mm}
\caption{LT$_{\rm BMP}$ predictions for the imaginary parts of the BMJ CFFs
$(\xB/\pi) \Im{\rm m}\, \cffF_{a+}(\xB,t,\Q^2)$ vs.~$\xB$ at $-t=0.375\, \GeV^2$ and $\Q^2=1.5\, \GeV^2$ for the GPD model (\ref{F^q(xi/x,xi)}):
$\cffF_{++}$ (dashed), $\cffF_{0+}$ (dash-dotted), and $\cffF_{-+}$  (short-dashed),  compared with
the LT$_{\rm KM}$ result for $\cffF_{++}$ (dotted).
}
\label{Fig:TLO}
\end{figure}

\begin{figure*}[t]
\begin{center}
\includegraphics[width=8.5cm]{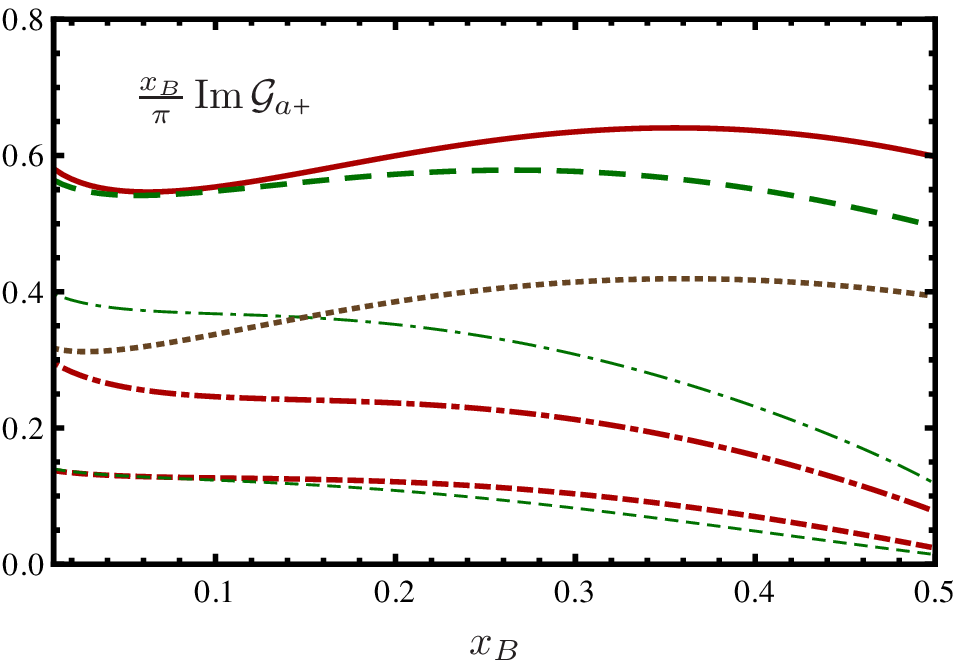}
\includegraphics[width=8.5cm]{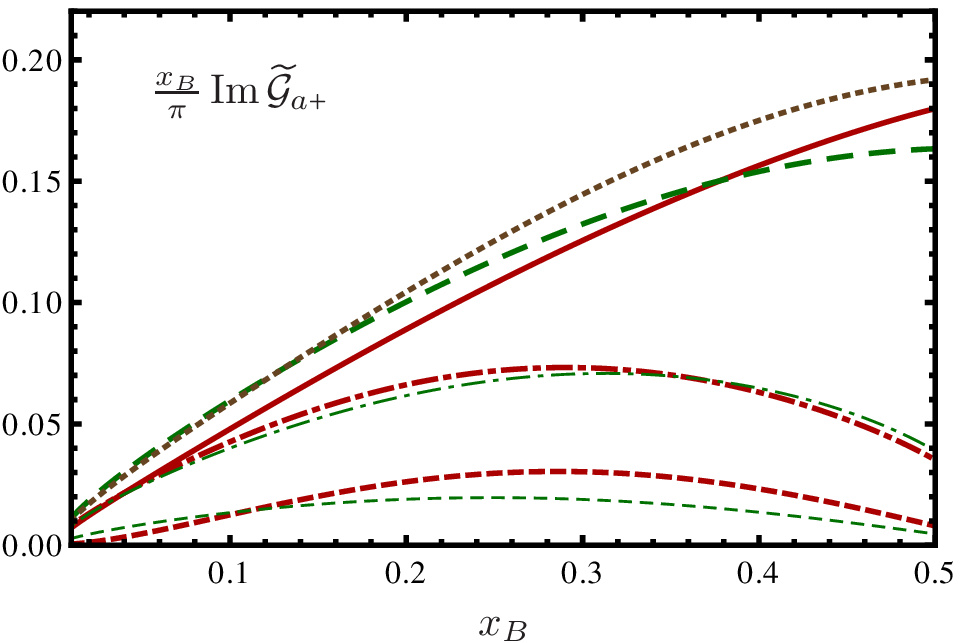}
\end{center}
\vspace{-5mm}
\caption{The imaginary part of `electric' CFFs $\cffG_{a+}(\xB,t,Q^2)$ (left panel) and $\cfftG_{a+}(\xB,t,Q^2)$ (right panel),
multiplied by $(\xB/\pi)$, in the photon helicity dependent CFF  basis (\ref{cffF}) with
$a=+$ (solid, dashed), $a=0$ (dashed-dotted), and $a=-$ (short dashed) versus $\xB$ at $t=-0.375\,\GeV^2$ and $Q^2=1.5\,\GeV^2$. They are evaluated from the GPD  model \GK with kinematical corrections  and compared to the leading twist-two BMP result (\ref{BMP-convention}) [dashed for $a=+$ and otherwise thin curves]
where  $\xi_{\rm BMP}=\xB(1+t/\Q^2)/(2-\xB+ \xB t/\Q^2)$ and to the leading twist-two KM result (\ref{KM-convention}) [dotted curves] where
$\xi_{\rm KM}=\xB/(2-\xB)$. }
\label{Fig:CFFs}
\end{figure*}

The resulting ambiguity is quite large because, first, the kinematic factors $\varkappa_0$ and $\varkappa$ are sizable
despite of being power-suppressed. For example, for $-t/\Q^2  \simeq 1/4$ one obtains $\varkappa/2 \sim 1/3$.
Second, $\xi_{\rm BMP} < \xi_{\rm KM}$, for practical purposes one can approximate $\xi_{\rm BMP} \approx (1+t/Q^2)\xi_{\rm KM}$ for $\xB \lesssim 0.4$.
Thus generally $F(\xi_{\rm BMP},\xi_{\rm BMP}) >  F(\xi_{\rm KM},\xi_{\rm KM})$ if the GPDs have Regge behavior, although this effect is moderated
for larger $t$ by the slope of the Regge-trajectory.
The qualitative picture is illustrated for our toy GPD model (\ref{F^q(xi/x,xi)}) in Fig.~\ref{Fig:TLO} where we show
the LT$_{\rm BMP}$ predictions for the imaginary parts of the BMJ CFFs  $\cffF_{++}$ (dashed), $\cffF_{0+}$ (dash-dotted) and  $\cffF_{-+}$
(short-dashes) vs.~$\xB$ for $t=-0.375\, \GeV^2$ and $\Q^2=1.5\, \GeV^2$.
The LT$_{\rm KM}$ result for  $\cffF_{++}$ is shown by dots for comparison.
Note that the upper value of $\xB$ is bounded by $t_{\rm min}(\xB,\Q^2)=-0.375\, \GeV^2$.
One sees that the  LT$_{\rm BMP}$ prediction for $\Im{\rm m}\, \cffF_{++}$ is much larger than LT$_{\rm KM}$, and the induced
longitudinal-to-transverse helicity flip CFF $\Im{\rm m}\, \cffF_{0+}$ for $\xB \lesssim 0.25$
is as large as the LT$_{\rm KM}$ helicity-conserving CFF, whereas the transverse helicity flip CFF
$\Im{\rm m}\,\cffFbmp_{-+}$ can be considered as small.

The ambiguity of the LT approximation is cured (to the $1/Q^2$ accuracy) by adding the higher-twist addenda to the BMP CFFs that was
studied in Sec.~\ref{Sec:Compton}.
To illustrate the effect, we employ a realistic GPD model that is compatible with experimental data within the conventional LT setting.
We have chosen the Goloskokov and Kroll model which we refer to as \GK, as used in \cite{Kroll:2012sm}.
It is based on the popular RDDA \cite{Radyushkin:1998es} and also involves a certain model for the $Q^2$ dependence
which we overtake in the numerical calculations presented below. Note, however, that the $Q^2$ evolution embedded in the \GK model
is not exactly the one predicted by the LO GPD evolution equations, especially in the small-$\xB$ region.
Technically, this model is rather convenient since it uses mostly integer
values for the profile parameters $b_i$ and PDF parameters $\beta_i$  so that all needed convolution integrals can be evaluated
analytically. To be precise, we will be using the negative sea quark GPD $E^{\rm sea}$ scenario.
Unfortunately, we were unable to find out
how the CFFs in Ref.~\cite{Kroll:2012sm}, evaluated at LO with
the convention (\ref{KM-convention}), are connected to observables.

As an example, we consider kinematical singularity-free `electric' CFF combinations $\cffG = \cffH +(t/4m^2)\cffE$,
cf.~Eq.~(\ref{electric-magnetic}), which are the dominant contributions for the  harmonics  of the interference term
(\ref{{cal C}^{I}_{unp}}) and the DVCS cross section (\ref{Def-CDVCSunp}) for unpolarized proton target.
The imaginary parts  $(\xB/\pi) \Im{\rm m}\, \cffG$ [left panel] and $(\xB/\pi) \Im{\rm m}\,\cfftG$ [right panel]
calculated using the \GK GPD model are shown in Fig.~\ref{Fig:CFFs} in the LT$_{\rm BMP}$ approximation and with full account of all (kinematic) twist-four
corrections. For the helicity-conserving CFFs $\cffG_{++}$ and $\cfftG_{++}$ we also show the LT$_{\rm KM}$ results for comparison (dotted curves).
For this plot we took again a rather low value for $\Q^2 = 1.5\,\GeV^2$ and a large value for $t = - 0.375\,\GeV^2$.

A qualitatively different $\xB$-dependence of the signature-even and -odd combinations is due to the built-in `pomeron-like' growth
of $H$ and $E$ at small $\xB$ whereas the increase in $\widetilde{H}$ and $\widetilde{E}$ is milder.
Hence $\xB\,\cffG_{a+}$ increases at $\xB\to 0$, whereas $\xB\,\cfftG_{a+}$, on the contrary, vanishes in the same limit.
Note that relevant GPD combinations are positive.

For the dominant CFF $\cffG_{++}$  we see that inclusion of the $1/Q^2$ addenda (solid curve) increases the LT$_{\rm BMP}$ result (dashed) somewhat,
which is in turn much larger than the commonly accepted LT$\approx$LT$_{\rm KM}$ approximation. Hence the two effects add up.
The difference between the LT$_{\rm BMP}$ expression and the full BMP result to the twist-four accuracy dies
out in the small-$\xB$ region. This is due to a  partial cancelation of the admixture of $\cffGbmp_{0+}$ and $\cffGbmp_{-+}$, as
can be seen from Eq.~(\ref{CFF-F2F}).
The large positive LT$_{\rm BMP}$ expression for $\cffG_{0+}$ (thin dash-dotted curves),
is significantly reduced so that the full result (thick dash-dotted curves) is much
smaller.
Finally the transverse helicity-flip BMJ CFF $\cffG_{-+}$ (short dashed curves), suppressed by $-t/\Q^2$,
turns out to be rather stable with respect to the twist-four addenda (and remains small) which, again, can be traced to a cancelation
of the corresponding contributions in Eq.~(\ref{CFF-F2F}).

For the signature-odd CFF $\cfftG_{++}$ the difference between the LT$_{\rm BMP}$ (thick dashed curve) and LT$_{\rm KM}$ (dotted curve)
approximations turns out to be smaller as compared to the signature-even CFF $\cffG_{++}$. This is mainly caused by a partial cancelation
of $1/Q^2$ corrections that arise from the transformation of bilinear spinors, cf.~Eq.~(\ref{BMP-convention1}), and photon helicity amplitudes, cf.~Eq.~(\ref{BMP-convention1}). Compared to CFF $\cfftG_{++}$, we find again that
the induced longitudinal helicity flip CFF $\cfftG_{0+}$ (dash-dotted curves) is rather sizeable while transverse helicity flip
CFF $\cfftG_{-+}$ is less important. The differences of the full BMP result and the LT$_{\rm BMP}$ approximation are mild.
In contrast to $\cffG_{a+}$,  the full BMP result for $\cfftGbmp_{++}$ is smaller than the LT$_{\rm KM}$ (for $\xB \lesssim 0.3$ also smaller than 
LT$_{\rm BMP}$) and the kinematical corrections to the CFF $\cfftG_{0+}$ are tiny.
The  reason is twofold: the partial cancelation of $1/Q^2$ corrections in this specific choice of CFF  and the corresponding convolution integrals are in general smaller than in the signature-even sector.

To summarize, we want to stress that the distinction of $1/Q^2$ corrections that are `implicitly' taken into account by the BMP choice
of the skewedness parameter $\xi_{\rm BMP} = \xi_{\rm BMP}(\xB,t,Q^2)$, and, thus, included in the LT$_{\rm BMP}$ approximation (\ref{BMP-convention}),
and `explicit' higher-twist corrections $\sim t/Q^2, m^2/Q^2$ to the BMP CFFs has no physical meaning. Only the sum of such corrections is well-defined
and unambiguous to the claimed $1/Q^2$ accuracy, although it can happen that one of them is numerically dominant in certain observables,
see examples below.

\subsection{From CFFs to DVCS observables}
\label{Sec:corrections1}

The power corrections to helicity-dependent CFFs that we have studied in the preceding sections do not necessarily propagate in a one-to-one correspondence
to the observables. E.g. in the (unpolarized) DVCS cross section the corrections to various CFFs $\cffF_{a+}$ add incoherently,
see Eqs.~(\ref{Res-Mom-DVCS-0-imp}) and (\ref{Def-CDVCSunp}), and for the harmonics of the interference term the corrections might partially cancel
or be amplified, so that there seems to be no simple general pattern.

For definiteness let us consider the $n=1$ odd harmonic $s^{\cal I}_{1,{\rm unp}}$ which governs the size of the electron
beam spin asymmetry (\ref{A_{LU}(phi)}), for which we already quoted the approximate expressions in Eq.~(\ref{Res-IntTer-unp}).
This example is sufficiently simple so that it can be discussed in analytic manner.
Including all corrections that have been omitted in Eq.~(\ref{Res-IntTer-unp}), we can write the exact BMJ result as
\begin{widetext}
\begin{align}
s^{\cal I}_{1,{\rm unp}}&= \frac{8 \tK \lambda \sqrt{1-y-\frac{y^2 \gamma^2}{4}} (2-y) y }{\Q(1+\gamma^2)}
\Im{\rm m}\!\biggl\{{\cal C}^{\cal I}_{\rm unp}\!\biggl(\!
 \biggl[\!1\!-\!\frac{\varkappa}{2\Q^2}\frac{\Q^2+t}{\sqrt{1+\gamma^2}}\!\biggr] \cffF_{++}
\!+\biggl[\!1\!-\!\frac{2+\varkappa}{2\Q^2} \frac{\Q^2+t}{\sqrt{1+\gamma^2}}\!\biggr]\cffF_{-+}
\!+ \frac{(\Q^2+t) \varkappa_0}{ \Q^2\sqrt{1+\gamma ^2}} \cffF_{0+}
\!\!\biggr)
\notag\\
&\phantom{=\frac{8 \tK \lambda \sqrt{1-y-\frac{y^2 \gamma^2}{4}} (2-y) y }{\Q(1+\gamma^2)}\Im{\rm m}\!\biggl\{}
+\!\frac{-t(\Q^2+t)}{\sqrt{1+\gamma ^2}\Q^4}
\Delta{\cal C}^{{\cal I}}_{\rm unp}\left(\cffF_{-+} +\frac{\varkappa}{2}[\cffF_{++} + \cffF_{-+}]-\varkappa_0\,\cffF_{0+}\right)
\biggr\},
\label{s^{I}_{1,{unp}}-exact}
\end{align}
where the function ${\cal C}^{\cal I}_{\rm unp}(\mathcal{F})$ is defined in Eq.~(\ref{{cal C}^{I}_{unp}}) and
the expression for $\Delta{\cal C}^{\cal I}_{\rm unp}(\mathcal{F})$ is given below.
Using the transformation rules in Eq.~(\ref{CFF-F2F}) we can rewrite this result, equivalently, in terms of the
BMP CFFs:
\begin{align}
s^{\cal I}_{1,{\rm unp}}&=
\frac{8 \tK \lambda \sqrt{1-y-\frac{y^2 \gamma^2}{4}} (2-y) y }{\Q(1+\gamma^2)}\Im{\rm m}\biggl\{
{\cal C}^{\cal I}_{\rm unp}\!\biggl(\!\!(1+\varkappa) \cffFbmp_{++}\! + \!\biggl[\! 1+\varkappa-\! \frac{\Q^2+t}{\Q^2\sqrt{1+\gamma^2}}\!\biggr]\cffFbmp_{-+}
\! -  2\varkappa_0\,\cffFbmp_{0+}
\!\!\biggr)
\notag\\
&\phantom{=\frac{8 \tK \lambda \sqrt{1-y-\frac{y^2 \gamma^2}{4}} (2-y) y }{\Q(1+\gamma^2)}\Im{\rm m}\biggl\{}
+\!\frac{-t(\Q^2+t)}{\Q^4\sqrt{1+\gamma^2}} \;\Delta{\cal C}^{{\cal I}}_{\rm unp}(\cffFbmp_{-+})
\biggr\}.
\label{s^{I}_{1,{unp}}-exact-BMP}
\end{align}
\end{widetext}
As already stated in Sec.~\ref{Sec:electroproduction}, the expression (\ref{{cal C}^{I}_{unp}}) for ${\cal C}^{\cal I}_{\rm unp}$ does not include the
kinematical addenda that appear in the second and third lines of Eqs.~(\ref{cffFbmp_{0+}}) and (\ref{cffFbmp_{-+}}). These terms are absorbed
in $\Delta{\cal C}^{{\cal I}}_{\rm unp}$ so that the resulting expression
\begin{align}
\Delta{\cal C}^{{\cal I}}_{\rm unp}&= \frac{2\xB (F_1+F_2) }{2-\xB+\frac{\xB t}{ \Q^2}} \Bigl[\xB (\cffHbmp_{-+}\!+\!\cffEbmp_{-+})+(1\!-\!\xB)\cfftHbmp_{-+}\Bigr]
\end{align}
is free from kinematical singularities.
Together  with the accompanying kinematical prefactor $- t(\Q^2+t)/\Q^4\sqrt{1+\gamma^2}$ this twist-four term can be considered
as a small correction.

The difference of the LT$_{\rm KM}$ and LT$_{\rm BMP}$ approximations can now be illuminated very clearly.
We find for the imaginary parts of the relevant CFF combinations
\begin{align}
  \text{LT}_{\rm KM}: ~&~ \pi {\cal C}^{\cal I}_{\rm unp}\!\biggl(\!
 \biggl[\!1-\!\frac{\varkappa}{2\Q^2}\frac{\Q^2+t}{\sqrt{1+\gamma^2}}\!\biggr] F(\xi_{\rm KM},\xi_{\rm KM})
\!\biggr),
\notag\\
  \text{LT}_{\rm BMP}: ~&~\pi {\cal C}^{\cal I}_{\rm unp}\!\Bigl(\! (1+\varkappa) F(\xi_{\rm BMP},\xi_{\rm BMP})\!\Bigr),
\end{align}
respectively. As we have discussed already, practically we have $F(\xi_{\rm KM},\xi_{\rm KM}) < F(\xi_{\rm BMP},\xi_{\rm BMP})$,
and the LT$_{\rm KM}$ prediction is further reduced by the kinematical factor
$$1-\varkappa(\Q^2+t)/2\Q^2 \sqrt{1+\gamma^2}$$
whereas the LT$_{\rm BMP}$ one is enhanced by the factor $1+\varkappa$
rather than $1+\varkappa/2$ that is present in Eq.~(\ref{BMP-convention}).
Thus the dominant $n=1$ odd harmonic is larger with the $\text{LT}_{\rm BMP}$ than the $\text{LT}_{\rm KM}$ convention.

%
\begin{figure*}[t]
\begin{center}
\includegraphics[width=16cm]{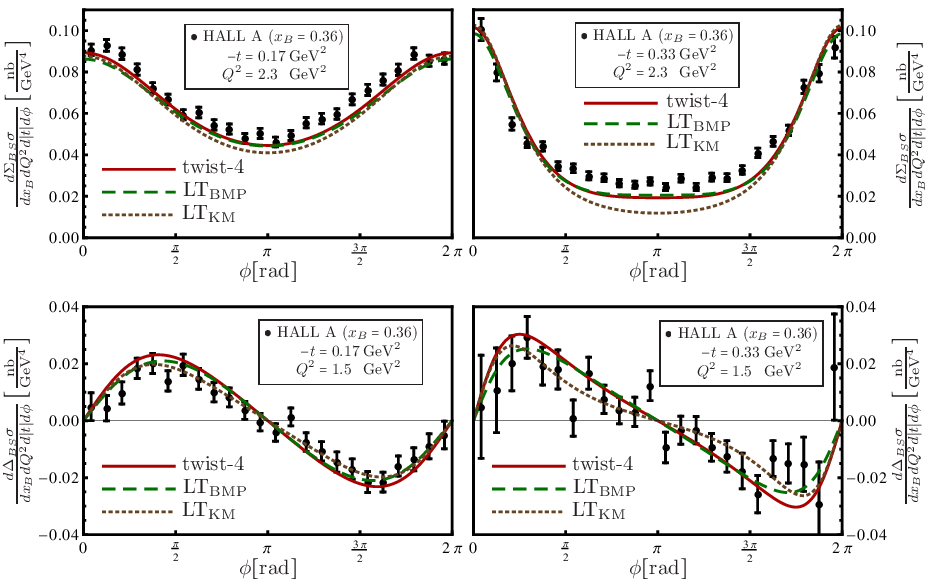}
\vspace{-1mm}
\caption{The unpolarized cross section (\ref{Sigma_{BS}}) for $\xB=0.36$ and $\Q^2=2.3\, \GeV^2$  [upper panels]
and electron helicity dependent cross section difference (\ref{Delta_{BS}}) for $\xB=0.36$ and $\Q^2=1.5\, \GeV^2$
[lower panels] from HALL A collaboration \cite{Munoz_Camacho:2006hx} vs.~\GK GPD model predictions, which are obtained
with the LT=LT$_{\rm KM}$ approximation (\ref{KM-convention}) [dotted curves], the LT$_{\rm BMP}$ approximation (\ref{BMP-convention}) [dashed curves],
and with full account of kinematic power corrections to the $1/Q^2$ accuracy [solid curves].
}
\label{Fig:HALLA}
\end{center}
\end{figure*}

Furthermore, if we include higher twist corrections, a partial cancelation of these $1/Q^2$ corrections
in the argument of ${\cal C}_{\rm unp}^{\cal I}$ might take place,  e.g., the  transverse CFFs $\cffFbmp_{-+}$ and longitudinal CFF
$\varkappa_0 \cffFbmp_{0+}$ contributions to the dominant `electric' CFF $\cffGbmp$ enter in Eq.~(\ref{s^{I}_{1,{unp}}-exact-BMP})
with different signs, see also corresponding lines in Figs.~\ref{Fig:cffF_{0+}} -- \ref{Fig:CFFs}.

The expression for the $n=1$ even harmonic is analogous to (\ref{s^{I}_{1,{unp}}-exact}), however,
in this case additional power suppressed contributions appear that depend on the photon polarization parameter $\varepsilon(y)$, defined in Eq.~(\ref{varepsilon}).
Moreover, the $n=0$ harmonic may play a certain role, too, and the behavior of the real part of CFFs can be rather model dependent.
For instance at larger values of $\xB$ it is determined by both valence and sea quarks as well as the $D$-term or pion-pole contributions,
while at small-$\xB$ the real part is `pomeron'-induced and is small compared to the imaginary part.
Somewhere in the transition region of intermediate $\xB$ the negative real part of the `pomeron' and the positive real part due to
 `reggeon' exchanges in $\cffH$ cancel each other.

\subsection{Fixed target kinematics (unpolarized proton)}

The HALL A collaboration provided high statistic cross section measurements in dependence of the electron
beam helicity \cite{Munoz_Camacho:2006hx}.  These data, in particular for the unpolarized cross section, suggest that the
DVCS cross section is larger than expected from popular GPD models and their description is widely regarded as  challenging,
see comments in \cite{Polyakov:2008xm}.
The unpolarized cross section HALL A data can be described, nevertheless, in a global twist-two fit,
if one assumes a large effective $\gpdtH$ and  $\gpdtE$ scenario \cite{Kumericki:2011zc}.

The unpolarized cross section (\ref{Sigma_{BS}}) data~\cite{Munoz_Camacho:2006hx}, corrected for QED radiative effects, are shown
in the two upper panels in Fig.~\ref{Fig:HALLA} for the smallest $-t =0.17\, \GeV^2$ [left panel] and the largest available $-t =0.33\, \GeV^2$ [right panel],
respectively. These data correspond to $\Q^2= 2.3\, \GeV^2$ and a rather large  $\xB=0.36$ value.
The data are compared with the QCD calculation using the \GK GPD model in three different approximations:
LT$_{\rm KM}$ (dotted curves), LT$_{\rm BMP}$ (dashed curves),  and with the full account of kinematic twist-four effects (solid curves).
The BH squared term is calculated using the formulae set from~\cite{Belitsky:2001ns} with Kelly's electromagnetic
form factor parametrization \cite{Kelly:2004hm}.
Because of this contribution, the differences of the predictions of the unpolarized cross section in
different models or approximations are washed out.

In the conventional LT$_{\rm KM}$ framework, the \GK GPD model underestimates the  data slightly
for the smallest $-t$ value and strongly for the large $-t$.
Note that $-t=0.17\, \GeV^2$ is very close to the kinematic boundary $t_{\rm min}= -0.158\, \GeV^2$, so that the both relevant
expansion parameters are small, $\sqrt{(t_{\rm min}-t)/\Q^2} \ll -t/Q^2 \sim  0.1$. As the result, the difference in LT predictions using KM (dotted)
and BMP (dashed) conventions is small as well and the effect of including extra $1/Q^2$ corrections (solid) appears to be tiny.
The power corrections for the large $-t =0.33\, \GeV^2$ are much larger. In particular changing $\text{LT}_{\rm KM} \to \text{LT}_{\rm BMP}$ produces
relative large enhancement of both the DVCS cross section and the interference term and the prediction becomes closer to the data, whereas
kinematical twist corrections proportional to $-t/\Q^2 \approx 0.14$ remain to be hardly visible.
Thus, for this observable, the $\text{LT}_{\rm BMP}$ approximation alone captures the main part of the total kinematic power correction.

The electron helicity dependent cross section difference (\ref{Delta_{BS}}) is shown in  Fig.~\ref{Fig:HALLA} in the two lower panels.
We take for this plot the data measured for the same values of the momentum transfer
$-t=0.17\, \GeV^2$ [left panel] and $t=-0.33\, \GeV^2$ [right panel] with $\xB=0.36$,
but for a different, the lowest available photon virtuality $\Q^2= 1.5\, \GeV^2$.
This helicity dependent cross section difference is well described with standard GPD models, see also \cite{Polyakov:2008xm},
and it mainly arises from the $n=1$ odd harmonic of the interference term  (the deviation from a pure $\sin\phi$ shape is induced by the
additional $\phi$-dependence of the electron propagators in the BH subprocess).
For $t=-0.17\,\GeV^2$ with $-t/\Q^2 \sim \sqrt{(t_{\rm min}-t)/\Q^2} \sim 0.1$, both the $\text{LT}_{\rm KM}$ vs.~$\text{LT}_{\rm BMP}$ difference
and the additional twist-four corrections are of the same order of magnitude and rather small.
For $t=-0.33\,\GeV^2$ with $-t/\Q^2 \approx 0.22$ and $\sqrt{(t_{\rm min}-t)/\Q^2} \approx  0.35$, the differences in the three
predictions are clearly visible and affect significantly the shape of the $\phi$-distribution.
Having in mind the experimental errors, all of the predictions are, nevertheless, compatible with the data.

%
\begin{figure}[t]
\begin{center}
\includegraphics[width=8cm]{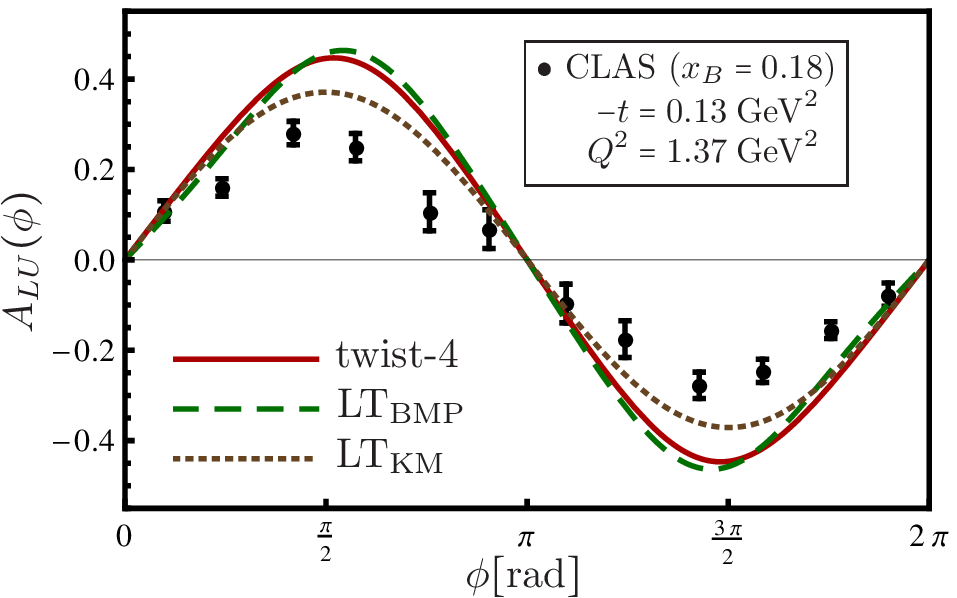}
\includegraphics[width=8cm]{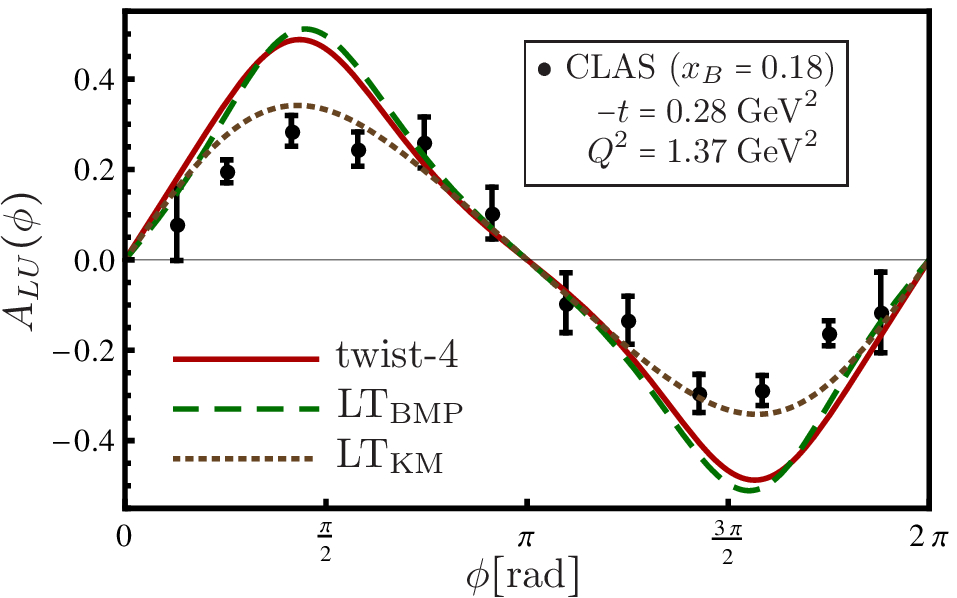}
\caption{The single electron beam spin asymmetry (\ref{A_{LU}(phi)}) measured by the CLAS collaboration for $\xB=0.18$, $\Q^2=1.37\,\GeV^2$
and two different $-t$ values $0.13\,\GeV^2$ (upper panel) and $0.28\,\GeV^2$ (bottom panel) \cite{Girod:2007aa}.
Curves are described in Fig.~\ref{Fig:HALLA}.}
\label{Fig:CLAS}
\end{center}
\end{figure}
%
The CLAS collaboration measured the electron beam spin asymmetry (\ref{A_{LU}(phi)})
over a rather large $-t$ interval \cite{Girod:2007aa}.
In the conservative KM analysis only data were included which satisfy
the criteria $|t|/\Q^2 \lesssim 0.25$ with $\Q^2 \gtrsim 1.5\, \GeV^2$ and the CLAS data were well described in a global fit.
The \GK model predictions for this observable are compared to the data in Fig.~\ref{Fig:CLAS}
for the relatively low $\Q^2 = 1.37\, \GeV^2$ and two values of the momentum transfer, $-t=0.13\, \GeV^2$ and $-t=0.28 \, \GeV^2$.
Typical model GPD predictions have the tendency to overshoot the data in the framework of the standard LT analysis,
as exemplified by the LT$_{\rm KM}$ (dotted) curves in Fig.~\ref{Fig:CLAS} (and, e.g., Fig.~5 in~\cite{Kroll:2012sm}).
Changing to LT$_{\rm BMP}$ (dashed curves) the discrepancy becomes larger whereas adding the remaining $1/Q^2$ power corrections (solid curves)
has marginal effect.
According to the left panel in Fig.~\ref{Fig:CFFs} the dominant CFF $\cffG_{++}$  which governs  the size of the $n=1$ odd harmonic in the
interference term, is very weakly affected by these corrections.
However, the DVCS cross section in the denominator of the asymmetry  increases and also the interference terms can change so that
the asymmetry may get slightly smaller.

%
\begin{figure}[t]
\begin{center}
\includegraphics[width=8cm]{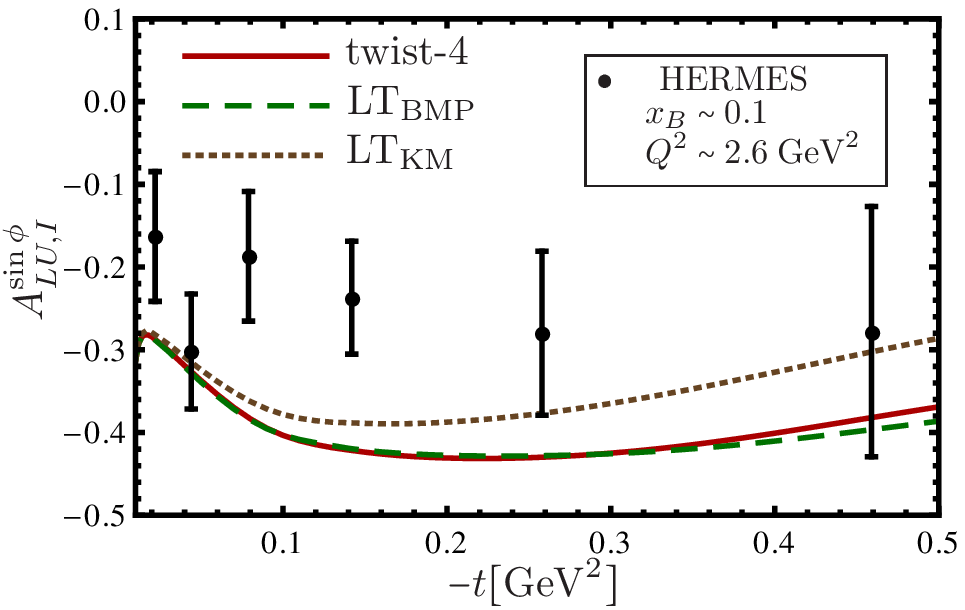}
\includegraphics[width=8cm]{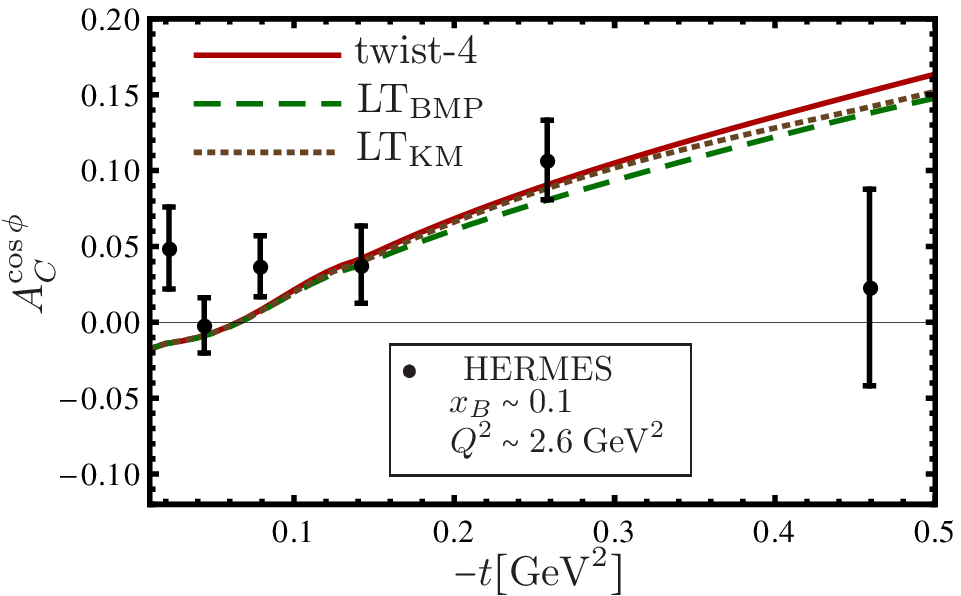}
\caption{
The single electron beam spin asymmetry (upper panel) in the charge odd sector and the
unpolarized beam charge asymmetry (bottom panel) measured by the HERMES collaboration \cite{Airapetian:2012mq}
at $\xB\sim 0.1$ and $\Q^2\sim 2.5\, \GeV^2$ in six $t$-bins.
Curves are described in Fig.~\ref{Fig:HALLA}.
}
\label{Fig:HERMES}
\end{center}
\end{figure}
%
The fixed target HERMES experiment had both $e^+$ and $e^-$ beams available and the collaboration provided measurements with an unpolarized
target of both beam spin asymmetry (\ref{A^{sin(n phi)}_{{LU},mp}}), including the charge-odd ones, and beam charge asymmetry (\ref{A^{cos(nphi)}_C}).
The main data set \cite{Airapetian:2001yk,Airapetian:2006zr,Airapetian:2009aa,Airapetian:2012mq} was extracted by using the missing mass technique, however, also fully exclusive measurements of the beam spin asymmetry were
performed with a recoil detector \cite{Airapetian:2012pg}.

In Fig.~\ref{Fig:HERMES} we display the data~\cite{Airapetian:2012mq} for the $n=1$ harmonics of the charge-odd electron beam spin asymmetry
$A^{\sin(\phi)}_{{\rm LU},{\cal I}}$ [upper panel] and charge asymmetry $A^{\cos(\phi)}_C$ [lower panel] for an unpolarized proton versus $-t$
for $\xB\approx 0.1$ and $\Q^2\approx 2.6\, \GeV^2$.
Note that the mean values of kinematical parameters for these data are correlated, in particular the mean $\langle\Q^2\rangle$ increases with growing $|t|$,
and thus the $-t/\Q^2$-ratio is for the highest available $-t$ value  $-t/\Q^2 \simeq 0.12$.
Furthermore, both asymmetries vanish at $t=t_{\rm min}$ which is also the case for the predictions, however,
it is not visible in the plots since our lowest value $-t$ is still larger than $-t_{\rm min}$.

Typically for standard GPD model predictions,  the beam spin asymmetry comes out to be too large (in absolute value)
and the prediction increases further for larger $-t$ values going over from LT$_{\rm KM}$ (dotted curve) to LT$_{\rm BMP}$ (dashed curve).
As  observed for CLAS kinematics, shown in Fig.~\ref{Fig:CLAS},
adding the remaining corrections (solid curve) implies only a very slight change of the predictions for the beam spin asymmetry.
Apart from the small changes of the dominant CFF $\cffG_{++}$, see solid and dashed curves in the left panel of Fig.~\ref{Fig:CFFs}, the net result is also influenced, presumably, by the excitation of higher odd and even harmonics in the interference and DVCS square term, respectively.
We remind that the denominator in the definition of asymmetries has a $\phi$-dependence,
and thus the  $n=1$ harmonics of the asymmetries are polluted by higher harmonics, see e.g.~Eqs.~(\ref{Sigma_{BS}})--(\ref{A^{sin(n phi)}_{{LU},mp}}).

The beam charge asymmetry is shown in Fig.~\ref{Fig:HERMES}, bottom panel.  As explained above, the real part of the dominant CFF $\cffG_{++}$ can be
small in the valence-to-sea quark transition region, which is consistent with the measurements. Nevertheless, it is not automatically guaranteed
that standard GPD models describe the HERMES data,  as the \GK model does, since the prediction depends very much on model details.
The \GK  model prediction proves to be very stable against power corrections (compare dotted, dashed and solid curves), but this
stability seems to be accidental rather than generic. We were not able to trace its precise origin.

\subsection{Fixed target kinematics (polarized proton)}

DVCS measurements on a polarized proton allow for a disentanglement of the various CFF species.
The HERMES collaboration provided the most complete set of DVCS measurements up to date in terms of asymmetry harmonics.
Apart from the measurements on an unpolarized target, a transversely polarized target for both $e^+$ and $e^-$ beams
was available \cite{Airapetian:2008aa,Airapetian:2011uq}
and measurements on a longitudinally polarized target were  performed with a positron beam~\cite{Airapetian:2010ab}.
The HERMES data allow at least in principle to access the imaginary and real parts of all CFFs, where,
however, suppressed contributions are very noisy, see the random variable map based on twist-two dominance hypothesis,
described in Ref.~\cite{Kumericki:2013br}. Such an analysis shows that besides the CFF $\cffH_{++}$ also the CFF $\cfftH_{++}$ is
constrained by measurements on a longitudinal polarized target, see also local CFF fits~\cite{Guidal:2010de}.

Proton spin dependent cross sections and single spin proton asymmetries are governed by the interference term and can be
utilized to address the imaginary parts of further CFF combinations. In particular for a longitudinally polarized proton the interference term
is governed by the $n=1$ odd harmonic, which is very sensitive to $\cfftG_{++}$ (or $\cfftH_{++}$).
single longitudinally polarized proton spin asymmetries $A_{{\rm UL},\cdots}(\phi)$ and their Fourier coefficients are defined in full analogy to the
single electron beam spin asymmetries in Eqs.~(\ref{A_{LU}(phi)})--(\ref{A^{sin(n phi)}_{{LU},mp}}),
i.e., replace the beam spin by the target spin.

%
\begin{figure}[t]
\begin{center}
\includegraphics[width=8cm]{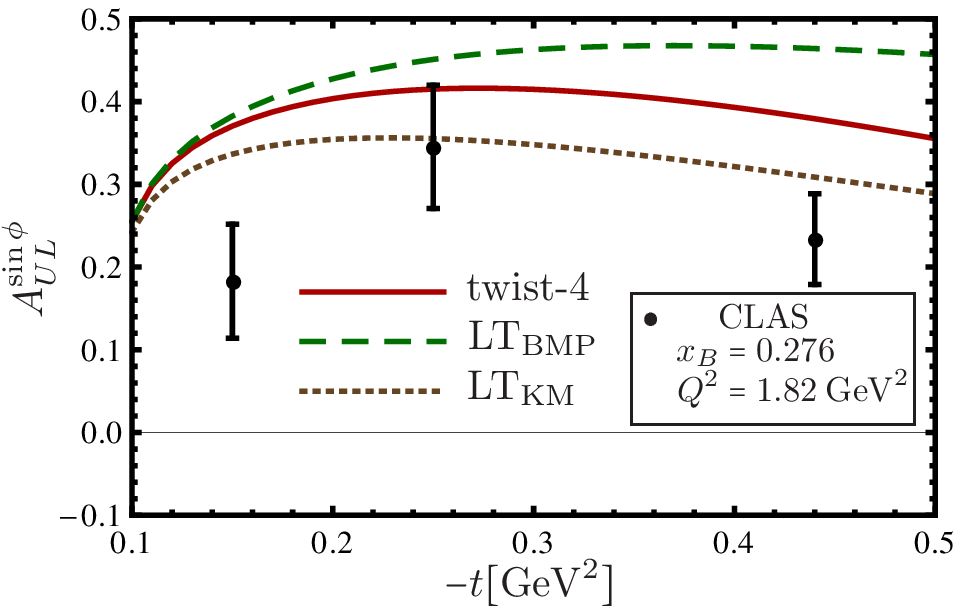}
\includegraphics[width=8cm]{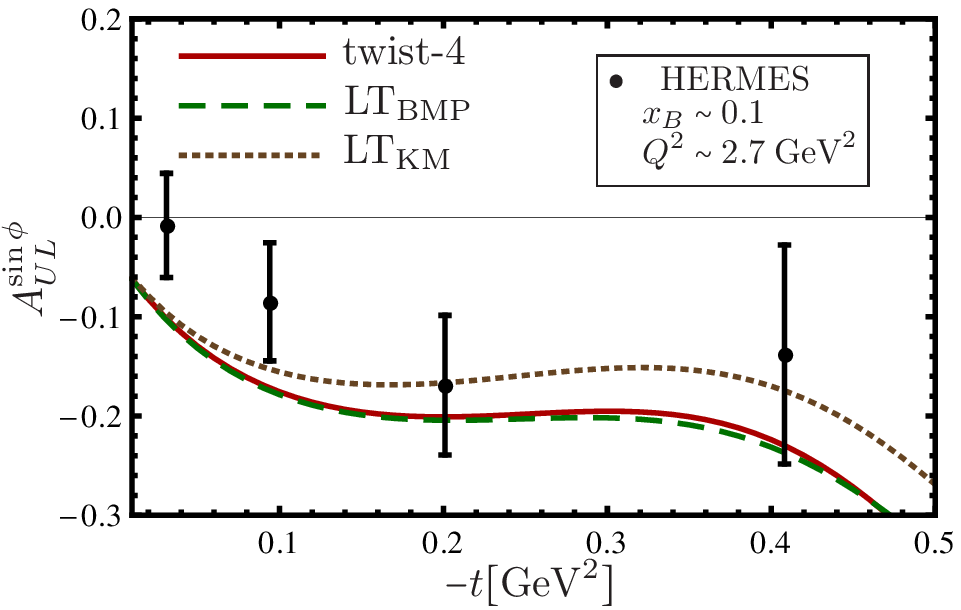}
\caption{
The longitudinal proton spin asymmetry from CLAS (upper panel)  \cite{Chen:2006na} , measured with an electron beam
at $\xB=0.276$ and $\Q^2=1.82\, \GeV^2$ and HERMES (lower panel) \cite{Airapetian:2010ab}, measured with a positron beam
at $\xB\sim 0.1$ and $\Q^2\sim 2.5\, \GeV^2$, versus $-t$. Curves are described in Fig.~\ref{Fig:HALLA}.
}
\label{Fig:AUL}
\end{center}
\end{figure}
%
The single longitudinally polarized proton spin asymmetry $A_{\rm UL}^{\sin(\phi)}$ was measured by
the CLAS collaboration \cite{Chen:2006na} with an electron beam and by the HERMES collaboration \cite{Airapetian:2010ab} with a positron beam.
These data are shown in the top and bottom panel in Fig.~\ref{Fig:AUL}, respectively.
(Again, this asymmetry vanishes at $t=t_{\rm min}$ but this point is outside the plotted region.)
For both the CLAS measurement at $\xB=0.276$ and $\Q^2=1.82\, \GeV^2$ with $-t/\Q^2 \le  0.24$
and HERMES measurements  the difference between LT$_{\rm KM}$ and LT$_{\rm BMP}$ is rather large
(compare dotted and dashed curve). Note that the robustness of $\cfftG_{0+}$, demonstrated in the right panel of Fig.~\ref{Fig:CFFs},
does not hold for the CFF $\cfftH_{++}$, which increases if we change from LT$_{\rm KM}$ to LT$_{\rm BMP}$.
A closer look reveals also that the longitudinal helicity flip CFF $\cfftH_{0+}$ plays an important role in
the dominant $n=1$ odd harmonic $s^{\cal I}_{1,{\rm LP}}$ of the interference term.
Adding the remaining kinematical higher-twist corrections (solid curve) reduces the difference between
LT$_{\rm KM}$ and LT$_{\rm BMP}$ predictions for CLAS kinematics considerably, but has very little effect for
HERMES kinematics, at least for the \GK model that we employ here.

Let us add that the target helicity flip CFFs  $\cffE_{++}$ and $\cfftE_{++}$ are much less constrained.
Because of the kinematical suppression factors that accompany these CFFs, mainly proportional to $-t/\M^2$,
and the pollution by contributions of proton helicity conserving CFFs, we expect that kinematical twist corrections are rather important
if one attempts to interpret transverse target observables in terms of GPDs $\gpdE$ or $\gpdtE$.

\subsection{Collider kinematics}
\label{sec:ColliderKinematics}
The dominant contribution in the small $\xB$ region arises from the `pomeron' exchange, which is included in the
small $\xi$ behavior of sea-quark GPD $H^{\rm sea}$  (and gluon GPD which enters explicitly at the NLO through the contribution of the box diagram).
It remains an open problem, related to the nucleon spin puzzle, whether also the GPD $\gpdE$ contains such a behavior.
Not much is known phenomenologically about the small $\xi$ behavior of GPD $\gpdtH$.
As a working hypothesis, we will assume that all of them and also the GPD $\xi \gpdtE$ are unimportant in the collider kinematics.

From Eqs.~(\ref{cffFbmp_{++}}), (\ref{cffFbmp_{0+}}), and (\ref{cffFbmp_{-+}}) we find with $t_{\min}\propto \xB^2 \approx 0$
for the CFFs is the BMP basis
\begin{align}
\cffHbmp_{++}\!= &\,\C_0\!\circledast \!\gpdH + \frac{-t}{Q^2}\!\left\{
\frac{1}{2}\C_{0}\!\circledast \!  \gpdH  - \C_{1}\!\circledast \!\gpdH  +  \xi^2 \partial^2_\xi\, \C_2\!\circledast \!\gpdH
\right\},
\notag
\\
\cffHbmp_{0+}\!= &-\!\frac{\sqrt{2} \sqrt{-t}}{Q}\,  \xi \partial_\xi\, \C_1\!\!\circledast \!\gpdH\,,
\quad
\cffHbmp_{-+}\!=  \frac{-t}{Q^2}\,\xi \partial_\xi^2 \xi\, \C_{1}^{(+)}\!\circledast \!\gpdH,
\label{cffHbmp-smallxB}
\end{align}
and analogous relations for CFFs $\cffEbmp_{a +}$ in terms of GPD $\gpdE$.
Note that the `pomeron' behavior of GPD $H$ implies the similar behavior of both photon helicity-conserving and helicity-flip amplitudes.

Going over to the BMJ CFF basis by means of the transformation rules in Eq.~(\ref{CFF-F2F}), where the kinematical
factors (\ref{varkappa}) can be safely approximated as
$$\varkappa = \frac{-2t}{\Q^2}\quad \mbox{and}\quad \varkappa_0 = \frac{\sqrt{-2t \Q^2}}{Q^2+t},$$
one obtains with Eq.~(\ref{cffHbmp-smallxB}) the following expressions
\begin{subequations}
\begin{align}
\cffH_{++}\!= &\,\C_0\!\circledast \!\gpdH + \frac{-t}{Q^2}\!\biggl[
\frac{3}{2}\C_{0}- \C_{1}+ 2\xi\partial_\xi \C_{1}    +  \xi^2\partial^2_\xi\, \C_2
\notag\\
&\, \phantom{\C_0\!\circledast \!\gpdH + \frac{-t}{Q^2}} - \frac{t}{Q^2+t}\mathbb{T}\biggr]\!\circledast \!\gpdH\,,
\\
\cffH_{0+}\!= &\, \frac{\sqrt{2} \sqrt{-t}}{Q}\biggl[ \C_0 +  \xi \partial_\xi\, \C_1\! - \frac{t}{Q^2+t}\mathbb{T} \biggr]\!\circledast \!\gpdH\,,
\\
\cffH_{-+}\!= &\, \frac{-t}{Q^2} \biggl[\C_{0}+2\xi\partial_\xi \C_{1} + \xi \partial_\xi^2 \xi\, \C_{1}^{(+)} - \frac{t}{Q^2+t}\mathbb{T}  \biggr]\!\circledast \!\gpdH,
\end{align}
where
\begin{align}
\mathbb{T} &= \frac{3}{2} \C_{0}- \C_{1}  + 2 \xi\partial_\xi\, \C_{1} + \xi^2 \partial_\xi^2\, \C_{2}  + \xi \partial_\xi^2\xi\, \C_1^{(+)}.
\label{mathbb{T}}
\end{align}
\end{subequations}

For this analysis we can assume that the GPD behaves (for $\alpha(t)>0$)  as
\begin{align}
F(\xi/x,\xi,t) &\stackrel{\xi\to 0}{=} (\xi/x)^{-\alpha}\, r(x,t)\,,
\end{align}
where $\alpha\equiv \alpha(t)$ is the effective leading Regge trajectory and  $r(x,t)$ is the residue function.
It is model-dependent and can be calculated similarly to perturbative QCD corrections in, e.g., Sec.~5 of \cite{Mueller:2013caa}.
For a RDDA model such as the one used in \GK, the $x$-dependence of the
residue function is given by a hypergeometric function
\begin{align}
r(x,t) &= r(t)\, {_2F_1}\biggl({\alpha/2,\alpha/2+1/2 \atop b+3/2}\big|x^2\biggr)\,,
\end{align}
where $b$ is the so-called profile parameter and $r(t)$ contains the residual $t$-dependence.
Note that the small-$\xi$ approximation (\ref{F^q(xi/x,xi)-smallxi}) of our toy GPD model (\ref{GPD-toy}) follows by setting $\alpha=1/2$ and $b=1$,
where the hypergeometric function reduces to a combination of elementary functions.

With this kind of models all kinematic twist corrections can be calculated analytically for general (positive) $b$ and $\beta$ values.
To this end the convolution integrals in the imaginary parts (\ref{Im-conv}) with the kernels (\ref{t-coef}) can be  obtained from
\begin{align}
   F(\xi,\xi) &\stackrel{\xi \to 0}{\approx}
r(t)  \frac{\Gamma\left(\frac{3}{2}+b\right) \Gamma(1+b-\alpha)}{\Gamma\left(1+b-\frac{\alpha }{2}\right)
\Gamma\left(\frac{3}{2}+b-\frac{\alpha }{2}\right)}  \xi^{-\alpha}\,,
\end{align}
and
\begin{widetext}
\begin{subequations}
\label{tF-smallxB}
\begin{align}
\int_{\xi}^1\frac{dx}{x} t_1(x) \frac{F(\xi/x,\xi)}{ F(\xi,\xi)}  &\stackrel{\xi \to 0}{\approx}
\frac12\Big[S_1\Bigl(b-\frac{\alpha}{2}\Bigr) -S_1\Bigl(b+\frac{1}{2}-\frac{\alpha}{2}\Bigr)\Big]
+\frac12\Big[S_1\Bigl(\frac{\alpha}{2} -\frac{1}{2}\Bigr)-S_1\Bigl(\frac{\alpha}{2}-1\Bigr)\Big],
\end{align}
\begin{align}
\int_{\xi}^1\frac{dx}{x} t^{(+)}_1(x)\frac{F(\xi/x,\xi)}{ F(\xi,\xi)} &\stackrel{\xi \to 0}{\approx}
\frac{1+b-\alpha}{(1+b-\frac{\alpha}{2}) (\alpha -1)}
- \int_{\xi}^1\frac{dx}{x} t_1(x) \frac{F(\xi/x,\xi)}{ F(\xi,\xi)} - \frac{1}{\xi}\cdot\text{const.}\,,
\label{tF-smallxB-b}\\[1mm]
\int_{\xi}^1\frac{dx}{x} t_2(x) \frac{F(\xi/x,\xi)}{ F(\xi,\xi)} &\stackrel{\xi \to 0}{\approx}
-\frac{1}{2} \int_{\xi}^1\frac{dx}{x} t_1(x) \frac{F(\xi/x,\xi)}{ F(\xi,\xi)}
+ \frac18\left[S_1\Bigl(b\!-\!\frac{\alpha }{2}\Bigr)-S_1\Bigl(\frac{1}{2}\!+\!b-\frac{\alpha }{2}\Bigr)-S_1\Bigl(\frac{\alpha }{2}\!-\!1\Bigr)+S_1\Bigl(\frac{\alpha\! -\!1}{2}\Bigr)\right]^2
\notag\\
&\phantom{{}\stackrel{\xi \to 0}{\approx}{}}  {+\frac12}\Big[S_2\big(1+2 b-\alpha\big)- S_2\big(\alpha -1\big)\Big],
\end{align}
\end{subequations}
\end{widetext}
where $S_k$ are the usual harmonic functions.  The term proportional to $1/\xi$ in (\ref{tF-smallxB-b}) is annihilated by the application of the
differential operator $\partial_\xi \xi$ and does not contribute to the final answer.


The set of formulae (\ref{tF-smallxB}) allows one to understand the behavior of twist corrections also for the special class
of GPD models that were conjectured in Refs.~\cite{Shuvaev:1999fm,Shuvaev:1999ce,Martin2009zzb}
and the GPD models obtained from a $t$-decorated PDF
by taking values $b=\alpha$ and $b\to\infty$, respectively. The \GK model corresponds to $b=2$. It turns out
that assuming the dominant effective `pomeron' trajectory with $0.9 < \alpha < 1.4$  and $b>1$,
the corrections can be quoted, generically, as
\begin{subequations}
\label{cffH-small-xB}
\begin{align}
\cffH_{++} &= \C_0\!\circledast \!\gpdH + \frac{-t}{Q^2} \left(1+\dots\right) \C_0\!\circledast \!\gpdH\,,
\\
\cffH_{0+} &= \sqrt{\frac{-t}{2Q^2}} \left(1+ 2\times \dots\right) \C_0\!\circledast \!\gpdH,
\label{cffH_{0+}-small-xB}
\\
\cffH_{-+} &=\frac{-t}{Q^2} \left(1+ \dots\right) \C_0\!\circledast \!\gpdH.
\end{align}
\end{subequations}
Here the ellipses contain terms that are numerically less important, including those that are additionally
suppressed in $-t/(\Q^2+t)$ and are determined by the convolution with the $\mathbb{T}$-kernel (\ref{mathbb{T}}).
Such corrections are roughly two times larger for $\cffH_{0+}$ compared to the other cases.
Comparing these expressions with the LT$_{\rm BMP}$ set in Eq.~(\ref{BMP-convention}), we see that they essentially coincide
for transverse CFFs $\cffH_{++}$ and $\cffH_{-+}$ so that in these
cases the remaining twist-four corrections (not included in  LT$_{\rm BMP}$) are small, however, they are significant and
reduce the longitudinal CFF  $\cffH_{0+}$.  This generic behavior is illustrated for the \GK model with $-t/\Q^2= 1/4$ in
Fig.~\ref{Fig:CFFs} [left panel]. In this case the term shown by the ellipses in (\ref{cffH_{0+}-small-xB})
is of order one, yielding $\cffH_{0+} \approx \sqrt{\frac{-2t}{Q^2}} \C_0\!\circledast \!\gpdH$.
Hence, the full result to the twist-four accuracy is reduced w.r.t. LT$_{\rm BMP}$ by the factor $1+t/Q^2 = 3/4$.

Since the intercept of the effective `pomeron' exchange is larger than one, the DVCS cross section
overwhelms at small-$\xB$ the BH cross section and the integration over $\phi$ suppresses the interference terms.
Hence, in collider experiments one has access directly to the DVCS cross section.
The unpolarized $t$-differential DVCS cross section within the Hand convention \cite{Hand:1963bb} is expressed
by the $n=0$ DVCS harmonic (\ref{Res-Mom-DVCS-0-imp}),
\begin{align}
\frac{d\sigma^{\rm DVCS}}{dt} &\simeq \frac{\pi \alpha_{\rm em}^2}{\Q^4 }
\xB^2\Big[ {\cal C}({\cal F}_{++},{\cal F}_{++}^\ast) + {\cal C}({\cal F}_{-+},{\cal F}_{-+}^\ast )
\nonumber\\
&\phantom{\simeq \frac{\pi \alpha_{\rm em}^2}{\Q^4 }
\xB^2\Big[} +
2 \varepsilon(y)\, {\cal C}({\cal F}_{0+},{\cal F}^\ast_{0+} ) \Big],
\label{XDVCS-approx}
\end{align}
where the $\mathcal{C}$-coefficient (\ref{Def-CDVCSunp}) can be approximated by
\begin{align}
{\cal C}({\cal F},{\cal F}^\ast) \approx
\left|{\cal H} \right|^2  - \frac{t}{4  \M^2} \left|{\cal E} \right|^2
\end{align}
and the photon polarization parameter (\ref{varepsilon}), i.e., the ratio of longitudinal to transverse photon flux, can be set to
$\varepsilon(y) = 2(1-y)/(2-2y+ y^2)$.

%
\begin{figure}
\begin{center}
\includegraphics[width=8cm]{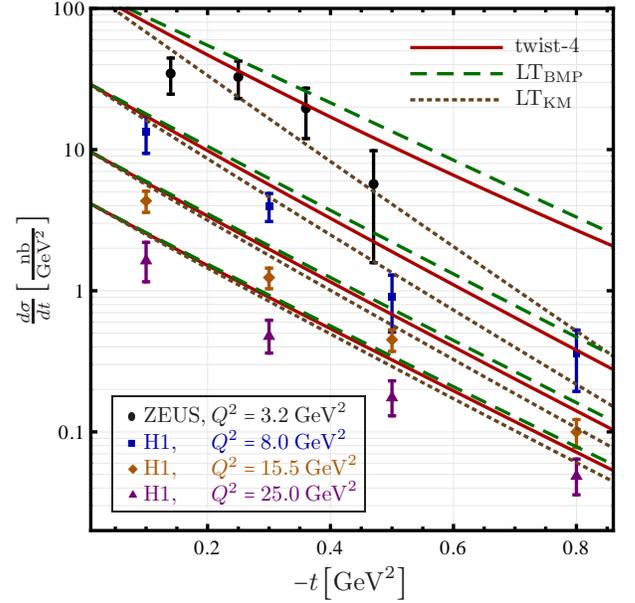}
\caption{The DVCS cross section versus $-t$ for various $\Q^2$ values from the H1 (squares, diamonds, triangles) \cite{Aktas:2005ty,Aaron:2009ac} and ZEUS (circles) \cite{Chekanov:2008vy} collaboration. Curves are described in Fig.~\ref{Fig:HALLA}.
}
\label{Fig:H1ZEUS}
\end{center}
\end{figure}
%
The H1 \cite{Aktas:2005ty,Aaron:2009ac} and ZEUS \cite{Chekanov:2008vy} data are shown in Fig.~\ref{Fig:H1ZEUS} together with the \GK model predictions versus $-t \le 0.8\,\GeV^2$ for different $\Q^2$
values in the range $3.2\,\GeV^2 \le \Q^2\le 25\,\GeV^2$.
In the LT$_{\rm KM}$ approximation (\ref{KM-convention}) [dotted curves] the \GK model describes the data well
(this RDDA model works at LO since GPD evolution is replaced by PDF evolution).
Going over to LT$_{\rm BMP}$  (\ref{BMP-convention})  (dashed curves) produces a huge correction for $\Q^2=3.2\, \GeV^2$ and even at
$\Q^2=8\, \GeV^2$ the effect is large. This is mainly caused by the fact that BMP skewedness parameter is smaller than the KM one
$\xi_{\rm BMP} = (1+t/\Q^2)\xi_{\rm KM}$, which produces a significant enhancement of the helicity-conserving bilinear CFF combination
\begin{align}
\frac{{\cal C}^{\rm BMP}(\cffF_{++},\cffF^\ast_{++})}{{\cal C}^{\rm KM}(\cffF_{++},\cffF^\ast_{++})}
& \stackrel{\rm LT}{\approx}
\frac{\left(\!1-\frac{t}{Q^2}\!\right)^2}{\left(\!1+\frac{t}{Q^2}\!\right)^{2\alpha}}\,.
\notag
\end{align}
Numerically, e.g., for $t=-0.8\, \GeV^2$ and $\Q^2=3.2\, \GeV^2$ this is an enhancement of roughly a factor three,
whereas for $\Q^2=8\, \GeV^2$ it is a factor $\sim 3/2$.
In addition, LT$_{\rm BMP}$ approximation (\ref{BMP-convention}) includes helicity-flip contributions (if translated to the BMJ basis), which
are commonly not considered in data analyzes.
This induced longitudinal-to-transverse helicity-flip CFF can be estimated, according to the above discussion, as
\begin{align}
\frac{{\cal C}^{\rm BMP}(\cffF_{0+},\cffF^\ast_{0+})}{{\cal C}^{\rm KM}(\cffF_{++},\cffF^\ast_{++})}
& \stackrel{\rm LO}{\sim} \frac{-2 t}{\Q^2\left(\!1+\frac{t}{Q^2}\!\right)^{2}},
\notag
\end{align}
and there is also a much smaller contribution bilinear in the transverse flip CFFs $\cffF_{-+}$, proportional to $t^2/\Q^4$.

Taken together, these two effects produce at $t=-0.8\, \GeV^2$ the enhancement of the LT$_{\rm BMP}$ predictions by roughly a factor of six (four)
as compared to LT$_{\rm KM}$ at  $\Q^2=3.2\, \GeV^2$ ($\Q^2=8\, \GeV^2$),
respectively, cf.~dotted and dashed curves in Fig.~\ref{Fig:H1ZEUS}.
The main effect of the remaining twist-four contributions is to reduce the longitudinal-to-transverse helicity-flip CFF, so that
the full kinematic higher-twist correction to the cross section is somewhat reduced as well, compare the solid and dashed curves.

\section{Conclusions}
\label{Sec:summary}

Our analysis has been based on the recent results in~\cite{Braun:2011zr,Braun:2011dg,Braun:2012bg,Braun:2012hq} (BMP) where the
DVCS tensor has been calculated in QCD to twist-four accuracy taking into account the descendants of the leading-twist operators.
We refer to these corrections as kinematic as they are expressed in terms of the leading-twist GPDs.
It has been checked that this addition restores gauge- and translation-invariance of the DVCS amplitudes at the considered order and their structure
is consistent with QCD collinear factorization. The final result is presented in Ref.~\cite{Braun:2012hq} as the expansion of the
DVCS tensor in terms of scalar invariant functions that can be identified with photon helicity dependent Compton form factors (CFFs)
in a certain reference frame. The twist expansion of the CFFs is organized in terms of two parameters
\begin{align}
    \frac{-t}{Q^2} \quad \text{and} \quad \frac{t_{\rm min}-t}{Q^2},
\notag
\end{align}
where the second one is related to the target transverse momentum in the BMP frame (\ref{xiPperp}).
In the case of a scalar target all target mass corrections are absorbed in $t_{\rm min} \propto x_B^2 m^2$ whereas for the
proton some additional terms in $m^2/Q^2$  arise due to spinor algebra; their structure is strongly constrained by the requirement
that certain harmonics in the cross section vanish for $t\to t_{\rm min}$.
In this work we present a detailed study of kinematic power corrections $\sim 1/Q, 1/Q^2$ to several key DVCS observables that
incorporates these developments.

Calculation of the observables starting from a given set of CFFs is by itself a nontrivial task. Instead of the direct calculation
in terms of BMP CFFs we use another set of CFFs, suggested by Belitsky, M{\"u}ller and Ji (BMJ)~\cite{Belitsky:2010jw,Belitsky:2012ch} at the intermediate
step. The transition between BMP and BMJ CFFs is a purely kinematic transformation that can be thought of as Lorentz transformation
to a different reference frame. We do this transformation exactly, and also use exact expressions for the observables in terms of the
BMJ CFFs available from Ref.~\cite{Belitsky:2012ch}. In this way the results for physical observables are the same as the ones that
one would obtain by a direct calculation employing the original BMP parametrization.

In order to discuss the impact of kinematic higher-twist corrections one has to formulate the leading-twist approximation
that would serve as the reference.
An important point that is often overlooked in phenomenological studies is that
this choice is not unique as the leading-twist calculations are intrinsically ambiguous.
The reason is that in the DVCS kinematics the four-momenta of the initial and final photons and protons do not lie in
one plane. Hence the distinction of longitudinal and transverse directions is convention-dependent. In the Bjorken
high-energy limit this is a $1/Q$ effect.
The freedom to redefine large `plus' parton momenta by adding smaller transverse components has two consequences.
First, the relation of the skewedness parameter $\xi$ appearing as an argument in GPDs
to the Bjorken variable $\xB$ may involve power suppressed contributions.
Second, such a redefinition generally leads to excitation of the subleading photon helicity-flip amplitudes.
Any attempt to compare the calculations with and without kinematic power corrections must start with specifying the
precise conventions, i.e.~the \emph{definition} of what is meant by `leading-twist' to the power accuracy.
Viewed in this context, the kinematic power corrections calculated in~\cite{Braun:2012bg,Braun:2012hq}
are convention-dependent as well. This dependence exactly cancels the convention-dependence of the leading twist
so that the full result is unambiguous (to the stated $1/Q^2$ accuracy).

The convention (\ref{KM-convention}) used by Kumeri{\v c}ki and M\"uller in global DVCS fits \cite{Kumericki:2009uq,Kumericki:2010fr,Kumericki:2011zc,Kumericki:2013br},
adopting the BMJ cross section formula from Ref.~\cite{Belitsky:2010jw,Belitsky:2012ch}, is in practical terms not very different
from the VGG convention used by Guidal, and also the convention used by Kroll, Moutarde, and Sabatie in \cite{Kroll:2012sm}.
We have, therefore, overtaken  Eq.~(\ref{KM-convention}) as the `standard' leading-twist LO approximation in our study.
The conventions used by BMP in~\cite{Braun:2012bg,Braun:2012hq} are quite different so that the
corresponding leading-twist approximations that we refer to as  LT$_{\rm BMP}$, defined in (\ref{BMP-convention}) vs.~LT$_{\rm KM}$
defined in (\ref{KM-convention}), are rather different as well. In particular the change in the definition of the skewedness
parameter has a large effect. It turns out that at least for some observables this difference presents
the main source (numerically) of kinematic corrections,
whereas the remaining higher-twist contributions to the BMP CFFs are rather mild.

Let us conclude about what we have learned from our studies for the phenomenological description of DVCS measurements.
Presently, the DVCS data are mainly discussed at LO and LT accuracy.
Changing to the LT$_{\rm BMP}$ convention (\ref{BMP-convention}) allows one to include the bulk of the calculated  higher twist corrections and
will increase the predicted value of (unpolarized) cross sections and longitudinal spin asymmetries with growing $-t/\Q^2$.
For standard GPD model predications this is desired with respect to the unpolarized HALL A cross section measurements,
shown in Fig.~\ref{Fig:HALLA}, however, it will further increase the tension with respect to CLAS and HERMES beam spin measurements,
see Figs.~\ref{Fig:CLAS} and \ref{Fig:HERMES}. Implementing the BMP convention (\ref{BMP-convention}) in the global
KM fitting framework \cite{Kumericki:2009uq,Kumericki:2010fr,Kumericki:2011zc,Kumericki:2013br},
a hybrid of GPD model and dissipative approach, can be straightforwardly done at leading twist and
should lead, effectively, to a reparametrization of the GPD extractions.
In future phenomenological studies it is highly advisable to implement besides the
kinematical corrections also perturbative next-to-leading order corrections and, certainly, GPD evolution must be taken properly into account.
This requires a change to global fitting routines that are based on appropriate GPD model parametrizations.
All this is partially done, and can be fully implemented in the KM routines that are based on Mellin-Barnes
integral representation. Surely, one can work in any other representation, too, however, in this case the relevant technology,
including a flexible GPD parametrization, has still to be developed.
Let us also mention with respect to global GPD fitting, which includes nowadays also Deeply Virtual Meson Production
in exclusive channels \cite{Meskauskas:2011aa,Lautenschlager:2013uya}, that it remains a challenge to work out the
kinematical corrections for Deeply Virtual Meson Production. Finally, we have learned that the
corrections are generically of order ${\cal O}(-t/\Q^2)$ (for some observables much smaller).  Hence, keeping
kinematical twist-four corrections under control requires an upper bound for the photon virtuality of the order of $\Q^2 \gtrsim - 4t$.
This constrain provides an important requirement for addressing the three-dimensional
picture of the proton in impact parameter space \cite{Burkardt:2002hr}, where one presumably needs to know the $-t$ dependence of GPDs up
to at least $\sim 1$-$2\, \GeV^2$.  Thus, large $\Q^2$ values are needed, which can be reached at proposed collider experiments
such as eRHIC \cite{Deshpande:2012bu}, for a comprehensive model study see Ref.~\cite{Aschenauer:2013qpa}.

To summarize our findings, the finite-$t$ kinematic power corrections to DVCS observables are significant and
must be taken into account in the data analysis aiming to extract GPDs at a quantitative level.
This result removes an important source of uncertainties in the QCD predictions for
intermediate photon virtuality square $Q^2\sim 1$-$5\,\GeV^2$ that are accessible in the existing
and planned experiments.

\section*{Acknowledgements}
This work was supported by the DFG, grant BR2021/5-2.
DM is grateful to the theory group of the Regensburg University
for the warm hospitality during his visits, where essential parts of
this work were done. He is also indebted to M.~Guidal, P.~Kroll, H.~Moutarde, and F.~Sabatie
for numerical comparisons and discussions.


\appendix

\section{Translation between BMP and BMJ conventions}
\label{BMP-BMJ}
Aim of this Appendix is to provide a detailed comparison of the notation and conventions
used in Refs.~\cite{Braun:2012hq} (BMP) and \cite{Belitsky:2012ch} (BMJ).
The final results of this translation are the expressions given below in Eqs.~(\ref{mathfrak++})~--~(\ref{mathfrak-+}) for the
BMJ helicity amplitudes that include finite-$t$ and target mass corrections calculated in~\cite{Braun:2012hq}.
The presentation is deliberately detailed as we think that a scrupulous comparison of
different conventions is important for further studies.
We retain the original notation in Refs.~\cite{Braun:2012hq,Belitsky:2012ch} whenever possible.

\subsection{BMP conventions and results}
\label{BMP-notation}

\subsubsection{BMP conventions}

The DVCS process (\ref{DVCS-process}) reads in BMP notation as \cite{Braun:2012hq,Braun:2012bg}
\begin{align}
\gamma^*(q)+ N(p,s) \longrightarrow \gamma(q')+ N(p',s')\,.
\label{DVCSprocess-BMP}
\end{align}
The DVCS tensor $\mathcal{A}_{\mu\nu}$ is defined by the following expression:
\begin{align}
\label{Amunu-def}
 &\mathcal{A}_{\mu\nu}(q,q',p)=\\
 &=i\! \int\!\! d^4 x\, e^{-i(z_1q-z_2 q')\cdot x }
\vev{p',s'|T\{j_\mu(z_1x)j_\nu(z_2x)\}|p,s},
\nonumber
\end{align}
where $j_\mu(z_1x)$ and $j_\nu(z_2x)$ are the electromagnetic currents at the indicated space-time positions, $z_1,z_2$ are real numbers and
it is assumed that $z_1-z_2=1$. Note that the tensor $\mathcal{A}_{\mu\nu}$ should not depend on $z_1+z_2$.
This property is referred to as translation invariance in Refs.~\cite{Braun:2012hq,Braun:2012bg} and has been verified to the required (twist-four)
accuracy by explicit calculation.

BMP use the photon momenta, $q$ and  $q'$, to define a longitudinal plane spanned by the two light-like
vectors
\begin{align}
n=q'\,, \qquad \tilde n=-q+(1-\tau)\, q'\,,
\end{align}
where $\tau= t/(Q^2+t)$ with $Q^2=-q^2$.
For this choice the momentum transfer to the target
$$\Delta = p'-p= q-q'\,, \qquad t=\Delta^2$$
is purely longitudinal and the both --- initial and final state --- proton momenta have a nonzero
transverse component
\begin{align}\label{Pperp}
P_\mu&=\frac{1}{2\xi}\left(\bar n_\mu-\tau n_\mu\right) + P_{\perp,\mu}\,,
\notag\\
| P_\perp |^2&=-m^2-\frac{t}{4}\frac{1-\xi^2}{\xi^2}\,.
\end{align}
Here, $P$ is  defined as {\em average} of nucleon momenta and the longitudinal momentum fraction in the
$t$-channel is defined w.r.t.~the light-like vector $n=q'$,
\begin{align}
P \equiv P^{\rm BMP} = \frac12(p+p')\,,
\quad \xi \equiv \xi^{\rm BMP} = -\frac{\Delta\cdot q'}{2P\cdot q'}\,.
\end{align}
$|P_\perp|^2$ can equivalently be written in terms of kinematic invariants as
\begin{align}
 |P_\perp |^2 &= \frac{1-\xi^2}{4\xi^2} (t_{\rm min}-t)\,, \;\; t_{\rm min} = -\frac{4m^2\xi^2}{1-\xi^2} \,,
\end{align}
where the BMP skewedness parameter $\xi$ can be expressed in terms of the Bjorken scaling variable as shown in Eq.~(\ref{xB2xiBMP}).

BMP write the DVCS amplitude $\mathcal{A}_{\mu\nu}$ as decomposition in scalar amplitudes in terms
of the photon polarization vectors that are chosen as follows:
\begin{align}
\varepsilon^0_\mu&=-\left(q_\mu-q'_\mu {q^2}/{(q\cdot q')}\right)/{\sqrt{-q^2}}\,,\notag\\
\varepsilon^\pm_\mu&=(P^\perp_\mu\pm i \bar P^\perp_\mu)/ {(\sqrt{2}|P_\perp|)}\,,
\label{varepsilon^a_mu}
\end{align}
where $P^\perp_\mu=g_{\mu\nu}^\perp P^\nu$, $\bar P^\perp_\mu=\epsilon_{\mu\nu}^\perp P^\nu$ and
\begin{align}
g_{\mu\nu}^\perp&=g_{\mu\nu}-(q_\mu q'_\nu+q'_\mu q_\nu)/(q\cdot q')+{q'_\mu}q'_\nu\,{q^2}/(q\cdot q')^2\,,
\notag\\
\epsilon_{\mu\nu}^\perp&=\epsilon_{\mu\nu\alpha\beta}{q^\alpha q'^\beta}/(q\cdot q')\,,\qquad {\epsilon^{\rm BMP}_{0123}=1}\,.
\end{align}
Normalization is such that $\varepsilon^+_\mu\varepsilon^{-\mu} = -1$\,,
$\varepsilon^0_\mu\varepsilon^{0\mu} = +1$. The pair $\varepsilon^\pm_\mu$ form an
orthonormal basis of two unit vectors in the transverse plane whereas $\varepsilon^0_\mu$
is a unit vector in longitudinal plane that is orthogonal to the real photon momentum
$q'$. Explicit construction uses a two-component spinor formalism and is explained in
Sec.~IIb in~\cite{Braun:2012bg}.

Using this basis, BMP write the DVCS amplitude $\mathcal{A}_{\mu\nu}$ (\ref{Amunu-def}) in terms of scalar (helicity) amplitudes defined as
\begin{align}
\!\mathcal{A}_{\mu\nu}={} &\varepsilon^+_{\mu} \varepsilon^-_{\nu} \mathcal{A}^{++}
+\varepsilon^-_{\mu} \varepsilon^+_{\nu} \mathcal{A}^{--}
+\varepsilon^0_{\mu} \varepsilon^-_{\nu} \mathcal{A}^{0+}
\notag\\
{}+{}&\varepsilon^0_{\mu} \varepsilon^+_{\nu} \mathcal{A}^{0-}\!
+\!\varepsilon^+_{\mu} \varepsilon^+_{\nu} \mathcal{A}^{+-}\!
+\!\varepsilon^-_{\mu} \varepsilon^-_{\nu} \mathcal{A}^{-+} 
\!.
\label{Amunu}
\end{align}
Note that a term proportional to $q'_\nu $ has been neglected since it does not contribute to any
observable.
Each helicity amplitude involves the sum over quark flavors, $\mathcal{A}=\sum e_q^2
\mathcal{A}_q$, and is written in terms
of the leading-twist GPDs $H^q,E^q,\widetilde H^q,\widetilde E^q$. For the GPD definitions
BMP follow Ref.~\cite{Diehl:2003ny}.
The results are written in terms of the vector and axial-vector bilinear spinors
\begin{align}\label{Dirac}
v^\mu=\bar u(p')\gamma^\mu u(p)\,, &&a^\mu=\bar u(p')\gamma^\mu \gamma_5u(p)\,
\end{align}
using shorthand notations
\begin{align}\label{Dirac2}
 v_\perp^\pm&=(v\cdot\varepsilon^\pm)\,,\quad a_\perp^\pm=(a\cdot\varepsilon^\pm)\,,
\notag\\
P_\perp^\pm&=(P\cdot \varepsilon^\pm) = - |P_\perp|/\sqrt{2}\,.
\end{align}
%

\subsubsection{BMP results for helicity amplitudes}
At leading twist only the  helicity-conserving amplitudes
$$\mathcal{A}^{\pm\pm}= \!\!\!\!\sum_{q=u,d,\dots}\!\!\!\! e_q^2 \mathcal{A}_q^{\pm\pm}$$
contribute. To the LO accuracy they read
\begin{align}
\mathcal{A}_q^{\pm\pm} =
\frac{v\cdot P}{2m^2}&  E^q\otimes{C}_0^- +\frac{v \cdot  q'}{q\cdot  q'}\, \xi\, M^q\otimes {C}_0^-
\notag\\
\pm \frac{a\cdot  \Delta}{4m^2}\, \xi\,&\widetilde E^q\otimes{C}_0^+ \pm \frac{a \cdot  q'}{q\cdot q'}\, \xi\,\widetilde H^q \otimes{C}_0^+\,,
\label{AnonflipLT}
\end{align}
where the shorthand notation
$$M^q(x,\xi,t)=H^q(x,\xi,t)+E^q(x,\xi,t)$$
for the `magnetic' GPD combination is used. The notation  $F\otimes C$ stands for the convolution of a
GPD $F$ with a coefficient function $C$:
$$
F\otimes C\equiv\int_{-1}^1\! dx\, F(x,\xi,t)\, C(x,\xi)\,,
$$
where the LO coefficient functions $C_0^\mp$ are given below in Eq.~(\ref{Coefff}).
The LT result (\ref{AnonflipLT}) extends to the following general decomposition
\begin{align}
\mathcal{A}_q^{\pm\pm}=
\frac{v\cdot P}{2m^2}\, \mathbb{V}^q_1+\frac{v \cdot  q'}{q\cdot  q'}\, \mathbb{V}^q_2
\pm \frac{a\cdot  \Delta}{4m^2}\, \mathbb{A}^q_1 \pm \frac{a \cdot  q'}{q\cdot q'}\, \mathbb{A}^q_2\,,
\label{Anonflip}
\end{align}
where in DVCS kinematics the bilinear spinors behave as
\begin{align}
\frac{v\cdot q'}{q\cdot q'} \sim \frac{a\cdot q'}{q\cdot q'} \sim
\frac{v\cdot P}{2m^2} \sim \frac{a\cdot \Delta}{4m^2} = \mathcal{O}(Q^0)\,.
\end{align}
The following expressions that include $\mathcal{O}(t/Q^2)$ and $\mathcal{O}(m^2/Q^2)$
corrections present the main result of Ref.~\cite{Braun:2012hq}:
\begin{widetext}
\begin{subequations}
\label{VAQ}
\begin{align}
\mathbb{V}^q_1
&=\Big(1-\frac{t}{2Q^2}\Big) E^q\otimes{C}_0^-
+\frac{t}{Q^2}E^q\otimes { C}_1^-
-\frac2{Q^2}\Big(\frac{t}\xi+2|P_\perp|^2\xi^2\partial_\xi\Big)\xi^2\partial_\xi
  E^q\otimes{C}_2^-
+\frac{8m^2}{Q^2}\xi^2\partial_\xi \xi\, M^q\otimes {C}_2^-\,,
\label{Vq1}\\
\mathbb{V}^q_2
&=\Big(1-\frac{t}{2Q^2}\Big)\xi\, M^q\otimes {C}_0^-
+\frac{t}{Q^2}\xi\,M^q\otimes{ C}_1^-
-\frac4{Q^2}\biggl[\Big(|P_\perp|^2\xi^2\partial_\xi+\frac{t}{\xi}\Big)\xi^2\partial_\xi-\frac{t}2\biggr]
 \xi\,M^q\otimes  {C}_2^-\,,
\label{Vq2}
\\
%
\mathbb{A}^q_1
&= \Big(1-\frac{t}{2Q^2}\Big) \xi\,\widetilde E^q\otimes{C}_0^+ +\frac{t}{Q^2}
\xi\,\widetilde E^q\otimes{C}_1^+
-\frac2{Q^2}\Big(\frac{t}\xi+2|P_\perp|^2\xi^2\partial_\xi\Big)\xi^2\partial_\xi
 \xi\, \widetilde E^q\otimes{C}_2^++\frac{8m^2}{Q^2}\xi^2\partial_\xi \widetilde
H^q\otimes{C}_2^+\,,
\label{Aq1}\\
\mathbb{A}^q_2
&= \Big(1-\frac{t}{2Q^2}\Big)\xi\,\widetilde H^q \otimes{C}_0^+
 +\frac{t}{Q^2}\xi\,\widetilde H^q\otimes {C}_1^+
-\frac4{Q^2}\biggl[\Big(|P_\perp|^2\xi^2\partial_\xi+\frac{t}{\xi}\Big)\xi^2\partial_\xi-\frac{t}2\biggr]
\xi\,\widetilde H^q\otimes{C}_2^+\,.
\label{Aq2}
\end{align}
\end{subequations}

The BMP results for longitudinal and transverse helicity-flip amplitudes read
\begin{align}
\mathcal{A}_q^{0\pm}=\frac2{Q}
\biggl\{&
\Bigl(v_\perp^\pm-4P_\perp^\pm \frac{v\cdot q'}{Q^2}
\xi^2\partial_\xi\Bigr)\xi\, M^q\otimes{C}_1^-
\pm\Bigl(a_\perp^\pm-4P_\perp^\pm \frac{a \cdot q'}{Q^2}
\xi^2\partial_\xi\Bigr)\xi\, \widetilde{H}^q\otimes{C}^+_1
\notag\\
&
+P_\perp^\pm\frac{v\cdot P}{m^2}\xi^2\partial_\xi\, E^q\otimes {C}_1^-
\pm P_\perp^\pm\frac{a\cdot \Delta}{2m^2}\xi^2\partial_\xi\,\xi
\,\widetilde E^q\otimes{C}_1^+\biggr\}
\label{A0}
\end{align}
and
\begin{align}
\label{Aflip}
\mathcal{A}_q^{\mp\pm}=
-\frac{8P_\perp^\pm}{Q^2} \biggl\{&\Big(v_\perp^\pm-2P_\perp^\pm\frac{v\cdot q'}{Q^2}\,\xi^2\partial_\xi\Big)\xi^2\partial_\xi
\,M^q\otimes{[x {C}_1^+]}
\pm\Big(a_\perp^\pm-2P_\perp^\pm
\frac{a\cdot q'}{Q^2}\xi^2\partial_\xi\Big)\xi^2\partial_\xi
\,\xi\,\widetilde H^q\otimes{C}_1^+
\notag\\
&
+P_\perp^\pm\frac{v\cdot P}{2m^2}\xi^3\partial_\xi^2
\,E^q\otimes{[x C_1^+]}
\mp P_\perp^\pm\frac{a\cdot \Delta}{4m^2}\xi^3
\partial_\xi^2\xi^2
\widetilde E^q\otimes{C}_1^+
\biggr\},
\end{align}
\end{widetext}
respectively.
The derivatives $\partial_\xi =\partial/\partial\xi$ and $\partial^2_\xi =\partial^2/\partial\xi^2$   act onto the full expression to the right,
i.e.~on both GPDs and coefficient functions, which are  given by the following expressions:
\begin{align}\label{Coefff}
{C}_0^\pm(x,\xi)&=\frac{1}{\xi+x-i\epsilon}\pm\frac{1}{\xi-x-i\epsilon}\,,
\notag\\
{C}^\pm_1(x,\xi)&=\frac{1}{x-\xi}\ln\Big(\frac{\xi+x}{2\xi}-i\epsilon\Big)\pm (x\leftrightarrow -x)\,,
\notag\\
{C}_2^\pm(x,\xi)&=
\biggl\{\frac{1}{\xi+x}\Big[\Li_2\Big(\frac{\xi-x}{2\xi}+i\epsilon\Big)-\Li_2(1)\Big]
\notag\\
&\phantom{=\biggl\{}\pm
(x\leftrightarrow -x)\biggr\}+\frac12 {C}_1^\pm(x,\xi)\,.
\end{align}
Note that ${C}_0^\pm$ have simple poles at $x=\pm \xi$ whereas ${C}_{1,2}^\pm$ have a
milder (logarithmic) singularity at the same points. This ensures that the kinematic power
corrections are factorizable, at least to the leading order in $\alpha_s$. The helicity-conserving amplitudes (\ref{Vq1})~--~(\ref{Aq2}) include leading
contributions $\mathcal{O}(1/Q^0)$ and the corrections $\mathcal{O}(1/Q^2)$, whereas all
terms in Eqs.~(\ref{A0}) and (\ref{Aflip}) are of the order $\mathcal{O}(1/Q)$ and
$\mathcal{O}(1/Q^2)$, respectively, as expected.

\subsubsection{BMP amplitudes in terms of BMJ spinor bilinears }
\label{App:BMP2BMJ-spinors}

The BMP amplitudes (\ref{Anonflip})~--~(\ref{Aflip}) can be expressed in terms of the BMJ  spinor
bilinears (\ref{BMJstructures}). We parameterize these amplitudes in analogy to the CFF decomposition in Eq.~(\ref{CFFs})
\begin{align}
\mathcal{A}_q^{a\pm}= \cffHbmp^q_{a\pm}h+\cffEbmp^q_{a\pm}e
\mp \cfftHbmp^q_{a\pm}\tilde h\mp \cfftEbmp^q_{a\pm}\tilde e\,.
\label{mathcal{A}^q_{apm}}
\end{align}
To find $\cffFbmp_{ab}\in \{\cffHbmp_{ab}, \cffEbmp_{ab},\cfftHbmp_{ab},\cfftEbmp_{ab}\}$
which, as we will explain below, differ from the CFFs  in Eq.~(\ref{cffF}),
one has to  express the BMP bilinear spinors, appearing in (\ref{Anonflip})~--~(\ref{Aflip}), in terms of the BMJ ones (\ref{BMJstructures}).
Note that the notation for particle momenta by BMJ and BMP is different as indicated in (\ref{DVCS-process}) and (\ref{DVCSprocess-BMP}), respectively, i.e.,
we have the correspondence
$$   q \leftrightarrow q_1\,,\quad q' \leftrightarrow q_2\,, \quad p \leftrightarrow p_1\,,\quad p' \leftrightarrow p_2\,. $$
In addition some care is needed since
\begin{align}
  P^{\rm BMJ} &= p_1 + p_2 = 2 P^{\rm BMP} = 2(p+p')\,,
\notag\\
  q^{\rm BMJ} &= \frac12(q_1+q_2) = \frac12 (q+q')^{\rm BMP}\,,
\notag
\end{align}
while $\Delta^{\rm BMJ}=\Delta^{\rm BMP}$ and the same bilinear spinors are used.
Making use of the free Dirac equation for the nucleon states, we find
\begin{align}
   \frac{v\cdot P}{2m^2} &= h - e\,,
&&
   \frac{v \cdot q'}{q\cdot q'} = -\frac{1}{\xi} h\,,
\notag\\
   \frac{a\cdot \Delta}{4m^2} &= -\frac{1}{\xi}\left(\! 1+ \frac{t}{Q^2}\!\right)\tilde e\,,
&&
   \frac{a\cdot q'}{q\cdot q'} = - \frac1{\xi}\tilde h -\frac{1}{\xi} \frac{4m^2}{Q^2}\tilde e \,,
\label{BMJ2BMPspinors1}
\end{align}
and
\begin{align}
  \frac{v_\perp^\pm}{\sqrt{2}}&=
 - |P_\perp|h - \frac{m^2}{|P_\perp|}\biggl[e - \frac{t}{4m^2} h\biggr]
 \mp \frac{m^2}{\xi|P_\perp|}\biggl[\tilde e - \frac{t}{4m^2}\, \tilde h\biggr],
\notag\\
 \frac{a_\perp^\pm}{\sqrt{2}} &=
  -\frac{m^2}{\xi^2 |P_\perp|}\biggl[\tilde e - \frac{t}{4m^2} \tilde h\biggr]
\mp \frac{m^2}{\xi |P_\perp|}\biggl[e - \frac{t}{4m^2} h\biggr].
\label{BMJ2BMPspinors2}
\end{align}
Using these relations and the original BMP results in (\ref{Anonflip})~--~(\ref{Aflip}) we obtain the desired expressions for the BMP CFFs $\cffFbmp_{ab}$
that we rewrite here in a more compact form in terms of charge parity-even GPDs (\ref{gpd-Ceven}) replacing
original BMP coefficients (\ref{Coefff}) by those defined in (\ref{T-coef}) such that
\begin{align}
\label{Coefff-1}
{C}_0^\pm(x,\xi)&=\pm (2\xi)^{-1}\phantom{\xi}\bigl[T_0(u)\pm T_0(1-u)\bigr]\,,
\notag\\
{C}^\pm_1(x,\xi)&=\pm (2\xi)^{-1}\phantom{\xi}\bigl[T_1(u) \pm T_1(1-u)\bigr] \,,
\notag\\
\bigl[x{C}_1^+(x,\xi)\bigr]&=+(2\xi)^{-1} \xi\bigl[T^{(+)}_{1}(u)-T^{(+)}_{1}(1-u)\bigr] \,,
\notag\\
{C}_2^\pm(x,\xi)&= \mp(2\xi)^{-1}\phantom{\xi} \bigl[T_2(u)\pm T_2(1-u)\bigr] \,,
\end{align}
where  $u= (\xi-i\epsilon+x)/2(\xi-i\epsilon)$.
To shorten the notation we use the convolution symbol (\ref{T-convolution}) where the summation over the quark flavors is included.
One obtains
\begin{widetext}
\begin{subequations}
\label{mathfrak++}
\begin{align}
\cffHbmp_{++} &=
  {T}_0\circledast H  + \frac{t}{Q^2}\Big[-\frac{1}{2}{T}_0 + {T}_1
+ 2 \xi  \mathbb{D}_\xi\,{T}_2 \Big]\circledast H
+ \frac{2t}{Q^2} \xi^2 \partial_\xi \xi  {T}_2 \circledast (H+E)\,,
\label{mathfrakH++}\\
\cffEbmp_{++} &=
{T}_0\circledast E  +\frac{t}{Q^2}\Big[-\frac12 {T}_0 + {T}_1
+ 2\xi \mathbb{D}_\xi\,{T}_2\Big]\circledast E
-\frac{8m^2}{Q^2}\xi^2\partial_\xi \xi\, {T}_2\circledast(H+E) \,,
\label{mathfrakE++}\\
\cfftHbmp_{++}&=
 {T}_0 \circledast \widetilde H   + \frac{t}{Q^2}\Big[-\frac{1}{2}{T}_0+  {T}_1
+ 2 \mathbb{D}_\xi\,\xi\, {T}_2  \Big] \circledast \widetilde H
+  \frac{2t}{Q^2}\, \xi \partial_\xi\,{T}_2 \circledast \widetilde H\,,
\\
\cfftEbmp_{++}&=
  {T}_0\circledast\widetilde E + \frac{t}{Q^2}\Big[-\frac12{T}_0 + {T}_1
+ 2\mathbb{D}_\xi\, \xi\, {T}_2\Big]\circledast \widetilde E
-\frac{8m^2}{Q^2}\,\xi \partial_\xi {T}_2 \circledast\widetilde H
\notag\\
&\phantom{={}}+\frac{4m^2}{\Q^2} \left[ {T}_0 + \frac{t}{Q^2}\bigg(- \frac{1}{2} {T}_0
+ {T}_1
+ 2 \mathbb{D}_\xi\,\xi\,{T}_2\bigg)  \right]\circledast\left[\gpdtH + \frac{t}{4m^2}\gpdtE\right]
\label{mathfraktE++}
\end{align}
\end{subequations}
for the helicity-conserved CFFs,
\begin{subequations}
\label{mathfrak0+}
\begin{align}
 \cffHbmp_{0+}&= - \frac{4 |\xi P_\perp|}{\sqrt{2}Q}\Big[
  \xi \partial_\xi {T}_1 \circledast H   + \frac{t}{Q^2} \partial_\xi \xi\,{T}_1 \circledast (H+E) \Big]
- \frac{t}{\sqrt{2}Q |\xi P_\perp|}\xi {T}_1 \circledast \Big[\xi\, (H+E) -\widetilde{H} \Big]\,,
\label{mathfrakH0+}\\
\cffEbmp_{0+}&=
- \frac{4 |\xi P_\perp| }{\sqrt{2}Q}\Big[\xi \partial_\xi {T}_1 \circledast E \Big]
+ \frac{4 m^2}{\sqrt{2}Q|\xi P_\perp|}\xi {T}_1 \circledast \Big[ \xi\, (H+E) - \widetilde{H}\Big]\,,
\\
\cfftHbmp_{0+}&= - \frac{4 |\xi P_\perp|}{\sqrt{2}Q}
\left(1+\frac{t}{Q^2}\right)\Big[\partial_\xi \xi {T}_1 \circledast \widetilde{H} \Big]
+
\frac{t}{\sqrt{2}Q|\xi P_\perp|} {T}_1 \circledast \Big[\xi\, (H+E) - \widetilde{H}\Big]\,,
\\
\cfftEbmp_{0+}&=- \frac{4 |\xi P_\perp|}{\sqrt{2}Q}
\left(1+\frac{t}{Q^2}\right) \Big[ \partial_\xi\xi {T}_1 \circledast \Big(\widetilde{E} +
\frac{4m^2}{Q^2} \widetilde{H} \Big)\Big]
-\frac{4 m^2 }{\sqrt{2}Q |\xi P_\perp|}{T}_1 \circledast \Big[\xi\,(H+E)- \widetilde{H} \Big]
\label{mathfraktE0+}
\end{align}
\end{subequations}
for the longitudinal-to-transverse helicity-flip CFFs, and
\begin{subequations}
\label{mathfrak-+}
\begin{align}
 \cffHbmp_{-+}&=\frac{4|\xi P_\perp|^2 }{Q^2}\Big[
\xi \partial^2_\xi \xi \,T_1^{(+)} \circledast H  + \frac{t}{Q^2} \partial^2_\xi \xi^2\, T_1^{(+)} \circledast (H+E) \Big]
+
\frac{2t}{Q^2} \xi
\Big[\xi\partial_\xi\xi \,T_1^{(+)} \circledast (H+E) + \partial_\xi\,\xi\,T_1 \circledast \widetilde H\Big],
\label{mathfrakH-+}\\
\cffEbmp_{-+}&=\frac{4 |\xi P_\perp|^2}{Q^2}\Big[\xi \partial_\xi^2\xi\,  T_1^{(+)} \circledast E\Big]
-\frac{8m^2}{Q^2}\xi \Big[
\xi\partial_\xi\xi\,T_1^{(+)} \circledast(H+E)
+ \partial_\xi\,\xi\,T_1 \circledast\widetilde H
\Big],
\\
\cfftHbmp_{-+}&= -\frac{4|\xi P_\perp|^2}{Q^2}
 \left(1+\frac{t}{Q^2}\right) \Big[
  \partial_\xi^2 \xi^2 \,{T}_1\circledast\widetilde H
\Big]
- \frac{2t }{Q^2}
\Big[\xi\partial_\xi\xi \,T_1^{(+)} \circledast (H+E) + \partial_\xi\,\xi\,T_1 \circledast \widetilde H\Big],
\\
\cfftEbmp_{-+}&= \frac{4 |\xi P_\perp|^2 }{Q^2}
 \left(1\!+\!\frac{t}{Q^2}\right) \Big[
  \partial^2_\xi\xi^2\, {T}_1 \circledast \Big(\widetilde{E} -\frac{4 m^2 }{Q^2}\widetilde{H}\Big) \Big]
+\frac{8m^2}{Q^2}
\Big[\xi\partial_\xi\xi \,T_1^{(+)} \circledast (H+E) + \partial_\xi\,\xi\,T_1 \circledast \widetilde H\Big].
\label{mathfraktE-+}
\end{align}
\end{subequations}
\end{widetext}
for transverse helicity-flip CFFs. In these expressions twist-five and higher power suppressed contributions ${\cal O}(1/Q^3)$ and ${\cal O}(1/Q^4)$
induced by the rewriting in terms of BMJ spinor bilinears are kept, i.e., they correspond literally to the  BMP result.

\subsection{BMJ  conventions}
\label{App:BMJ-conventions}

BMJ define the DVCS tensor in the notation of (\ref{DVCS-process}) as
\begin{align}
\label{Tmunu}
\lefteqn{T_{\mu\nu}(q_1,q_2,p_1)=} \\
&=
i\int\!\! d^4 x \, {\rm e}^{i(q_1 + q_2) \cdot x/2}
\vev{p_2,\!s_2 | T \{ j_\mu (x/2) j_\nu (- x/2) \} | p_1,\!s_1}.
\nonumber
\end{align}
Setting $z_2=-z_1=1/2$ in the BMP definition (\ref{Amunu-def}), one realizes  that both tensors are consistent,
except that $\mu$ refers to the outgoing photon rather to the incoming one, i.e.,
\begin{align}
 T_{\nu\mu}^{\rm BMJ}(q_1,q_2,p_1) \equiv  \mathcal{A}^{\rm BMP}_{\mu\nu}(q,q',p)\,.
\label{A2T-tensors}
\end{align}
Note that to leading accuracy in $1/Q$ both tensors would look the same without additional interchange
of Lorentz indices  since BMJ, compared to BMP, use the  opposite sign convention
for the Levi-Civita tensor
$$
    \epsilon^{\rm BMJ}_{0123} = -1\,, \quad \mbox{i.e.}, \quad
    \epsilon^{\rm BMJ}_{\alpha\beta\gamma\delta} = -\epsilon^{\rm BMP}_{\alpha\beta\gamma\delta}.
$$

In the BMJ reference frame the nucleon target is at rest, $p_1^\mu=(m,0,0,0)$, and the incoming photon
momentum is specified as
\begin{align}
q_1^\mu=\frac{Q}{\gamma} \bigl(1,0,0,-\sqrt{1+\gamma^2}\bigr)\,,\quad \gamma \equiv \epsilon^{\rm BMJ} = \frac{2 m \xB}{Q}\,.
\end{align}
To avoid confusion with polarization vectors, we denote here the original variable $\epsilon\equiv \epsilon^{\rm BMJ}$ as $\gamma$.
The polarization vectors of the initial photon are defined as
\begin{align}
\epsilon_1^\mu(0)&=\frac{1}{\gamma}\bigl(-\sqrt{1+\gamma^2},0,0,1\bigr),
\notag\\
\epsilon_1^\mu(\pm)&=\frac{e^{\mp i\phi}}{\sqrt{2}}\bigl(0,1,\pm i,0\bigr),
\end{align}
where the phase is given by the azimuthal angle $\phi$ of the final state nucleon.

The essential difference to BMP is that BMJ defines helicity amplitudes
\begin{align}
\label{mathcal{T}^{BMJ}_{ab}}
\mathcal{T}^{\rm BMJ}_{a\pm} &=
 (-1)^{a-1}\epsilon_2^{\nu\ast}(\pm) T_{\nu\mu} \epsilon_1^{\mu}(a)\,,
\end{align}
where $a \in\{\pm 1,0\},$ in the specified target rest frame and, thus, the BMP and BMJ amplitudes differ from each other by $1/Q^2$
suppressed terms.  BMJ define the CFFs (\ref{cffF}) using the parametrization of the helicity amplitudes
of the form
\begin{align}
\mathcal{T}^{\rm BMJ}_{a\pm}  =  \cffH_{a\pm}\,h +  \cffE_{a\pm}\,e \mp   \cfftH_{a\pm}\,\tilde h \mp  \cfftE_{a\pm}\,\tilde e\,,
\label{mathcal{T}^{BMJ}_{ab}-1}
\end{align}
in terms of the bilinear spinors (\ref{BMJstructures}). Let us add that the corresponding sets of BMJ
polarization vectors can be constructed from the four momenta
\begin{widetext}
\begin{align}
\label{eps_1(0)}
\epsilon_1^{\mu}(0)&=
-\frac{1}{{ Q}\sqrt{1 + \gamma^2}}\, q_1^\mu - \frac{2 \xB}{{ Q}\sqrt{1 + \gamma^2}}\, p_1^\mu,
\\
\label{eps_1(1)}
 \epsilon_1^{\mu}(\pm)&= \frac{\sqrt{1 + \gamma^2}}{\sqrt{2}\widetilde K } \left[ \Delta^\mu -
\frac{\gamma^2\left({ Q}^2 -t\right) -2\xB t}{2{ Q}^2\left(1 + \gamma^2\right)} q_1^\mu+
\xB \frac{{ Q}^2  - t + 2 \xB t}{{ Q}^2\left(1 + \gamma^2\right)}
p_1^\mu \right]  \mp  \frac{\xB}{\sqrt{2} \widetilde K} \frac{i\epsilon_{Pq\Delta}^{\phantom{pq\Delta}\mu}}{Q^2}  \,,
\end{align}
for the initial and
\begin{align}
\label{eps_2(1)}
\epsilon_2^{\mu}(\pm)&=  \frac{1+\frac{ \gamma^2}{2} \frac{{Q}^2 + t}{Q^2 + \xB t}}{\sqrt{2}\widetilde K } \left[ \Delta^\mu -
\frac{\gamma^2\left({ Q}^2 -t\right) -2\xB t}{2{ Q}^2\left(1 + \gamma^2\right)} q_1^\mu+
\xB \frac{{ Q}^2  - t + 2 \xB t}{{ Q}^2\left(1 + \gamma^2\right)} p_1^\mu \right]
\nonumber\\
&\phantom{{}={}}
+  \frac{\widetilde{K}}{\sqrt{2}\left(1+\gamma^2\right)\left(Q^2+\xB t\right)}\left[\gamma^2\,q_1^\mu- 2\,\xB\,  p_1^\mu \right]
\mp  \frac{\xB}{\sqrt{2} \widetilde K} \frac{i\epsilon_{Pq\Delta}^{\phantom{pq\Delta}\mu}}{{ Q}^2}
\end{align}
\end{widetext}
for the final state photons. Here, a kinematical variable $\widetilde{K}$ is employed that is related to $|P_\perp|$ in the BMP notation:
\begin{align}
 \widetilde{K} =  \xB\, \frac{Q^2+t}{Q^2} |P_\perp|^{\rm BMP}.
\end{align}
Another representation is \cite{Belitsky:2012ch}:
\begin{align}
\tK =\sqrt{\frac{\xB\left(1-\xB+\frac{\xB \M^2}{\Q^2}\right)(t_{\rm min}-t)(t-t_{\rm max})}{\Q^2}}\,,
\label{tK}
\end{align}
where $t_{\rm min}$ and  $t_{\rm max}$ as function of $\xB$ and $\Q^2$ are given in Eq.~(\ref{t_{min/max}}).

\subsection{Mapping of BMJ  and BMP helicity amplitudes}
\label{App:map-helicity-amplitudes}
In order to  use the BMP results from Sec.~\ref{BMP-notation} for the evaluation of the differential leptoproduction cross section \cite{Belitsky:2012ch},
one needs to express the helicity dependent BMJ CFFs $\cffF_{ab}$ in terms of the BMP CFFs $\cffFbmp_{ab}$  in (\ref{mathfrak++})~--~(\ref{mathfrak-+}).
The relation between the corresponding DVCS tensors (\ref{A2T-tensors}) implies that the BMP and BMJ helicity amplitudes
(\ref{mathcal{T}^{BMJ}_{ab}}) are related as
\begin{align}
\mathcal{T}^{\rm BMJ}_{a\pm} &=
 (-1)^{a-1}\epsilon_1^{\mu}(a) \mathcal{A}^{\rm BMP}_{\mu\nu} \epsilon_2^{\nu\ast}(\pm)\,.
 \label{ABMP2TBMJ}
\end{align}
The BMJ polarization vectors (\ref{eps_1(0)}), (\ref{eps_1(1)}) for the initial state photon $\epsilon_{1,\mu}(a) $
can be written in terms of the BMP polarization vectors $\varepsilon^{0,\pm}_\mu$, cf.~(\ref{varepsilon^a_mu}), as follows:
\begin{align}
\epsilon_{1,\mu}(0)&=
- (1+\varkappa) \varepsilon^0_\mu - \varkappa_0 \Big[\varepsilon^+_\mu+ \varepsilon^-_\mu\Big] \,,
\notag\\
 \epsilon_{1,\mu}(\pm)&= \varepsilon^\mp_\mu +
 \frac{\varkappa}{2}\Big[\varepsilon^+_\mu+ \varepsilon^-_\mu\Big] +  \varkappa_0\, \varepsilon^0_\mu\,,
\end{align}
where the kinematical factors
\begin{align}
\varkappa_0 = \frac{\sqrt{2} Q \widetilde K}{\sqrt{1 + \gamma^2}(Q^2+t)}\,, \quad
\varkappa =\frac{{ Q}^2  - t + 2 \xB t}{\sqrt{1 + \gamma^2}(Q^2+t)}-1
\end{align}
are of order $\varkappa_0 = \mathcal{O}(1/Q)$ and  $\varkappa = \mathcal{O}(1/Q^2)$.
In turn, the BMP (\ref{eps_2(1)}) and BMJ (\ref{varepsilon^a_mu}) polarization vectors for the final state photon coincide up to terms
proportional to $q'_\mu$,
\begin{align}
 \epsilon_{2,\mu}(\mp) = \varepsilon^\pm_\mu + \mathcal{O}(q'_\mu)  \simeq \varepsilon^\pm_\mu \quad
 \left[\text{or }  \epsilon_{2,\mu}^\ast(\pm)\simeq \varepsilon^\pm_\mu\right],
\end{align}
which are irrelevant as they drop out because of current conservation.
Using these expressions and the parametrization of the BMP tensor in (\ref{Amunu}),
we immediately read off Eq.~(\ref{ABMP2TBMJ}) the desired relations
\begin{align}
\label{TArelation}
 \mathcal{T}_{0+} &=-\left(1+\varkappa\right) \mathcal{A}^{0+}+\varkappa_0 \Big[\mathcal{A}^{++} + \mathcal{A}^{-+}\Big],
\nonumber \\
\mathcal{T}_{\pm+} &=
\mathcal{A}^{\pm+} + \frac{\varkappa}{2}\Big[ \mathcal{A}^{++} + \mathcal{A}^{-+}\Big] - \varkappa_0\,  \mathcal{A}^{0+},
\end{align}
and three more similar relations follow from the interchange of the final photon helicity $+\leftrightarrow -$.

Since we use the same expression for the parametrization of helicity amplitudes in terms of bilinear spinors,
compare Eqs.~(\ref{mathcal{A}^q_{apm}}) and (\ref{mathcal{T}^{BMJ}_{ab}-1}),
identical relations hold also between the BMJ and BMP CFFs.
The result is quoted in (\ref{CFF-F2F}).

\section{Double distribution representation for BMP helicity amplitudes}
\label{App:DD}

The studies of GPDs require building theoretical models that satisfy several nontrivial
constraints. In this context the approach based on the so-called double distributions (DDs) representation \cite{Mueller:1998fv,Radyushkin:1997ki}
has several advantages and is receiving a lot of attention, see e.g.~Ref.~\cite{Radyushkin:2013hca}
for a recent discussion. For this reason the expressions for BMP helicity amplitudes  (\ref{VAQ})~--~(\ref{Aflip})
directly in terms of DDs can be of considerable interest for the future data analysis.
Such expressions, in fact, arise naturally at intermediate steps of the calculation.
They have not been given in Ref.~\cite{Braun:2012hq} because of space limitations.
In this Appendix we follow the notation and conventions of Ref.~\cite{Braun:2012hq}, cf. App.~\ref{BMP-notation}.

The representation of GPDs in terms of DDs is not unique. For the present task
the following parametrization of the nucleon matrix element of light-ray vector- and axial-vector operators
turns out to be the most convenient:
\begin{widetext}
\begin{align}\label{DDr}
\vev{p'|\bar q\left(z_1n\right) \slashed{n} q\left(z_2n\right)|p}
&=\int\!\!\!\!\int\!\! d\betaRad d\alphaRad\, e^{i\betaRad P_+z_{12}+i\frac12\Delta_+(z_1+z_2-z_{12}\alphaRad)}
\biggl\{\bar u(p')\slashed{n}u(p) h^q(\betaRad,\alphaRad,t)+
\frac{i\bar u(p')u(p)}{z_{12}m}
\Phi^q(\betaRad,\alphaRad,t)\biggr\}\,,
\notag\\
\vev{p'|\bar q\left(z_1n\right) \slashed{n}\gamma_5 q\left(z_2n\right)|p}
&=\int\!\!\!\!\int\!\! d\betaRad d\alphaRad\, e^{i\betaRad P_+z_{12}+i\frac12\Delta_+(z_1+z_2-z_{12}\alphaRad)}
\biggl\{\bar u(p')\slashed{n}\gamma_5 u(p) \tilde h^q(\betaRad,\alphaRad,t)+
\frac{i\bar u(p')\gamma_5 u(p)}{z_{12}m}
\widetilde \Phi^q(\betaRad,\alphaRad,t)\biggr\}\,.
\end{align}
\end{widetext}
Here, the integration goes over the domain $|\betaRad|+|\alphaRad|\leq 1$,  $n$ is an auxiliary light-like vector, $z_{12} = z_1-z_2$, and the double distribution variables are
related to Radyushkin`s notation as $\betaRad \equiv \beta$ and $\alphaRad \equiv \alpha$ (as in Sect.~\ref{Sec:GPDModel}, $\alpha$ and $\beta$
are commonly used to parameterize the small-$x$ and large-$x$ behavior of PDFs, respectively).

These expressions define four DDs $h^q$, $\widetilde h^q$,  $\Phi^q$, $\widetilde \Phi^q$
in terms of which the `standard' GPDs~\cite{Diehl:2003ny} can be expressed as follows:
\begin{align}
(H^q\!+\!E^q)(x,\xi,t)&=\int\!\!\!\!\int\!\! d\betaRad d\alphaRad\, \delta(x-\betaRad-\xi\alphaRad)\,h^q(\betaRad,\alphaRad,t)\,,
\notag\\
\partial_xE^q(x,\xi,t)&=-\int\!\!\!\!\int\!\! d\betaRad d\alphaRad\,\delta(x-\betaRad-\xi\alphaRad)\,\Phi^q(\betaRad,\alphaRad,t)\,,
\notag\\
\widetilde H^q(x,\xi,t)&=\int\!\!\!\!\int\!\! d\betaRad d\alphaRad\,\delta(x-\betaRad-\xi\alphaRad)\,\tilde h^q(\betaRad,\alphaRad,t)\,,
\notag\\
\partial_x\widetilde E^q(x,\xi,t)&=-\frac1\xi \int\!\!\!\!\int\!\! d\betaRad d\alphaRad\,\delta(x\!-\!\betaRad\!-\!\xi\alphaRad)\,
\widetilde \Phi^q(\betaRad,\alphaRad,t)\,.
\end{align}
The DDs $\Phi,\widetilde \Phi$  can be represented in a somewhat more conventional form as
\begin{align}
\Phi(\betaRad,\alphaRad,t)&=\partial_\betaRad f(\betaRad,\alphaRad,t)+\partial_\alphaRad g(\betaRad,\alphaRad,t)\,,
\notag\\
\widetilde \Phi(\betaRad,\alphaRad,t)&=\partial_\betaRad \tilde f(\betaRad,\alphaRad,t)+\partial_\alphaRad \tilde g(\betaRad,\alphaRad,t).
\end{align}
where $f,g, \tilde f, \tilde g$ are new functions which are also referred to as DDs, that are not defined uniquely.
Using this representation one obtains
\begin{align}\label{EtE}
E&=-\int\!\!\!\!\int\!\! d\betaRad d\alphaRad\, \delta(x-\betaRad-\xi\alphaRad)\,(f+\xi g)\,,
\notag\\
\xi \widetilde E&=-\int\!\!\!\!\int\!\! d\betaRad d\alphaRad\, \delta(x-\betaRad-\xi\alphaRad)\,(\tilde f+\xi \tilde g)\,.
\end{align}
Time reversal invariance implies that all GPDs are even functions of $\xi$~\cite{Diehl:2003ny}.
As a consequence the DDs $h$, $\tilde h$ and $\Phi$ are even functions of $\alphaRad$
and $\widetilde \Phi$ is odd:
\begin{align}
h^q(\betaRad,\alphaRad,t)&=h^{q}(\betaRad,-\alphaRad,t)\,, \qquad
        \Phi^q(\betaRad,\alphaRad)= \Phi^q(\betaRad,-\alphaRad)\,,
\notag\\
\tilde h^q(\betaRad,\alphaRad,t)&=\tilde h^{q}(\betaRad,-\alphaRad,t)\,, \qquad
         \widetilde\Phi^q(\betaRad,\alphaRad)=-\widetilde\Phi^q(\betaRad,-\alphaRad)\,.
\end{align}
Next, only charge conjugation even $C=+1$ combinations of the GPDs can contribute to DVCS. They are
\begin{align}
H^{q^{(+)}}
(x,\xi,t) &= H^q(x,\xi,t) - H^q(-x,\xi,t)
\notag\\
\widetilde H^{q^{(+)}}%
 (x,\xi,t) &= \widetilde H^q(x,\xi,t) + \widetilde H^q(-x,\xi,t)
\end{align}
and similar for $E, \widetilde E$. In the forward limit these distributions are reduced to `singlet' quark parton
distributions $H^{q^{(+)}}
(x,\xi,t) = q(x)+\bar q(x)$, as opposed to $C=-1$ combinations that are related to
`valence' quark densities $H^{q^{(-)}}
(x,\xi,t) = q(x)-\bar q(x)$.

Going over to the DD representation, this means that only the following $C=+1$ combinations can appear:
\begin{align}
  h^q_{-}(\betaRad,\alphaRad,t) &= \frac12\big[h^q(\betaRad,\alphaRad,t) - h^q(-\betaRad,\alphaRad,t)\big],
\notag\\
 \Phi^q_+(\betaRad,\alphaRad,t)&= \frac12\big[\Phi^q(\betaRad,\alphaRad,t)+\Phi^q(-\betaRad,\alphaRad,t)\big],
\notag\\
  \tilde h^q_{+}(\betaRad,\alphaRad,t) &= \frac12\big[\tilde h^q(\betaRad,\alphaRad,t) +\tilde h^q(-\betaRad,\alphaRad,t)\big],
\notag\\
 \widetilde \Phi^q_+(\betaRad,\alphaRad,t)&=\frac12\big[\widetilde\Phi^q(\betaRad,\alphaRad,t)+\widetilde\Phi^q(-\betaRad,\alphaRad,t)\big].
\end{align}
They correspond to matrix elements of the (anti)symmetrized over quark positions combinations of vector- and axial-vector operators,
$O_V(z_1,z_2)-O_V(z_2,z_1)$ and $O_A(z_1,z_2)+O_A(z_2,z_1)$,
that contribute, as well known, to the expansion of the product of electromagnetic  currents.
The subscript `$\pm$' indicates the symmetry behavior under the simultaneous sign change of the both arguments:
$(\betaRad,\alphaRad)\to (-\betaRad,-\alphaRad)$, e.g. $h^q_{-}(\betaRad,\alphaRad,t)=-h^q_{-}(-\betaRad,-\alphaRad,t)$, etc.

The calculation of finite-$t$ and target mass corrections for DVCS for the nucleon follows closely the
procedure explained in~\cite{Braun:2012bg} for the scalar target, but becomes considerably more cumbersome.
To this end it is convenient to define the following variable
\begin{align}\label{def-w}
w=\frac12\left(\frac{\betaRad}\xi+\alphaRad+1\right)\,.
\end{align}
We obtain for the helicity amplitudes (\ref{VAQ})~--~(\ref{Aflip}) in the DD representation:
\begin{widetext}
\begin{align}
\label{AHDD}
\mathbb{V}_1^q&=2\int\!\!\!\!\int\!\! d\betaRad d\alphaRad\,\biggl\{\Phi_{+}^q\,\ln(w-i\epsilon)-\frac{1}{Q^2}
\left[2\M^2\, h_-^q \,\betaRad\partial_w +\Phi_{+}^q\left(|P_\perp|^2\betaRad^2\partial_w-t\left(1+\frac\betaRad\xi-w\right)
\right)\right]
S_+(w)\biggr\},
\notag\\
\mathbb{A}_1^q&=2\int\!\!\!\!\int\!\! d\betaRad d\alphaRad\,\biggl\{\widetilde\Phi_{-}^q\,\ln(w-i\epsilon)-\frac{1}{Q^2}
\left[2\M^2\, \tilde h_+^q \,\left(2+\frac1\xi\betaRad\partial_w\right) +\widetilde\Phi_{-}^q\left(|P_\perp|^2\betaRad^2\partial_w-t\left(1+\frac\betaRad\xi-w\right)
\right)\right]
S_-(w)\biggr\},
\notag\\
\mathbb{V}_2^q&=\int\!\!\!\!\int\!\! d\betaRad d\alphaRad\, h^q_{-}\biggl\{\frac1{w-i\epsilon}-\frac1{Q^2}\biggl[
|P_\perp|^2(\betaRad\partial_w)^2-2t\left(1+\frac{1}\xi\betaRad\partial_w-\frac12\partial_w\big(w-1\big)\right)
\biggr]S_+(w)\biggr\},
\notag\\
\mathbb{A}_2^q&=\int\!\!\!\!\int\!\! d\betaRad d\alphaRad\, \tilde h^q_{+}\biggl\{\frac1{w-i\epsilon}-\frac1{Q^2}\biggl[
|P_\perp|^2(\betaRad\partial_w)^2-2t\left(1+\frac{1}\xi\betaRad\partial_w-\frac12\partial_w\big(w-1\big)\right)
\biggr]S_-(w)\biggr\},
\\[2mm]
\mathcal{A}_q^{0,\pm}&=-\frac2{Q}\int\!\!\!\!\int\!\! d\betaRad d\alphaRad\,\biggl\{
\betaRad P_\perp^\pm\left[\frac{v\cdot P}{\M^2}\Phi^q_+\pm \frac{a\cdot \Delta}{2\M^2} \widetilde \Phi_-^q\right]
-h_-^q\left[v^\pm- \frac{v\cdot q' }{q\cdot q'}P_\perp^\pm \,\betaRad\partial_w \right]
\mp\tilde h_+^q\left[a^\pm- \frac{a\cdot q' }{q\cdot q'}P_\perp^\pm\, \betaRad\partial_w \right]
\biggr\}\frac{\ln(w\!-\!i\epsilon)}{w-1},
\\
\mathcal{A}_q^{\mp\pm}&=-\frac{\sqrt{2}|P_\perp|}{Q^2}\int\!\!\!\!\int\!\! d\betaRad d\alphaRad\,\biggl\{\left[
-\betaRad \Phi_+^q\frac{v\cdot P}{\M^2}P_\perp^\pm+2 h_-^q\left(v^\pm-\frac{v\cdot q'}{2q\cdot q'}P_\perp^\pm\betaRad\partial_w\right)
\right]\betaRad\partial_w\,\frac{2w-1}{w-1}\ln(w-i\epsilon)
\notag\\
&\phantom{=-\frac{\sqrt{2}|P_\perp|}{Q^2}\int\!\!\!\!\int\!\! d\betaRad d\alphaRad}
\pm
\left[\betaRad \widetilde \Phi_-^q\frac{a\cdot\Delta}{2\M^2}P_\perp^\pm+
2 \tilde h_+^q\left(a^\pm-\frac{a\cdot q'}{2q\cdot q'}P_\perp^\pm\betaRad\partial_w\right)
\right]\betaRad\partial_w\,\frac{1}{w-1}\ln(w-i\epsilon)
\biggr\},
\label{AflipHDD}
\end{align}
\end{widetext}
where
\begin{align}
S_\pm(w)&=\frac1{w-1}\biggl[\frac12\ln (w-i\epsilon)\pm\Bigl(\Li_2(w+i\epsilon)-\Li_2(1)\Bigr)\biggr].
\end{align}
Note that the leading-twist coefficient functions $\sim 1/(w-i\epsilon)$ and $\sim \ln (w-i\epsilon)$ in the helicity-conserving
amplitudes $\mathbb{V}_k$ and  $\mathbb{A}_k$ have singularities at $w=0$.
The twist-four contributions  $\sim S^\pm(w)$ have singularities at $w=0$ as well and in addition
the logarithmic branching point at $w=1$ due to $\Li_2(w+i\epsilon)$.
Since
$$
w(-\betaRad,-\alphaRad)=1-w(\betaRad,\alphaRad)
$$
and thanks to symmetry properties of the DDs under the transformation $(\betaRad,\alphaRad)\to(-\betaRad,-\alphaRad)$,
the two points $w=0$ and $w=1$ are, however, equivalent. It is possible to rewrite the results in Eq.~(\ref{AHDD})
in the form where all singularities are at the point $w=0$ only.

It can be shown that the twist-four contributions (\ref{AHDD})~--~(\ref{AflipHDD}) are well defined (finite),
provided the integrals for the leading-twist contributions converge. The danger is that
derivatives with respect to $w$ might produce stronger singularities as compared to the leading terms.
Notice that these derivatives are always accompanied by the prefactor $\betaRad$. Using
Eq.~(\ref{def-w}) one can rewrite $\betaRad\partial_w$ in terms of the derivative with respect to the asymmetry parameter $\xi$:
$$
\betaRad\partial_w f(w)=-2\xi^2\partial_\xi f(w)\,,
$$
and move all $\xi$-derivatives out of the integral~\footnote{
This trick is not necessary for the term $\partial_w(w-1) S_\pm(w)$ which, as can easily be seen,
has a pole $\sim 1/w$ and a logarithmic singularity at $w=1$. Both are integrable and do not cause problems.}.
In this way one sees that the singularities of higher-twist coefficient functions are not enhanced
as compared to the leading twist ones.
The $\betaRad,\alphaRad$-integrals converge and define smooth functions of $\xi$ (away from $\xi=0$).

In order to recast the results in the DD representation, Eqs.~(\ref{AHDD})--(\ref{AflipHDD}), in terms of GPDs one can rewrite, e.g.
\begin{align}
\int\!\!\!\!\int\!\! d\betaRad d\alphaRad\, \Phi(\alphaRad,\betaRad) \betaRad F(w)
= \xi^2\partial_\xi\frac1\xi \int\!\!\!\!\int\!\! d\betaRad d\alphaRad\, (f+\xi g) F(w)\,.
\end{align}
Inserting
$$
1=\int_{-1}^{1}\!dx\,\delta(x-\betaRad-\xi\alphaRad)
$$
under the $\betaRad,\alphaRad$-integral and changing the order of integrations one arrives after
some algebra to the expressions in Eqs.~(\ref{VAQ})--(\ref{Aflip}) of Ref.~\cite{Braun:2012hq}.

We add that in standard GPD models, used in phenomenology, the original DD distribution representation \cite{Mueller:1998fv,Radyushkin:1997ki}
\begin{align}
F(x,\xi,t) = \int\!\!\!\!\int\!\! d\betaRad d\alphaRad\, \delta(x-y-\xi z) f(y,z,t)
\label{DD-F}
\end{align}
for $F\in\{\gpdH,\gpdE,\gpdtH, \gpdtE\}$ is employed, where $f\in \{h,e,\widetilde{h},\widetilde{e}\}$ denote
the corresponding DDs. Plugging  this representation (\ref{DD-F}) into  convolution formulae as they appear in the kinematic twist corrections
(\ref{mathfrak++})~--~(\ref{mathfrak-+}) (or in perturbative higher order corrections), they can be simply translated into the `standard' DD representation
by means of the equality
\begin{align}
\int_{-1}^1\!\! \frac{dx}{2\xi}
&(\xi \partial_\xi)^p T_i\!\!\left(\!\frac{\xi+x- i\epsilon}{2(\xi-i\epsilon\!)}\right)\gpdF^{q^{(+)}}(x,\xi,t) =
\notag\\
&=\frac{1}{2\xi}\int\!\!\!\!\int\!\! d\betaRad d\alphaRad\, \left[(-y\partial_y)^p \C_{i}(w)\right]  f^{q^{(+)}}(y,z,t)\,,
\label{DD-F-conv}
 \end{align}
where $w(y,z)$ is defined in Eq.~(\ref{def-w}) and $f^{q^{(+)}}(y,z,t)$ are charge parity even DD functions with the symmetry properties
spelled out above. Note that the homogeneous differential operator $(\xi \partial_\xi)^p$ acts in (\ref{mathfrak++})~--~(\ref{mathfrak-+}) also on
the integral measure and that some care is needed with respect to the imaginary parts of the coefficient functions, which is inherited from the $\xi-i\epsilon$ prescription, and translates in our notation (\ref{T-coef}) into $T_i(w+i\epsilon)$.

Finally, we add that the DD-representation (\ref{DD-F}) is not complete
for GPD $H$ or $E$, however, it is complete for $H+E$. To fix this, a so-called $D$-term, which we write here as
\begin{align}
D^q(x,\xi,t) = \theta(|x|\le|\xi|)\,{\rm sign}(\xi)\, \varphi_D^q\!\!\left(\!\frac{\xi+x}{2\xi\!},t\!\right),
\end{align}
is added or subtracted to the DD-representation (\ref{DD-F}),
$$
H^q \to H^q +D^q\,,\quad E^q \to E^q - D^q\,.
$$
This term is antisymmetric in $x$, i.e., $\varphi_D(u,t)= -\varphi_D(1-u,t)$. In the similar manner the pion-pole contribution,
appearing in GPD $\gpdtE$, is modeled as \cite{Mankiewicz:1998kg,Frankfurt:1999xe}
\begin{align}
\gpdtE_\pi^q(x,\xi,t) = \theta(|x|\le |\xi|)\, \frac{1}{|\xi|}\varphi^q_\pi\!\!\left(\!\frac{\xi+x}{2\xi\!},t\!\right),
\end{align}
which is symmetric in $x$, i.e., $\varphi_\pi(u,t)= \varphi_\pi(1-u,t)$. In our convolution formulae the integrals read
\begin{align}
T_i\circledast D &= 2 \sum_{q} e_q^2 \int_0^1\!du\, T_i(u) \varphi^q_D(u,t)\,,
\notag\\
T_i\circledast \gpdtE_\pi &= \frac{2}{\xi} \sum_{q} e_q^2 \int_0^1\!du\, T_i(u) \varphi^q_D(u,t)\,.
\label{Dpi-term}
\end{align}
Consequently, the $D$-term and pion-pole contribution are annihilated by the differential operators $\partial_\xi$ and $\partial_\xi \xi$ respectively.

\section{Analyticity}
\label{App:analyticity}

In this appendix we show that the convolution formulae for BMP CFFs, given in Eqs.~(\ref{mathfrak++})~--~(\ref{mathfrak-+}),
can easily be converted into dispersion relation (DR) integrals.
Such a dispersion representation is interesting in its own right and can be used in practice to evaluate CFFs
numerically starting from a given GPD model.
Without loss of generality, to simplify the derivation we employ here the DD-representation (\ref{DD-F}) together with its $D$-term and
pion-pole addenda.

First we demonstrate that the convolution integrals satisfy the DR
\begin{align}
&\int_{-1}^1\!\! \frac{dx}{2\xi}T_i\!\!\left(\!\frac{\xi+x- i\epsilon}{2(\xi-i\epsilon\!)}\right)\gpdF^{q^{(+)}}(x,\xi,t)
\notag\\
&\;\;=\int_{0}^1\!dx\, \frac{x+\xi + \sigma(x-\xi)}{\xi^2-x^2-i\epsilon} \int_x^1\!\frac{dr}{r}\,t_i(r)\gpdF^{q^{(+)}}(x/r,x,t)\,.
\label{DR-1}
 \end{align}
This is evident for the imaginary part, where we can equate
$$\Im{\rm m} \frac{x+\xi + \sigma(x-\xi)}{\xi^2-x^2-i\epsilon}
= \pi \delta(\xi-x)\;\;\mbox{for}\;\; x\ge 0\,,\;\; \xi\ge 0, $$
and the imaginary part on the l.h.s.~is thus by definition  equal to the r.h.s., see Eq.~(\ref{F^q(xi/x,xi)-smallxi}).
To show that (\ref{DR-1}) holds true for the real part, we first exploit the symmetry,
$$\gpdF^{q^{(+)}}(-x/r,-x,t) = -\sigma \gpdF^{q^{(+)}}(x/r,x,t),$$
to rewrite (\ref{DR-1}) in the following form
\begin{align}\int_{-1}^1\!\! \frac{dx}{2\xi}
&T_i\!\!\left(\!\frac{\xi+x- i\epsilon}{2(\xi-i\epsilon\!)}\right)\gpdF^{q^{(+)}}(x,\xi,t)
\notag\\
&=\int_{-1}^1\!dx\, \frac{1}{\xi-x-i\epsilon} \int_{|x|}^1\!\frac{dr}{r}\,t_i(r)\gpdF^{q^{(+)}}(x/r,x,t)\,.
\label{DR-2}
 \end{align}
Next we plug the DD-representation (\ref{DD-F}) into Eq.~(\ref{DR-2}), where the l.h.s.~is
given by Eq.~(\ref{DD-F-conv}) with $p=0$, and the r.h.s.~reads after integration over $x$
as follows,
\begin{align}
&\int_{-1}^1\!dx\, \frac{1}{\xi-x-i\epsilon} \int_{|x|}^1\!\frac{dr}{r}\,t_i(r)\gpdF^{q^{(+)}}(x/r,x,t)
\notag\\
&=\int\!\!\!\!\int\!\! d\betaRad d\alphaRad\, \int_{0}^1\! dr
\frac{\theta(1-|y|-r z)}{\xi(1-r z)-y r-i\epsilon}
t_i(r)
f^{q^{(+)}}(y,z,t)\,,
\label{DR-3}
\end{align}
where the $\theta$-function does not imply any further restrictions.
Employing the definition $t_i(r)= \Im{\rm m} T_i((1+r)/2r)/(2\pi r)$, the $r$-integral can be
written after the transformation of variables $u=(1+r)/2r$ in form of a DR integral
\begin{align}
\int_{0}^1\! dr \frac{t_i(r)}{\xi(1-r z)-y r-i\epsilon}
&
=
\frac{1}{2\pi\xi}\int_{1}^\infty\!du\,\frac{\Im{\rm m} T_i(u)}{u-w-i\epsilon}
\notag\\
&
= \frac{1}{2\xi}  T_i(w+i\epsilon)
\,,
\label{DR-4}
\end{align}
where $w(y/\xi,z)$ is specified in Eq.~(\ref{def-w}) and we defined, without loss of generality, the coefficient functions in
such a manner that they have only cuts on the real axis for $u\ge 1$. Thus, this DR integral yields the
functions  $T_i(w+ i \epsilon)$.
Plugging Eq.~\eqref{DR-3} into Eq.~\eqref{DR-4} and using the DD convolution formula~\eqref{DD-F-conv} for $p=0$ establishes
Eqs.~\eqref{DR-2} and~\eqref{DR-1}.
We add that the kernel $T_1^{(+)}(u)$
needs a subtraction yielding an ambiguous term $c/\xi$ that, as we have discussed in Sec.~\ref{sec:ColliderKinematics},
is at the end irrelevant.

Second, we explicitly show that the action of differential operators, appearing in the convolution formulae (\ref{mathfrak++})~--~(\ref{mathfrak-+}),
acting on both the real and imaginary part is compatible with the DR-representation.
The application of the homogeneous differential operator $\xi^{\frac{1+\sigma'}{2}}\partial_\xi \xi^{\frac{1-\sigma'}{2}}$ with $\sigma'=\pm 1$ on the Cauchy integral
kernel reads after partial integration
\begin{align}
&\xi^{\frac{1+\sigma'}{2}}\partial_\xi \xi^{\frac{1-\sigma'}{2}} \int_0^1 \frac{x+\xi + \sigma(x-\xi)}{\xi^2-x^2-i\epsilon} \tau(x)
\notag\\
& =
 \int_0^1 \frac{x+\xi + \sigma(x-\xi)}{\xi^2-x^2-i\epsilon} x^{\frac{1+\sigma'}{2}}\partial_x x^{\frac{1-\sigma'}{2}}  \tau(x),
\end{align}
where we assumed that the test function $\tau(x)$ vanishes at $x=1$ and that $x^{(1+\sigma)/2}\,x\tau(x)$
vanishes at $x=0$. With the same assumptions for the small-$x$ behavior and supposing that $\partial_x\tau(x)$ vanishes at $x=1$,
we can reshuffle the relevant homogeneous differential operator of second order $\xi^{\frac{1+\sigma'}{2}}\partial^2_\xi \xi^{\frac{3-\sigma'}{2}}$ as well.
Finally note that the differential operator (\ref{mathbbDxi}) can be written as
\begin{align}
& 2\mathbb{D}_\xi =
\left[2\partial_\xi-\partial^2_\xi \xi \right]  + \xi^2 \left[1- \frac{4 m^2}{t}\right]    \partial^2_\xi \xi
\end{align}
and the terms proportional to $\xi^2$ can be treated algebraically,
\begin{align}
& \xi^2\, \frac{x+\xi + \sigma(x-\xi)}{\xi^2-x^2-i\epsilon} =  x^2\,  \frac{x+\xi + \sigma(x-\xi)}{\xi^2-x^2-i\epsilon}
\notag\\
&\phantom{\xi^2\, \frac{x+\xi + \sigma(x-\xi)}{\xi^2-x^2-i\epsilon} =} + x(1+\sigma)+\xi(1-\sigma)\,.
\end{align}
This allows us to establish the equality
\begin{align}
&\xi^{\frac{1+\sigma}{2}}2 \mathbb{D}_\xi \xi^{\frac{1-\sigma}{2}} \int_0^1\!dx\, \frac{x+\xi + \sigma(x-\xi)}{\xi^2-x^2-i\epsilon} \tau(x)
\notag\\
&=
\int_0^1\!dx\, \frac{x+\xi + \sigma(x-\xi)}{\xi^2-x^2-i\epsilon} x^{\frac{1+\sigma}{2}} 2\mathbb{D}_x x^{\frac{1-\sigma}{2}}  \tau(x)
\notag\\
&\phantom{=}\;+ 2(1+\sigma)\left[1- \frac{4 m^2}{t}\right] \int_0^1\!dx\, x\, \tau(x) ,
\label{DR-C1}
\end{align}
where for even-signature an additional `subtraction' term appears. In a similar manner, the last term in Eqs.~(\ref{mathfrakH++})
and (\ref{mathfrakE++})  for  $\cffHbmp_{++}$ and $\cffEbmp_{++}$ can be rewritten as, respectively,
\begin{align}
&\xi^2 \partial_\xi\xi\,\int_0^1\!dx\frac{2x\, \tau(x)}{\xi^2 -x^2-i\epsilon}
\notag\\
&\quad=\int_0^1\!dx\frac{2x\, x^2\partial_x x\,\tau(x)}{\xi^2 -x^2-i\epsilon}-2 \int_0^1\!dx \tau(x)\,.
\label{DR-C2}
\end{align}
Note that the additional constants in Eqs.~(\ref{DR-C1}) and (\ref{DR-C2}) cancel each other in the `magnetic' combination $\cffHbmp_{++} + \cffEbmp_{++}$.

Since convolution integrals can be converted into DR-integrals, and the application of differential operators can be reshuffled from
the real to the imaginary parts up to a possible constant, we can rewrite the BMP convolution formulae (\ref{mathfrak++})~--~(\ref{mathfrak-+}) as DRs.
Thereby, the existence of DR-integrals is ensured if we require that the GPDs for $x=\xi$ vanish in the limit $\xi\to 1$,
i.e.~convolution integrals for the imaginary parts are suppressed by one additional power $(1-\xi)$,  and that GPDs possess the common `Regge' behavior,
see discussions in Secs.~\ref{Sec:GPDModel} and \ref{sec:ColliderKinematics}.  The subtraction constants are  calculated by means of Eq.~(\ref{Dpi-term})
and additional contributions, calculated from the imaginary parts, can only appear in $\cffHbmp_{++}$ and $\cffEbmp_{++}$ in form of a $D$-term addition.
For the helicity conserving CFFs (\ref{mathfrak++}) we find the DR-integral
\begin{widetext}
\begin{align}
\cffFbmp_{++}(\xi,t,\Q^2)  = \frac{1}{\pi}\int_0^1 \frac{x+\xi + \sigma(x-\xi)}{\xi^2-x^2-i\epsilon} \Im{\rm m} \cffFbmp_{++}(x,t,\Q^2) +
\left(\delta_{\cffFbmp\cffHbmp}-\delta_{\cffFbmp\cffEbmp}\right) \mathfrak{D}_{++}(t,\Q^2) + \frac{1}{\xi}\delta_{\cffFbmp\cfftEbmp}\, \mathfrak{P}_{++}(t,\Q^2)\,,
\end{align}
where the imaginary part is calculated from (\ref{mathfrak++}), e.g., by means of the convolution integral (\ref{Im-conv}), and
the non-vanishing subtraction constants read
\begin{align}
\mathfrak{D}_{++}(t,\Q^2) =& 2 \int_0^1\!\frac{du}{1-u} \left\{1-\frac{t}{2Q^2} \left(1-2\ln u\right) \right\}\sum_q e^2_q\varphi_D^q(u,t)
\notag\\
&-4 \int_0^1\!d\xi\, \xi\, \int_\xi^1\!\frac{dx}{x}\, t_2(x) \sum_q e^2_q\left[\frac{4m^2}{\Q^2} H^{q^{(+)}}+ \frac{t}{\Q^2} E^{q^{(+)}}\right](\xi/x,\xi,t)\,,
\label{mathfrak{D}_{++}}\\
\mathfrak{P}_{++}(t,\Q^2) =& 2\Big(1+\frac{t}{\Q^2}\Big)\int_0^1\!\frac{du}{1-u}
\left\{1-\frac{t}{2\Q^2} - \frac{t}{Q^2} \ln u \right\}
\sum_q e^2_q\varphi_\pi^q(u,t)\,,
\end{align}
\end{widetext}
where the coefficient function $t_2(x)$ is defined in Eq.~(\ref{t2-coef}).

For `electric' helicity flip CFFs, defined in Eq.~(\ref{electric-magnetic}),
and for the even-signature  CFF combinations (\ref{constraint-2}) one immediately realizes from the explicit expressions in (\ref{mathfrak0+}) and (\ref{mathfrak-+})
that the kinematical factors can be stripped off,
\begin{align}
\frac{\cffGbmp_{0+}}{|\xi P_\perp|{\phantom{^2}}}\,,  \;  \frac{\cffEbmp_{0+}+\xi \cfftEbmp_{0+}}{|\xi P_\perp|}{\phantom{^2}}\,  (\sigma=+1)\,,
&\quad
\frac{\cfftGbmp_{0+}}{|\xi P_\perp|}\,  (\sigma=-1) \,,
\notag\\
\frac{\cffGbmp_{-+}}{|\xi P_\perp|^2}\,,\; \frac{\cffEbmp_{-+}+\xi \cfftEbmp_{-+}}{|\xi P_\perp|^2} \,  (\sigma=+1)\,,
&\quad
\frac{\cffGbmp_{-+}}{|\xi P_\perp|^2} \,  (\sigma=-1) \,,
\notag
\end{align}
and that  such CFFs satisfy unsubtracted DRs. It is evident from the BMP results, quoted in Eqs.~(\ref{mathfrak0+}) and (\ref{mathfrak-+}), and the equality
$t+\xi^2 (4 m^2- t)= -4|\xi P_\perp|^2$, cf.~Eq.~(\ref{xiPperp}), that two further  helicity flip CFF combinations,
\begin{align}
\frac{\cfftHbmp_{0+}+\xi (\cffHbmp_{0+} + \cffEbmp_{0+})}{|\xi P_\perp|}\,, \;
\frac{\cfftHbmp_{-+}-\xi (\cffHbmp_{-+} + \cffEbmp_{-+})}{|\xi P_\perp|^2} \,   (\sigma=-1)\,,
\notag
\end{align}
exist that are free of kinematical singularities. It can be easily shown that they are independent from the above quoted combinations
and satisfy  unsubtracted signature-odd DRs.
  We emphasize that the $D$-term and pion-pole contribution drop out in all
longitudinal and transverse flip BMP  CFFs (\ref{mathfrak0+}) and (\ref{mathfrak-+}), i.e., in all terms proportional to
$|\xi P_\perp|$ or $|\xi P_\perp|^2$  as well as in the addenda. We also note that the two additional terms in the third line of Eqs.~(\ref{cffFbmp_{0+}}) and (\ref{cffFbmp_{-+}}),
satisfy unsubtracted  odd-signature DRs which can be converted into subtracted even-signature DRs
(after multiplication with a factor $\xi$), where the subtraction constant is calculated from the  `magnetic' GPD $H+E$.

We finally add that for a scalar target only three CFFs appear.
For the twist-four results, given in Eqs.~(120) and (121) of Ref.~\cite{Braun:2012bg}, one can immediately show that BMP helicity amplitudes satisfy DRs
in which the kinematical factors are removed.
In the BMP basis a $D$-term induced subtraction constant (modified by the imaginary part, cf.~Eq.~(\ref{mathfrak{D}_{++}}))
{\em only} appears for the  helicity conserved CFF.
After a transformation to another CFF basis, e.g., the BMJ basis (\ref{CFF-F2F}), this subtraction constant propagates, however, to the
DRs for helicity flip CFFs, as emphasized  in Ref.~\cite{Moiseeva:2008qd}.

\begin{table}[h]
\renewcommand{\arraystretch}{1.2}
\vspace{4mm}
\begin{tabular}{|l|c|c|}
  \hline
  $\C_0(u)$\phantom{\bigg]}& $\dfrac{1}{\bu}$ &  1 \\
  $\C_{1}^{(+)}(u)\phantom{\bigg]}$ &  $\dfrac{(\bu-u)\ln\bu}{u}$  & $\dfrac{(j+1)_2+2}{(j)_4}$ \\
   $\C_{1}^{(-)}(u)\phantom{\bigg]}$ &  $-\dfrac{\ln\bu}{u}$  &$\dfrac{1}{(j+1)_2}$ \\
  $\C_{2}(u)\phantom{\bigg]}$ & $\dfrac{{\rm Li}_2(1)-{\rm Li}_2(u)}{\bu} + \dfrac{\ln\bu}{2u}$  & $\dfrac{(j+1)_2+2}{2[(j+1)_2]^2}$ \\
  \hline
\end{tabular}
\caption{Coefficient functions (first row), their expressions in $u$ variable (second row), and as conformal moments (third row).
Note that $\C_{1}\equiv\C_{1}^{(-)}$.}
\label{tab:C}
\renewcommand{\arraystretch}{1.0}
\end{table}
\section{Conformal moments of coefficient functions}
\label{App:confmom}

To implement the kinematical corrections in an existing GPD fitting code \cite{Kumericki:2007sa,Kumericki:2009uq},
which is based on a Mellin-Barnes integral representation, the conformal moments of the coefficients (\ref{T-coef}) are needed.
For non-negative integer $n$ the conformal partial waves are restricted to the region $0\le u\le 1$ and
are given in terms of Gegenbauer polynomials with index $3/2$, normalized here as
\begin{align}
\widehat{p}_n(u)= 2 u\bar{u}\, C_{n}^{\frac{3}{2}}(u-\bar{u})
\mbox{ \ with \ }
\widehat{p}_n(\bar{u}) = (-1)^n  \widehat{p}_n(u)\,,
\end{align}
where $\bu\equiv 1-u$.
The conformal moments are calculated by the convolutions
\begin{subequations}
\begin{align}
&\int_0^1\!du\, \frac{1}{\bar{u}}\,\widehat{p}_n(u) =  1\,,
\\
&
\int_0^1\!du\, \ln \bar{u}\;\;  \widehat{p}_n(u)  =  - \frac{(n+1)_2}{(n)_4}\,,
\\
&
\int_0^1\!du\,  \frac{\ln \bar{u}}{u}\,\widehat{p}_n(u)  =   -\frac{1}{(n+1)_2}\,,
\\
& \int_0^1\!du\,  \frac{\Li_2(1)-\Li_2(u)}{\bar{u}}\,\widehat{p}_n(u)  = \frac{(n+1)_2+1}{[(n+1)_2]^2}\,,
\end{align}
\end{subequations}
where $(n)_a= n\cdots (n+a-1)$ denotes the Pochhammer symbol.
The transformation $\bar{u} \to u$ in the coefficients corresponds to a multiplication with the factor $(-1)^n$.
For complex-valued conformal moments this sign alternating factor is replaced by $-\sigma$ with the signature factor $\sigma$.
The conformal moments of the coefficients (\ref{T-coef}) and the auxiliary ones are listed in Tab.~\ref{tab:C}. This table allows one
to translate easily the twist corrections (\ref{mathfrak++})--(\ref{mathfrak-+}) into the space of conformal moments.
We add that the derivatives w.r.t.~$\xi$ in the expressions (\ref{mathfrak++})--(\ref{mathfrak-+}) act on the integrand in the Mellin-Barnes integrals,
i.e., on $\xi^{-j-1} f_j(\xi,t,\Q_0^2)$, where $f_j(\xi,t,\Q_0^2)$ are conformal GPD moments.
For integer $j=n$ they are given by polynomials in $\xi^2$ of order $(n\pm 1)/2$ [for signature-even $n\in \{1,3,\cdots\}$] or
$n/2$ [for signature-odd $n\in \{0,2,\cdots\}$], respectively.
Finally note that a transformation of BMP CFFs to the basis employed
in the code used to calculate DVCS observables,  e.g., to the BMJ basis given in (\ref{CFF-F2F}), is needed.


\end{document}